\documentclass[10pt]{article}
\usepackage{array}
\usepackage{faktor}
\usepackage{wrapfig, amsmath}
\usepackage{amsfonts}
\usepackage{amssymb}
\usepackage{bm}
\usepackage{upgreek}

\usepackage{graphicx}
\usepackage{epsfig,cancel,amsthm, amssymb, todonotes, amsmath}
\usepackage{color,tikz}
\usetikzlibrary{decorations.markings,arrows, calc}
\usepackage{color}
\usepackage{graphicx,framed,verbatim,caption,amsbsy}
\usepackage[colorlinks=true, pdfstartview=FitV, linkcolor=darkblue, citecolor=darkblue, urlcolor=darkblue]{hyperref}
\usepackage{enumitem}
\usepackage[rflt]{floatflt}
\usepackage{float}

\definecolor{shadecolor}{rgb}{0.9, 0.9, 0.86}
\definecolor{darkgreen}{rgb}{0.2, 0.5,  0}
\definecolor{darkblue}{rgb}{0.1,0.1,0.45}

\def\&{\vspace{-5pt}&}

\def\e{{\rm e}}
\def\dd{{\rm d}}
\newcommand\norm[1]{\left\lVert#1\right\rVert}

\usepackage{tikz}
\usepackage{pgf}
\usepackage{pgflibraryarrows}
\usepackage{pgflibrarysnakes}
\usetikzlibrary{decorations.text}
\usepgfmodule{shapes}
\usetikzlibrary{decorations.pathmorphing}
\usetikzlibrary{decorations.markings}
\usetikzlibrary{patterns}
\usetikzlibrary{automata}
\usetikzlibrary{positioning}

\tikzset{->-/.style={decoration={
 markings,
 mark=at position #1 with {\arrow{>}}},postaction={decorate}}}

\def \eqref#1{(\ref{#1})}
\def \& {&\hspace{-10pt}}
\def\Ai{ {\mathrm {Ai}}}
\def\E{ {\mathrm {E}}}

\renewcommand{\d}{\mathrm d}
\newtheorem{theorem}{Theorem}[section]

\newtheorem{lemma}[theorem]{Lemma}
\newtheorem{remark}[theorem]{Remark}
\newtheorem{problem}[theorem]{Riemann--Hilbert Problem}

\newtheorem{proposition}[theorem]{Proposition} 
\newtheorem{corollary}[theorem]{Corollary}

\newtheorem{assumption}[theorem]{Assumption}

\def\d{{\rm d}}
\def\M{{\rm M}}
\def\g{{\rm g}}
\def\G{{\rm G}}
\def\S{{\rm S}}
\def\P{{\rm P}}
\def\T{{\rm T}}
\def\R{{\rm R}}
\def\Y{{\rm Y}}
\def\I{{\rm I}}
\def\q{{\rm g}}
\def\L{{\rm L}}
\def\h{{\rm h}}
\def\W{{\rm W}}
\def\X{{\rm X}}

\def\i{\mathrm i}
\DeclareMathOperator{\im}{Im}

\DeclareMathOperator{\supp}{supp}
\DeclareMathOperator{\re}{Re}

\makeatletter
\@addtoreset{equation}{section}
\makeatother

 \usepackage{authblk}

\date{}                     

\title{Asymptotic expansion for multiplicative statistics in a Hermitian matrix model connected to the lower tail of the KPZ equation}
\author[1]{Carla Mariana da Silva Pinheiro}
\affil[1]{\textit{Instituto de Ciências Matemáticas e de Computação -  Universidade de São Paulo;} \texttt{carla.pinheiro@usp.br}}
\begin{document}
\maketitle

\begin{abstract}
We explore the multiplicative statistics for a unitary random matrix ensemble with a parameter-dependent deformation inserted in the probability measure. Such deformations had been studied for a bounded or decaying parameter. In the present work, we extend the previous results for a growing parameter under a controlled rate, and show that the underlying statistics relate to the lower tail study for the KPZ equation.
\end{abstract}

\tableofcontents

\newpage

\section{Introduction and statement of the results}

\label{sec:intro}

Eigenvalues from random matrix models are known to be the key to the understanding of complicated abstract objects, such as fermions \cite{DMS2018}, Coulomb-gas and big-data processes \cite{mehta2004}. But they also model everyday life elements such as coffee stains \cite{takeuchi2011} and buses timetables. The wide range of applications led to a big interest in the study of statistics for eigenvalues in random matrix models and, in  particular, of unitary Hermitian ensembles. In order to build an unitary ensemble, one starts with the space $\mathcal{H}_n$ of $n \times n$ Hermitian matrices. This means that the eigenvalues $\{\lambda_j\}_{j \geq 1}$ are real and distinct. This space can be equipped with a probability measure
$$\upmu_n(H) \dd H = \frac{1}{\bar{Z}_n} \e^{-nV(H)} \dd H,
$$
for a certain potential $V$ and where $\dd H$ is a Lebesgue measure. For now, it is enough to assume that $V$ is such that the integral over $\mathcal{H}_n$ converges. More assumptions will be discussed in Section \ref{sec:eqmeas}. In unitary ensembles the eigenvalues form a determinantal point process and the Weyl formulas \cite{guionnet2009} allow us to recover the density on the space of eigenvalues
\begin{equation}
p_n(\lambda) = \frac{1}{Z_n} \prod_{1 \leq k <j \leq n} (\lambda_j-\lambda_k)^2 \prod_{j=1}^n \e^{-n V(\lambda_j)},
\label{eq:measure2}
\end{equation}
where $Z_n$ is the normalization constant, also known as \textit{partition function}. In particular, the relevant statistics are encoded in a very elegant way by the Christoffel–Darboux kernel $K_n(\lambda_i, \lambda_j)$ of orthogonal polynomials (see Equation \eqref{eq:defCD}), in the sense that the $N$-th correlation point function is given by
$$\rho_N(\lambda_1, \cdots, \lambda_N) = \det (K_n(\lambda_i, \lambda_j))_{i, j = 1}^N.
$$

On the other side, integrable kernels are a very rich and interesting object on their own, and are defined as follows. A kernel $K(\lambda,\mu)$ is of integrable type if for a sufficiently nice curve $\Gamma$ there exist a set of functions $\{f_j, h_j\}_{j=1}^n \in L^{\infty}(\Gamma)$ satisfying
$$\sum_{j=1}^n f_j(\lambda)h_j(\lambda)=0,
$$
such that $K(\lambda,\mu)$ admits the representation
$$K(\lambda,\mu) = \frac{\sum_{j=1}^n f_j(\lambda)h_j(\mu)}{\lambda-\mu}. 
$$

%
Some classical examples include the Airy, the Bessel and the Sine kernel. Lately, deformations of such kernels had been studied in two different basic approaches: by inserting a deformation at the kernel level (\cite{CC2019}, \cite{CS2025}, \cite{CCR2021}, \cite{BCT2022}) or at matrix model level (\cite{GG21}). At the kernel level, for instance, one considers a non-decreasing, non-negative function $\sigma$ and set the deformed kernel to be $K_{\sigma}(\lambda, \mu) = \sqrt{\sigma(\lambda)}K(\lambda, \mu) \sqrt{\sigma(\mu)}$.

Deformed kernels had been shown to describe finite temperature fermions (\cite{BB2019}, \cite{DMS2018}), and are useful in the study of statistics for \textit{thinned processes}. The thinned process $\tilde{\Lambda}$ is built by conditioning over the original point process $\Lambda$ in the following way: given a function $\tilde{\sigma}: \mathbb{R} \to [0,1]$, each point $\lambda_j$ is eliminated with probability $1-\tilde{\sigma}(\lambda_j)$. Then, the gap probabilities of the point process governed by the deformed kernel is given by the multiplicative statistics associated to the original point process $\Lambda$ of the eigenvalues $\{\lambda_j\}_{j \geq 1}$,
\begin{equation}
 L_n = \mathbb{E} \left[\prod_{j \geq 1}(1-\tilde{\sigma}(\lambda_j))\right],
 \label{eq:defLn}
\end{equation}
where the expectation is over the set of configurations of points in the point process $\Lambda$. In particular, the multiplicative statistics for deformations of the Airy kernel are connected to solutions to the Kardar-Parisi-Zhang (KPZ) equation. The KPZ equation \cite{KPZ1986} is a stochastic partial differential equation given by
\begin{equation}
\partial_T \h(T,\X) = \frac{1}{2} \partial_{\X}^2 \h(T,\X)+\frac{1}{2}(\partial_{\X} \h(T,\X))^2+\W (T,\X),
\label{eq:KPZ}
\end{equation}
where $T, \X$ are time and space variables respectively, and $\W$ stands for the white noise. It was shown that the probability distribution of the solution with narrow wedge initial data is completely characterized by $\h(T,0)$ \cite{C2012}, and relates to the Fredholm determinant of the deformed Airy kernel \cite{BG2016},
\begin{equation}
\L(s,T) = \mathbb{E}_{\Ai}\left[\prod_{j \geq 1} \frac{1}{1+\e^{T^{1/3}(\lambda_j+s)}} \right],
\label{eq:defCCl}
\end{equation}
where the expectation is over the set of configurations of the Airy point process, $T>0$ is the time parameter and $s \in \mathbb{R}$ is a new shifting parameter. In the present work we consider a deformation at the random matrix model level by $\sigma_n(z):=\sigma_n(z;x)$, where 
\begin{equation}
\sigma_n(z)^{-1} = 1+\e^{x-n^{2/3}Q(z)},
\label{def:sigma}
\end{equation}
for some real parameter $x$ and some function $Q$ to be defined later, and study the multiplicative statistics
$$ L_n^Q(x) = \mathbb{E} \left[\prod_{j \geq 1}{\sigma_n}(\lambda_j)\right],
$$
as $n\to \infty$ where the expectation is taken over the set of configurations of the point process of the eigenvalues in the non-deformed matrix model. The case $x \leq x_0$ for a fixed $x_0>0$ is discussed in \cite{GG21}. Our results extend the analysis to the case $x=x_0 n^{\alpha}$ for $\alpha \in (0, 2/9)$, that is, we consider $x \to \infty$ with a certain rate as $n \to \infty$.

As for the deformation, we work under the following assumptions:
\begin{assumption}\hfill
\label{asump2}
\begin{itemize}
\item There exist a neighborhood $\mathcal{R}$ of the real line such that $Q(z)$ is analytic for all $z \in \mathcal{R}$. In particular, $Q$ is analytic in a neighborhood of the origin.
\item $Q(z)$ is real-valued for all $z \in \mathbb{R}$, with a simple zero at $z=0$ and such that $Q(z)>0$ for $z \in(-\infty,0)$ and $Q(z)<0$ for $z \in(0,\infty)$.
\end{itemize}
\end{assumption}

An important role is played by the first derivative at zero
\begin{equation}
t:= -Q'(0)>0.
\end{equation}

The first result shows that, in the limit of large number of particles, the only relevant contribution for the multiplicative statistics comes from a neighborhood of the origin. Before introducing the result, we need a few more definitions. Let $\uppi_k(z):=\uppi_k^{(n,x)}(z)$ be the monic orthogonal polynomial of degree $k$ with respect to the weight $\omega_n(z;x) = \e^{-nV(z)}\sigma_n(z)$,
$$\int_{\mathbb{R}}\uppi_k^{(n,x)}(s)\uppi_j^{(n,x)}(s)\omega_n(s)\dd s = \frac{\delta_{kj}}{\upgamma^{(n)}_k(x)^2}.
$$
Then, we define the  kernel of orthogonal polynomials (also knows as \textit{Christoffel–Darboux kernel}) by
\begin{equation}
K_n^Q (\lambda, \mu;x) = \sum_{k=1}^{n-1}\upgamma_k^{(n)}(x)^2\uppi_k^{(n,x)}(\lambda)\uppi_k^{(n,x)}(\mu).
\label{eq:defCD}
\end{equation}

\begin{proposition}
\label{theo1}
Let $Q$ be under Assumption \ref{asump2} and let $\alpha$ fall under one of the two cases of Assumption \ref{asump1}. Take $t_0\in(0,1)$ a real constant and $t \in [t_0, 1/t_0]$. Let $\omega_n(z;x) = \e^{-nV(z)}\sigma_n(z)$ and set $K_n^Q(\lambda, \mu;x)$ to be the Christoffel–Darboux kernel of orthogonal polynomials with respect to the deformed weight $\omega_n$. Then, there exist $m, \bar{\epsilon}, \tilde{\epsilon}>0$ such that
$$
\log L_n^Q(x) = - \int_{-\infty}^x\int_{-\tilde{\epsilon}n^{\alpha-\frac{2}{3}}}^{ \bar{\epsilon}n^{\frac{2}{3}\alpha-\frac{2}{3}}} K_n^Q(\lambda, \lambda;x') \frac{\omega_n(\lambda;x')}{1+\e^{-x'+n^{2/3}Q(\lambda)}}\dd \lambda \dd x' +O(\e^{-m n^{\alpha}}),
$$
uniformly in both $x=x_0n^{\alpha}$ and $t \in [t_0, 1/t_0]$, as $n \to \infty$.
\end{proposition}

A closer inspection of the kernel gives the principal result of the present work. First, we recall that Claeys and Cafasso \cite{CC2019} investigated the asymptotics for multiplicative statistics of a deformation of the Airy kernel $\log \L(s, T)$ defined in equation \eqref{eq:defCCl} through the analysis of the following Riemann-Hilbert problem. Let $\Psi_{cc}(\zeta):=\Psi_{cc}(\zeta; s,T)$, depending on two parameters $s \in \mathbb{R}$ and $T>0$, be the $2 \times 2$ matrix-valued function such that for some fixed $\zeta_0>0$,
\begin{problem}\hfill\label{RHPCC}
\begin{enumerate}
\item $\Psi_{cc}(\zeta)$ is analytic on $\mathbb{C}\setminus \Sigma$, where $\Sigma = \mathbb{C}\setminus \left(\mathbb{R}\cup (\zeta_0+\i \mathbb{R})\right)$, with continuous boundary values $\Psi_{cc,\pm}$ satisfying the jump condition
\begin{equation}
\Psi_{cc,+}(\zeta)=
\Psi_{cc,-}(\zeta)
\times 
\begin{cases}
\begin{pmatrix}
1 & \sigma_0(\zeta) \\ 0 & 1
\end{pmatrix}, & \zeta\in (\zeta_0, \infty), \\
\begin{pmatrix}
1 & 0 \\ \sigma_0(\zeta)^{-1} & 1
\end{pmatrix}, & \zeta\in \zeta_0+i\mathbb{R}, \\
\begin{pmatrix}
0 & \sigma_0(\zeta) \\ -\sigma_0(\zeta)^{-1} & 0
\end{pmatrix}, & \zeta\in (-\infty,\zeta_0), 
\end{cases}
\end{equation}
where $\sigma_0(\zeta)= (1+\e^{T^{1/3}(s+\zeta)})^{-1}$.
\item As $\zeta\to \infty$,
$$
\Psi_{cc}(\zeta)= \left(I+\frac{\Psi_{cc}^{(1)}}{\zeta}+O \left(\frac{1}{\zeta^2}\right)\right)\zeta^{\sigma_3/4}U_0^{-1}
\e^{-\left(\frac{2}{3}\zeta^{3/2}\right)\sigma_3},
$$
where
$$U_0 = \frac{1}{\sqrt{2}}\begin{pmatrix}
1 & \i \\ \i & 1
\end{pmatrix}, \qquad \text{and} \qquad \sigma_3 = \begin{pmatrix}
1 & 0 \\ 0 & -1
\end{pmatrix}.
$$
\end{enumerate}
\end{problem}

The same Riemann-Hilbert problem plays an important role in the characterization of the multiplicative statistics for the largest eigenvalue in the matrix model under consideration in the present work.

\begin{theorem}
\label{theo2}
Let $t_0\in (0,1)$, $x_0>0$ and $\alpha$ under Assumption \ref{asump1}. Set $\Psi_{cc}(\zeta)$ the solution to the Riemann-Hilbert problem \ref{RHPCC} and
$$\bar{\sigma}_0(\zeta) = \frac{1}{1+\e^{x+t\zeta}}\frac{1}{1+\e^{-x-t\zeta}}, \qquad \Xi_0(\zeta) = \begin{pmatrix}
1 & 0 \\ (1+\e^{x+t\zeta})\chi_{(-\infty, \zeta_0)} & 1
\end{pmatrix},
$$
where $\zeta_0>0$ comes from the formulation of the Riemann-Hilbert problem for $\Psi_{cc}(\zeta)$. Then
\begin{equation} \partial_x \log L_n^Q(x) = -\frac{1}{2\pi\i}\int_{\mathbb{R}} \bar{\sigma}_0(\zeta) \left[\Xi_0(\zeta)^{-1}\Psi_{cc}(\zeta)^{-1}\frac{\dd}{\dd\zeta}\{\Psi_{cc}(\zeta)\Xi_0(\zeta)\}\right]_{21}\dd \zeta+ O(x^3n^{-2/3}),
\label{eq:partxLQ}
\end{equation}
uniformly for $x = x_0 n^{\alpha}$ and $t \in [t_0, 1/t_0]$.
\end{theorem}

\begin{corollary}
\label{corol}
Let $t_0\in (0,1)$, $x_0>0$ and $\alpha$ under Assumption \ref{asump1}. Then
$$ \partial_x \log L_n^Q(x) = -\frac{2 t^4}{3\pi^4}\left(\sqrt{1+\pi^2 x/t^3}-1\right)^3-\frac{t^4}{\pi^4}\left(\sqrt{1+\pi^2 x/t^3}-1\right)^2+ O(x^3n^{-2/3}),
$$
uniformly for $x = x_0 n^{\alpha}$ and $t \in [t_0, 1/t_0]$.
\end{corollary}

\begin{remark} Given Theorem \ref{theo2}, the Corollary \ref{corol} is a direct consequence of one of the main results by Claeys and Cafasso \cite{CC2019}. In fact, Proposition 5.11 from \cite{CC2019} gives that for fixed $T>0$
\begin{align*}
\partial_s \log \L(s,T) =& -\frac{T^{1/3}}{2\pi \i} \int_{\mathbb{R}}\left(\frac{1}{1+\e^{-T^{1/3}(\zeta+s)}}\right)'\left[\Xi_0(\zeta)^{-1}\Psi_{cc}(\zeta)^{-1}\frac{\dd}{\dd\zeta}\{\Psi_{cc}(\zeta)\Xi_0(\zeta)\}\right]_{21}\dd \zeta\\
=& -\frac{2 T^{4/3}}{3\pi^4}\left(\sqrt{1+\pi^2 s T^{-2/3}}-1\right)^3-\frac{T^{4/3}}{\pi^4}\left(\sqrt{1+\pi^2 s T^{-2/3}}-1\right)^2 +O(s^{-1/2}T^{-1/3}),
\end{align*}
as $s \to \infty$. The correspondence $s=x/t$ and $T=t^3$ leads to Equation \eqref{eq:partxLQ}. In other words, Theorem \ref{theo3} says that as $n \to \infty$ and $x = O(n^{\alpha})$, for $\alpha$ under assumptions \ref{asump1}, the multiplicative statistics for the largest eigenvalue associated to the random matrix model deformed by $\sigma_n(z;x)$ converges to the ones associated to the solution to the KPZ equation considered in \cite{CC2019}.
\end{remark}

At last, we also obtained the leading terms of the asymptotic expansion for the normalizing constant $\upgamma^{(n)}_{n-1}(x)$ of the monic orthogonal polynomials with respect to the weight $\omega_n(z;x) = \sigma_n(z)\e^{-nV(z)}$ defined above.

\begin{theorem}
\label{theo3}
Let $Q$ be under Assumption \ref{asump2}, $x_0>0$ and $\alpha$ under Assumption \ref{asump1}. Take $t_0\in(0,1)$ a real constant and $t \in [t_0, 1/t_0]$. Let $\ell_V$ and $c_V$ be the constants associated to the equilibrium measure characterized in Section \ref{sec:eqmeas}. Then as $n \to \infty$,
$$\upgamma^{(n)}_{n-1}(x)^2 = \e^{-2n\ell_V} \left(\frac{a}{8\pi}-\frac{a[\Psi_{cc}^{(1)}]_{21}}{4\pi\i c_V^{1/2}}+O(n^{-\bar{\beta}})\right),
$$
uniformly in both $x=x_0n^{\alpha}$ and $t$, where $\bar{\beta} = \min\{\frac{1}{3}+\frac{\alpha}{2}, \frac{1}{3}+\frac{2}{3}\tau\}$ for $\tau\in(0,1)$ given in Theorem \ref{theo:smallnorm} and where $[\Psi_{cc}^{(1)}]_{21}$ comes from the asymptotic expansion for the model Riemann-Hilbert problem \ref{RHPCC}.
\end{theorem}

\begin{remark}
The main difference in the computations of Theorem \ref{theo3} when compared to the previous literature, relies on the fact that the auxiliary function $\g$ for the global parametrix (see Section \ref{sec:RHapproach}) has leading terms decaying slower than the contribution from the local parametrix. More precisely, as $x \to \infty$, the decay of $\g$ overthrows the decay order of the contribution from the local Riemann-Hilbert problem (see Section \ref{sec:param0}). However, such terms cancel out nicely, and we are left with only the contribution from the local parametrix.
\end{remark}

\subsection{Outline of the paper}

As already mentioned, the statistics for the point process comes from the kernel for orthogonal polynomials. In this sense, our approach relies on the study of the large $n$ asymptotics for orthogonal polynomials through Riemann-Hilbert problems. The work is organized as follows. Sections \ref{sec:eqmeas} and \ref{sec:modellp} present the necessary tools for the analysis, while Sections \ref{sec:RHapproach} and \ref{sec:multiest} present the Riemann-Hilbert analysis for orthogonal polynomials and the proof of the main results.

More precisely, in Section \ref{sec:eqmeas} important mathematical objects related to the potential $V$ are explored in further details, such as the equilibrium measure $\mu_V$, the $\phi$-function and the conformal map $\varphi$. Such definitions allow us to introduce the last set of assumptions on the growing rate of $x$. In Section \ref{sec:modellp}, we recall some results for the Riemann-Hilbert problem associated to the KPZ equation explored by Claeys and Cafasso in \cite{CC2019}, and study its connection to the local Riemann-Hilbert problem that appears in Section \ref{sec:param0}. In Section \ref{sec:RHapproach}, starting from the Riemann-Hilbert problem for orthogonal polynomials established by Fokas in '92 \cite{fokas1992}, we perform a series of transformations in order to simplify the original problem. The new one is then approximated by a \textit{global parametrix} away from the endpoints of the support of the equilibrium measure, and by \textit{local parametrices} around these endpoints. The remaining analysis involves the small norm study of the connection between the original problem and the approximated ones. At last, in Section \ref{sec:multiest} we extract asymptotics for the multiplicative statistics from the Riemann-Hilbert results.

The main challenges when compared with the existing literature rely on the fact that the jumps of the problem with deformation $\sigma_n$ can not be properly approximated by the jumps of the local parametrix, and the local solution around the origin is not bounded in $x$. The first issue is solved by working with the conjugated problem $Y \e^{-\frac{x}{2}\sigma_3}$ instead of the original problem $Y$, while the last one is solved by a careful inspection of the results by Claeys and Cafasso in \cite{CC2019} when extracting the asymptotics in Section \ref{sec:multiest}.

\section*{Acknowledgement}
The author thanks Thomas Bothner for the guidance and great literature suggestions. Their discussions during the research visit to Bristol in December 2024 were decisive for the development of the results in this paper. The author also thanks Alfredo Deaño and Giuseppe Orsatti for inspiring conversations, and Pierre Lazag for very helpful comments in the preliminary version of this paper. C.M.S.P. acknowledges the support of the São Paulo Research Foundation (FAPESP), grants $\#$2021/10819-3 and $\#$2023/14157-0.

\section{Equilibrium measure and related functions}
\label{sec:eqmeas}
Associated to the potential $V$ in Equation \eqref{eq:measure2} we have an equilibrium measure $\mu_V$ (see \cite{Saff_book}) defined as the unique minimizer of the operator
$$\mathbf{I} := \int_{\mathbb{R}^2} \log |x-y|^{-1}\dd \mu(x)\dd\mu(y) + \int_{\mathbb{R}} V(x) \dd \mu (x),
$$
over the space of all probability measures on the real line. In the large dimension limit, the limit for the kernel rely on general features of $\mu_V$. When the density of $\mu_{V}(x)$ vanishes as $x^{-1/2}$ in the edge of the support, the limit behavior is given by Bessel kernel \cite{TW1994}, while for an annihilation of order $x^{1/2}$ one recovers the Airy kernel \cite{forrester92}. Moreover, it is a conjecture in physics \cite{Brezin1990} that if the density of the equilibrium measure vanishes as $x^{\frac{4k+1}{2}}$ in the edge of the support, one recovers the Claeys-Vanlessen kernel related to the $(2k)$-th equation in the Painlev\'e I hierarchy \cite{CV2007}.

Throughout this work, the potential $V$ is assumed to be a non-constant real polynomial of even degree and positive leading coefficient. In particular, it guarantees the existence and uniqueness of the equilibrium measure. Moreover, it will be assumed that $\mu_V$ is one-cut, that is, $\mu_V$ is compactly supported in a single interval, which can be taken as $[-a,0]$ for some $a>0$, without loss of generality.

The assumptions on $V$ also imply that $\mu_V$ is regular in the following sense \cite{lubinsky2009}: its density is a non-vanishing analytic function in $(-a,0)$ and at the endpoints of the support it vanishes as a square-root. Furthermore, there exist a constant $\ell_V \in \mathbb{R}$ for which $\mu_V$ satisfies the Euler-Lagrange equations:
\begin{align*}
2\int \log |x-y|\dd \mu_V(y) -V(x)-\ell_V &= 0, \quad x \in \supp \mu_V \\
2\int \log |x-y|\dd \mu_V(y) -V(x)-\ell_V &< 0, \quad x \in \mathbb{R}\backslash\supp \mu_V.
\end{align*}

Now set $C^{\mu_V}$ to be the Cauchy transform of the equilibrium measure, that is,
$$C^{\mu_V}(s) = \int \frac{\dd\mu_V(s')}{s'-s}, \quad s \in \mathbb{C}\backslash \supp \mu_V.
$$

From the properties of $\mu_V$ it follows that there exist a polynomial $\mathbf{h}_V$, non-vanishing in $(-a,0)$, such that
\begin{equation}
\left(C^{\mu_V}(s) +\frac{V'(s)}{2}\right)^2 = \frac{1}{4}s (s+a) \mathbf{h}_V(s)^2.
\label{eq:cauchy}
\end{equation}

An important quantity that arises in this context is the $\phi$-function
\begin{equation}
\phi(z) := \int_0^z C^{\mu_V}(s) +\frac{V'(s)}{2} \dd s, \qquad z \in \mathbb{C}\backslash (-\infty, 0].
\label{eq:defphi}
\end{equation}

Standard analysis of the equilibrium measure imply the following properties of $\phi(z)$:
\begin{lemma}[\cite{GG21}]
\label{lemma:GG21}
The $\phi$-function associated to the potential $V$ satisfies the following:
\begin{enumerate}
\item $\phi$ is analytic on $\mathbb{C}\backslash (-\infty, 0]$ and has boundary values $\phi_{\pm}$ as $z$ approaches $(-\infty, 0)$ satisfying the jump relations
\begin{align*}
\phi_+(z)+\phi_-(z) &=0,  & z \in (-a,0) \\
\phi_+(z)-\phi_-(z) &=-2\pi i \left(\mu_V((z,0))\chi_{(-a,0)}(z)+ \chi_{(-\infty, -a)}(z)\right), & z \in (-\infty,0).
\end{align*}
\item For every $z \in \mathbb{R}\backslash [-a,0]$,
$$\re \phi_+(z) = \re \phi_-(z)>0.
$$
\item As $z \to \infty$, $\phi$ as the following asymptotic expansion
$$\phi(z) = \frac{V(z)}{2}+\ell_V-\log z +\frac{\phi_{\infty}}{2}+O(z^{-1}).
$$
\item As $z \to 0$, it satisfies
$$\phi(z) = \frac{1}{3}\mathbf{h}_V(0) a^{1/2}z^{3/2}(1+O(z)),
$$
where $\mathbf{h}_V$ is the polynomial defined by Equation \eqref{eq:cauchy}.
\item For some fixed $\delta>0$, the function
\begin{equation}
\varphi (z):= \left(\frac{3}{2}\phi(z)\right)^{2/3}
\label{def:varphi}
\end{equation}
is a conformal map from a neighborhood of the origin to $B_{2\delta}(0)$, and $\varphi(z) = c_Vz+\tilde{c}_Vz^2+O(z^3)$ as $z\to 0$.
\end{enumerate}
\end{lemma}

\begin{remark}
Notice that the assumptions on the potential and the equilibrium measure are exactly the same as in \cite{GG21}. Consequently, for a proof of Lemma \ref{lemma:GG21} we refer Proposition 8.1 and Proposition 8.2 from \cite{GG21}. The main differences will start to appear once we define the conformal map $\varphi$ and start dealing with the growing nature of $x$ and the new features of $Q$ - elements over which the assumptions differ from the existing literature.
\end{remark}

Take a neighborhood $\mathcal{U}^0$ of the origin where $Q$ is analytic. We are interested in understanding the properties of the function $H(z):= Q(\varphi^{-1}(z))$ for $|z|\leq \delta$. From Lemma \ref{lemma:GG21}, $\varphi$ is conformal in $B_{2\delta}(0)$, therefore analytic and it has a series expansion $\varphi(z) = \sum_{k=0}^{\infty}\mathrm{a}_kz^{k}$ valid for all $z \in B_{2\delta}(0) \supset \mathcal{U}^0$. Moreover, from the properties of $\phi$ it is straightforward that $\mathrm{a}_0=0$ and $\mathrm{a}_1= c_V>0$. By series inversion techniques it follows that $\varphi^{-1}:\mathrm{Range}(\varphi) \to \mathcal{U}^0$ also has a power series expansion given by
\begin{equation}
\varphi^{-1}(w) = \sum_{k=1}^{\infty}A_k w^k,
\label{eq:expAphi}
\end{equation}
where the $\{A_k\}_{k \geq 1}$ can be recovered from $\{\mathrm{a}_k\}_{k\geq 1}$ by plugging $w=\varphi(z)$ on Equation \eqref{eq:expAphi}. The first three terms read as
\begin{align*}
A_1&=\frac{1}{\mathrm{a}_1}, & A_2 &=-\frac{\mathrm{a}_2}{\mathrm{a}_1^3}, & A_3&= \frac{2\mathrm{a}_2^2-\mathrm{a}_1\mathrm{a}_3}{\mathrm{a}_1^5}.
\end{align*}

The function $Q(z)$ is assumed to be analytic in this neighborhood of the origin, so that it has a power series expansion $Q(z)=\sum_{k=0}^{\infty}\mathrm{q}_kz^k$ valid for all $z \in \mathcal{U}^0$. From Assumption \ref{asump2}, $\mathrm{q}_0=0$ and $\mathrm{q}_1=-t$. Altogether, it gives that $H(z)$ is analytic for all $z \in \mathcal{U}^0$ with a power series
$$H(z) = \sum_{k=0}^{\infty} \mathrm{h}_k z^k,
$$
where the first terms are given by $\mathrm{h}_0 = 0$, $\mathrm{h}_1 = \frac{\mathrm{q}_1}{\mathrm{a}_1}$ and $\mathrm{h}_2 = \frac{\mathrm{q}_2\mathrm{a}_1- \mathrm{q}_1\mathrm{a}_2}{\mathrm{a}_1^3}$. With this in hands, the last assumptions on $Q(z)$ and $x$ are stated as follows.

\begin{assumption}
\label{asump1} Let $\epsilon>0$ and take $Q$ a function under Assumption \ref{asump2}, such that the expansion $Q(z)=\sum_{k=0}^{\infty}\mathrm{q}_kz^k$ is valid for all $z \in \mathcal{U}^0$. Moreover, let $\varphi$ be the conformal map with expansion \eqref{eq:expAphi} around the origin. Fix a constant $x_0>0$ and set $x=x_0 n^{\alpha}$. The analysis is split into the two following cases:
\begin{itemize}
\item \textbf{Case 1}: Assume $\alpha \in [\epsilon, \frac{4}{21}-\epsilon]$.
\item \textbf{Case 2}: Assume $\alpha \in [\epsilon, \frac{2}{9}-\epsilon]$ and $\mathrm{q}_2 =- \frac{t\tilde{c}_V}{c_V}$.
\end{itemize}
\end{assumption}

\section{Model local problem}
\label{sec:modellp}
The main results in the present work rely on the asymptotic analysis of the Riemann-Hilbert problem for the orthogonal polynomials. This analysis involves the construction of approximate solutions known as parametrices and a small norm study of the connection between the original problem and the approximated ones. In order to build the local parametrix around the origin, it will be necessary to understand a model problem explored by Claeys and Cafasso in \cite{CC2019}. The current section is devoted to the study of the connection between the Claeys and Cafasso Riemann-Hilbert problem and the parametrix needed for Section \ref{sec:param0} in the analysis of the Riemann-Hilbert problem for orthogonal polynomials. 
 
\subsection{The Riemann-Hilbert problem for the lower tail of the KPZ equation}
\label{sec:model}
Claeys and Cafasso \cite{CC2019} investigated the asymptotics for multiplicative statistics of a deformation of the Airy kernel through the analysis of the Riemann-Hilbert problem \ref{RHPCC} presented in Section \ref{sec:intro}. In what follows, we summarize the relevant definitions and properties developed in \cite{CC2019}. The authors perform the change in variables $\zeta = s(z-1)$ and define the following transformation:
\begin{equation}
S(z) = s^{-\sigma_3/4} \E \Psi_{cc}(s(z-1))\e^{s^{3/2}(g(z)+\frac{V(z_0)}{2})\sigma_3},
\label{eq:defS} 
\end{equation}
where $z_0 = \zeta_0/s+1$, $g$ is an auxiliary function, $V(z) := s^{-3/2}\log(1-\sigma(sT^{1/3}z))$ and
\begin{equation}
\E = \begin{pmatrix}
1 & i(g_1-\frac{1}{4})s^2\\ 0 & 1
\end{pmatrix},
\label{eq:defE}
\end{equation}
for a certain $g_1$ depending on $g$. The main properties of $g(z)$ are investigated in Section 3 of \cite{CC2019}, and are summarized as follows:
\begin{proposition}[Proposition 3.5, \cite{CC2019}]
Set $V(z) := s^{-3/2}\log(1-\sigma(sT^{1/3}z))$ and let $g(z)$ be the auxiliary function in Equation \eqref{eq:defS} satisfying
\begin{align*}
g_+(z)+g_-(z) &= V(z)-V(z_0), & z\in (-\infty, z_0) \\
g(z) &= \frac{2}{3}z^{3/2}-z^{1/2}-\frac{V(z_0)}{2}+g_1 z^{-1/2}+O(z^{-3/2}), & z\to \infty.
\end{align*}
Then, $V$ is negative, strictly decreasing in $z$, $V(z) \to 0 $ as $z\to - \infty$ and $V(z) \to - \infty$ as $z\to \infty$. Moreover,
\begin{align}
|\e^{-s^{3/2}(2g(z)-V(z)+V(z_0))}| & \leq \e^{-\frac{4}{3} s^{3/2}(z-z_0)^{3/2}}, & z> z_0,\\
|\e^{s^{3/2}(2g(z)-V(z)+V(z_0))}| & \leq 2\e^{-\frac{2\sqrt{2}}{3}s^{3/2}|z-z_0|^{3/2}}, & z\in z_0 +i \mathbb{R}.
\end{align}
\end{proposition}

For the asymptotic analysis in Section \ref{sec:multiest} we also need some of the matrix-valued functions in \cite{CC2019}. In particular, their global parametrix is given by $(z-z_0)^{\sigma_3/4}U_0^{-1},
$ and their local parametrix for $z \in (z_0-\varepsilon, z_0+\varepsilon)$ is given by
$$\left(\frac{z-z_0}{s\mu(z)}\right)^{\sigma_3/4}\Phi_{\Ai}^{cc}(s\mu(z))\e^{s^{3/2}(g(z)-V(z)/2+V(z_0)/2)\sigma_3},
$$
where $\mu$ is a conformal map such that $\frac{2}{3}\mu(z)^{3/2} = g(z)-V(z)/2+V(z_0)/2$, and for $\Ai(\zeta)$ the Airy function,
$$\Phi_{\Ai}^{cc}(\zeta) = -\sqrt{2\pi}\begin{pmatrix}
\Ai'(\zeta) & -w \Ai'(w^2 \zeta) \\
\i \Ai(\zeta) & -\i w^2 \Ai(w^2 \zeta)
\end{pmatrix}.
$$

At last, their small norm problem is such that $R_{cc}(z) = I +O\left(\frac{1}{s^{3/2}(|z|+1)} \right)$.

\subsection{A Riemann-Hilbert problem for the local parametrix}

For a positive real parameter $\kappa$ set $h_{\kappa}(z)$ to be a function under the following set of assumptions:
\begin{assumption}
\label{asump3}\hfill
\begin{itemize}
\item $h_{\kappa}(z) \in C^{\infty}(\Sigma)$, where $\Sigma = \mathbb{C}\setminus [\mathbb{R}\cup (\zeta_0+\i \mathbb{R})]$ and in the neighborhood $|z| \leq \kappa^{\nu}$ the expansion
$$h_{\kappa}(z)=tz+O(|z|^2/\kappa),
$$
holds, for $t>0$ a fixed constant.
\item There exist $\eta >0$ such that for all $z \in (-\infty,\zeta_0)$ it holds that
$$|h_{\kappa}(z)| \leq -\eta |z|.
$$
\item For a fixed $\epsilon>0$ and for all $z \in \zeta_0 +i\mathbb{R}$, it holds that
$$\re h_{\kappa}(z) \leq c |\im z|^{3/2-\epsilon}.
$$
\item For all $z \in (\zeta_0, \infty)$,
$$h_{\kappa}(z)\geq - c z^{3/2-\epsilon}.
$$
\end{itemize}
\end{assumption}

For our problem, we need to consider $h_{\kappa}(z)=-\kappa H(z/\kappa),$ where $H(z) = Q(\varphi^{-1}(z))$ in a neighborhood of the origin. The following result establishes that $H$ has the desired properties:
\begin{lemma} \label{lemma:H}
There exist constants $\tilde{\delta}> \delta > 0$ and $\epsilon_0 \in (0, 1/2)$ such that the following holds. Set $\Sigma^{\delta} = \cup_{j=1}^4\Sigma^{\delta}_j$ where $\Sigma^{\delta}_0 = (\delta, \infty)$, $\Sigma^{\delta}_1 = \delta +i \mathbb{R}_+$, $\Sigma^{\delta}_2 = (-\infty, \delta)$ and $\Sigma^{\delta}_3 = \delta -i \mathbb{R}_+$. There exist neighborhoods $\mathcal{S}_j$ of $\Sigma^{\delta}_j$ such that $\mathcal{S}_j\cap \mathcal{S}_k \subset B_{\tilde{\delta}}(0)$ and
\begin{itemize}
\item $H(z)$ is independent of $x, \kappa$, real valued on the real line, analytic for all $z \in \mathcal{S} \cup B_{\tilde{\delta}}(0)$, and
$$H(z) = Q(\varphi^{-1}(z)), \qquad |z| \leq \tilde{\delta}.
$$
\item For $z \in \mathcal{S}\backslash B_{\tilde{\delta}}(0)$, the following estimates hold
\begin{align*}
|H(z)| & \leq -\eta |z|, & z \in \mathcal{S}_2 \backslash B_{\tilde{\delta}}(0)\\
H(z) & \leq c |z|^{3/2-\epsilon_0}, & z \in \mathcal{S}_0 \backslash B_{\tilde{\delta}}(0)\\
\re H(z) & \geq -|\im z|^{3/2-\epsilon_0}, & z \in \mathcal{S}_1 \cup \mathcal{S}_3 \backslash B_{\tilde{\delta}}(0)
\end{align*}
\end{itemize}
\end{lemma}

The analyticity property follows from the assumptions of $Q$ together with the fact that $\varphi$ is a conformal map. The desired decays come from the construction of the analytic continuation.

\begin{remark}
Notice that for $\kappa$ large enough, $\zeta_0/\kappa < \delta$ and therefore Lemma \ref{lemma:H} implies that $h_{\kappa}(z)$ is well defined in a neighborhood of $\Sigma$.
\end{remark}

The local parametrix in Section \ref{sec:param0} will be shown to match $\Phi_{\kappa}(\zeta)$ for $\kappa=n^{2/3}$, where $\Phi_{\kappa}(\zeta)$ solves the following Riemann-Hilbert problem 
\begin{problem}\hfill
\begin{enumerate}
\item $\Phi_{\kappa}(\zeta)$ is analytic on $\mathbb{C}\setminus \Gamma$, with continuous boundary values $\Phi_{\kappa,\pm}$ satisfying the jump condition
\begin{equation}
\Phi_{\kappa,+}(\zeta)=
\Phi_{\kappa,-}(\zeta)
\times 
\begin{cases}
\begin{pmatrix}
1 & \e^x\sigma_{\kappa}(\zeta) \\ 0 & 1
\end{pmatrix}, & \zeta\in \Sigma_0, \\
\begin{pmatrix}
1 & 0 \\ \e^{-x}\sigma_{\kappa}(\zeta)^{-1} & 1
\end{pmatrix}, & \zeta\in \Sigma_1 \cup \Sigma_3, \\
\begin{pmatrix}
0 & \e^x\sigma_{\kappa}(\zeta) \\ -\e^{-x}\sigma_{\kappa}(\zeta)^{-1}& 0
\end{pmatrix}, & \zeta\in \Sigma_2, 
\end{cases}
\end{equation}
where $\sigma_{\kappa}(\zeta)= (1+\e^{x+h_{\kappa}(\zeta)})^{-1}$ for a function $h_{\kappa}$ satisfying Assumption \ref{asump3}.
\item As $\zeta\to \infty$,
$$
\Phi_{\kappa}(\zeta)=\e^{-\frac{\log(1+\e^{x+t\zeta_0})}{2}\sigma_3}\e^{\frac{x}{2}\sigma_3}\E\left(I+o(1)\right)\zeta^{\sigma_3/4}U_0^{-1}
\e^{-\left(\frac{2}{3}\zeta^{3/2}\right)\sigma_3}\e^{-\frac{x}{2}\sigma_3},
$$
where $\E$ is the same as in Equation \eqref{eq:defE} under the correspondence $s=x/t$.
\end{enumerate}
\end{problem}
Let $\Omega_{\pm}$ denote the regions with boundary $\Sigma_1 \cup (0, \zeta_0) \cup \i \mathbb{R}_+$ (respectively, $\Sigma_3 \cup (0, \zeta_0) \cup \i \mathbb{R}_-$) depicted in Figure \ref{figomega} and take $\Psi_{\kappa}$ to be defined by the transformation
\begin{equation}
\Phi_{\kappa}(\zeta) = \Psi_{\kappa}(\zeta) \times \begin{cases}
\begin{pmatrix}
1 & 0 \\ \pm\e^{-x}\sigma_{\kappa}(\zeta)^{-1} & 1
\end{pmatrix}, & \zeta\in \Omega_{\pm}, \\
I, & \text{elsewhere}, 
\end{cases}
\end{equation}
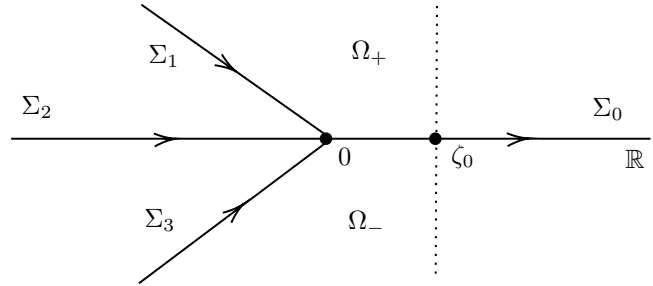
\begin{figure}[H]
    \centering
\tikzset{
pattern size/.store in=\mcSize, 
pattern size = 5pt,
pattern thickness/.store in=\mcThickness, 
pattern thickness = 0.3pt,
pattern radius/.store in=\mcRadius, 
pattern radius = 1pt}
\makeatletter
\pgfutil@ifundefined{pgf@pattern@name@_l9ittg0fj}{
\pgfdeclarepatternformonly[\mcThickness,\mcSize]{_l9ittg0fj}
{\pgfqpoint{0pt}{0pt}}
{\pgfpoint{\mcSize+\mcThickness}{\mcSize+\mcThickness}}
{\pgfpoint{\mcSize}{\mcSize}}
{
\pgfsetcolor{\tikz@pattern@color}
\pgfsetlinewidth{\mcThickness}
\pgfpathmoveto{\pgfqpoint{0pt}{0pt}}
\pgfpathlineto{\pgfpoint{\mcSize+\mcThickness}{\mcSize+\mcThickness}}
\pgfusepath{stroke}
}}
\makeatother

 
\tikzset{
pattern size/.store in=\mcSize, 
pattern size = 5pt,
pattern thickness/.store in=\mcThickness, 
pattern thickness = 0.3pt,
pattern radius/.store in=\mcRadius, 
pattern radius = 1pt}
\makeatletter
\pgfutil@ifundefined{pgf@pattern@name@_9eql0y1kc}{
\pgfdeclarepatternformonly[\mcThickness,\mcSize]{_9eql0y1kc}
{\pgfqpoint{0pt}{0pt}}
{\pgfpoint{\mcSize+\mcThickness}{\mcSize+\mcThickness}}
{\pgfpoint{\mcSize}{\mcSize}}
{
\pgfsetcolor{\tikz@pattern@color}
\pgfsetlinewidth{\mcThickness}
\pgfpathmoveto{\pgfqpoint{0pt}{0pt}}
\pgfpathlineto{\pgfpoint{\mcSize+\mcThickness}{\mcSize+\mcThickness}}
\pgfusepath{stroke}
}}
\makeatother
\tikzset{every picture/.style={line width=0.75pt}} 

\begin{tikzpicture}[x=0.75pt,y=0.75pt,yscale=-1,xscale=1]

\draw    (126.22,94.33) -- (448.69,94.33) ;
\draw    (190.67,167.33) -- (287.67,94.32) ;
\draw    (191.93,26.67) -- (287.67,94.32) ;

\draw  [dash pattern={on 0.84pt off 2.51pt}]  (341,27.33) -- (340.33,165.33) ;
\draw    (239.17,130.83) -- (241.97,128.45) ;
\draw [shift={(243.5,127.16)}, rotate = 139.76] [color={rgb, 255:red, 0; green, 0; blue, 0 }  ][line width=0.75]    (10.93,-3.29) .. controls (6.95,-1.4) and (3.31,-0.3) .. (0,0) .. controls (3.31,0.3) and (6.95,1.4) .. (10.93,3.29)   ;
\draw    (203.67,94.67) -- (207.34,94.45) ;
\draw [shift={(209.33,94.33)}, rotate = 176.63] [color={rgb, 255:red, 0; green, 0; blue, 0 }  ][line width=0.75]    (10.93,-3.29) .. controls (6.95,-1.4) and (3.31,-0.3) .. (0,0) .. controls (3.31,0.3) and (6.95,1.4) .. (10.93,3.29)   ;
\draw    (235,57.33) -- (238.13,59.39) ;
\draw [shift={(239.8,60.49)}, rotate = 213.38] [color={rgb, 255:red, 0; green, 0; blue, 0 }  ][line width=0.75]    (10.93,-3.29) .. controls (6.95,-1.4) and (3.31,-0.3) .. (0,0) .. controls (3.31,0.3) and (6.95,1.4) .. (10.93,3.29)   ;
\draw    (381.67,94.5) -- (385.34,94.28) ;
\draw [shift={(387.33,94.17)}, rotate = 176.63] [color={rgb, 255:red, 0; green, 0; blue, 0 }  ][line width=0.75]    (10.93,-3.29) .. controls (6.95,-1.4) and (3.31,-0.3) .. (0,0) .. controls (3.31,0.3) and (6.95,1.4) .. (10.93,3.29)   ;
\draw  [fill={rgb, 255:red, 0; green, 0; blue, 0 }  ,fill opacity=1 ] (282.33,94.32) .. controls (282.33,92.85) and (283.53,91.65) .. (285,91.65) .. controls (286.47,91.65) and (287.67,92.85) .. (287.67,94.32) .. controls (287.67,95.79) and (286.47,96.99) .. (285,96.99) .. controls (283.53,96.99) and (282.33,95.79) .. (282.33,94.32) -- cycle ;
\draw  [fill={rgb, 255:red, 0; green, 0; blue, 0 }  ,fill opacity=1 ] (337.5,94.33) .. controls (337.5,92.86) and (338.69,91.67) .. (340.17,91.67) .. controls (341.64,91.67) and (342.83,92.86) .. (342.83,94.33) .. controls (342.83,95.81) and (341.64,97) .. (340.17,97) .. controls (338.69,97) and (337.5,95.81) .. (337.5,94.33) -- cycle ;
\draw  [draw opacity=0][pattern=_l9ittg0fj,pattern size=7.050000000000001pt,pattern thickness=1.5pt,pattern radius=0pt, pattern color={rgb, 255:red, 206; green, 203; blue, 203}] (341,27.33) -- (340.17,94.33) -- (287.67,94.32) -- (255,71.65) -- (191.93,26.67) -- cycle ;
\draw  [draw opacity=0][pattern=_9eql0y1kc,pattern size=7.050000000000001pt,pattern thickness=1.5pt,pattern radius=0pt, pattern color={rgb, 255:red, 202; green, 200; blue, 200}] (340.33,165.33) -- (342.83,94.33) -- (285,96.99) -- (243.5,127.16) -- (190.67,167.33) -- cycle ;

\draw (434.66,98.32) node [anchor=north west][inner sep=0.75pt]    {$\mathbb{R}$};
\draw (289.67,97.72) node [anchor=north west][inner sep=0.75pt]    {$0$};
\draw (346.67,96.07) node [anchor=north west][inner sep=0.75pt]    {$\zeta _{0}$};
\draw (296.33,43.4) node [anchor=north west][inner sep=0.75pt]    {$\Omega _{+}$};
\draw (294.67,129.07) node [anchor=north west][inner sep=0.75pt]    {$\Omega _{-}$};
\draw (418,72.4) node [anchor=north west][inner sep=0.75pt]    {$\Sigma _{0}$};
\draw (194,45.4) node [anchor=north west][inner sep=0.75pt]    {$\Sigma _{1}$};
\draw (130,71.4) node [anchor=north west][inner sep=0.75pt]    {$\Sigma _{2}$};
\draw (192,128.4) node [anchor=north west][inner sep=0.75pt]    {$\Sigma _{3}$};

\end{tikzpicture}
\caption{Domains for deformation of the contour.}
\label{figomega}
\end{figure}

Then, $\Psi_{\kappa}$ solves the following problem:
\begin{problem}\hfill
\begin{enumerate}
\item $\Psi_{\kappa}(\zeta)$ is analytic on $\mathbb{C}\setminus \{\mathbb{R} \cup (\zeta_0 + i \mathbb{R})\}$, with continuous boundary values $\Psi_{\kappa,\pm}$ satisfying the jump condition
\begin{equation}
\Psi_{\kappa,+}(\zeta)=
\Psi_{\kappa,-}(\zeta)
\times 
\begin{cases}
\begin{pmatrix}
1 & \e^x\sigma_{\kappa}(\zeta) \\ 0 & 1
\end{pmatrix}, & \zeta\in (\zeta_0, \infty), \\
\begin{pmatrix}
1 & 0 \\ \e^{-x}\sigma_{\kappa}(\zeta)^{-1} & 1
\end{pmatrix}, & \zeta\in \zeta_0+i\mathbb{R}, \\
\begin{pmatrix}
0 & \e^x\sigma_{\kappa}(\zeta) \\ -\e^{-x}\sigma_{\kappa}(\zeta)^{-1}& 0
\end{pmatrix}, & \zeta\in (-\infty, \zeta_0), 
\end{cases}
\end{equation}
where $\sigma_{\kappa}(\zeta)= (1+\e^{x+h_{\kappa}(\zeta)})^{-1}$.
\item As $\zeta\to \infty$,
$$
\Psi_{\kappa}(\zeta)=\e^{-\frac{\log(1+\e^{x+t\zeta_0})}{2}\sigma_3}\e^{\frac{x}{2}\sigma_3}\E\left(I+o(1)\right)\zeta^{\sigma_3/4}U_0^{-1}
\e^{-\left(\frac{2}{3}\zeta^{3/2}\right)\sigma_3}\e^{-\frac{x}{2}\sigma_3}.
$$
\end{enumerate}
\end{problem}

As $\kappa \to 0$, it is expected that the Riemann-Hilbert problem for $\Psi_{\kappa}(\zeta)$ should approach the Riemann-Hilbert problem for
$$ \Psi_0(\zeta) = \e^{\frac{x}{2}\sigma_3}\e^{-\frac{\log(1+\e^{x+t\zeta_0})}{2}\sigma_3}\E\Psi_{cc}(\zeta)\e^{-\frac{x}{2}\sigma_3},
$$
where $\Psi_{cc}(\zeta; x/t, t^3)$ is the model problem in Section \ref{sec:model} under the correspondence $s=x/t$ and $T=t^3$. Now we investigate the error problem relating $\Psi_0(\zeta)$ and $\Psi_{\kappa}(\zeta)$.


\begin{lemma}
\label{RHPDelta}
Set
$$ \Delta(\zeta) = \Psi_{\kappa}(\zeta) \Psi_0(\zeta)^{-1}.$$
Then, $\Delta(\zeta)$ solves the following Riemann-Hilbert problem
\begin{enumerate}
\item $\Delta(\zeta)$ is analytic on $\mathbb{C}\setminus \{\mathbb{R} \cup (\zeta_0 + i \mathbb{R})\}$ and satisfies the jump condition
\begin{equation}
\Delta_{+}(\zeta)=
\Delta_{-}(\zeta)
\times 
\begin{cases}
\Psi_{0,+}(\zeta)\begin{pmatrix}
\frac{\sigma_0}{\sigma_{\kappa}} & 0 \\ 0 & \frac{\sigma_{\kappa}}{\sigma_0}
\end{pmatrix}\Psi_{0,+}(\zeta)^{-1}, & \zeta \in (- \infty, \zeta_0), \\
\Psi_{0,+}(\zeta) \begin{pmatrix}
1 & \e^x(\sigma_{\kappa}-\sigma_0 )\\ 0 & 1
\end{pmatrix}\Psi_{0,+}(\zeta)^{-1}, & \zeta \in (\zeta_0, \infty), \\ 
\Psi_{0,+}(\zeta)\begin{pmatrix}
1 &0 \\ \e^{-x}(\sigma_{\kappa}^{-1}-\sigma_0^{-1}) & 1
\end{pmatrix}\Psi_{0,+}(\zeta)^{-1}, & \zeta \in \zeta_0+i \mathbb{R}.
\end{cases}
\end{equation}
\item As $\zeta\to \infty$,
$$
\Delta(\zeta)=\left(I+O \left(\frac{1}{\zeta}\right)\right).
$$
\end{enumerate}
\end{lemma}
\textbf{Proof:} For the jumps, notice that
\begin{align*}
\Delta_+(\zeta) &= \Psi_{\kappa,+}(\zeta) \Psi_{0,+}(\zeta)^{-1}\\
&= \Psi_{\kappa,-}(\zeta) \Psi_{0,-}(\zeta)^{-1} \left[\Psi_{0,+}(\zeta)J_0^{-1}J_{\kappa}\Psi_{0,+}(\zeta)^{-1}\right].
\end{align*}
On the other side,
\begin{align}
J_0^{-1}J_{\kappa} = \begin{cases}
\begin{pmatrix}
0 & - \e^x\sigma_0 \\ \e^{-x}\sigma_0^{-1} & 0
\end{pmatrix}\begin{pmatrix}
0 & \e^x\sigma_{\kappa} \\ -\e^{-x}\sigma_{\kappa}^{-1} & 0 \end{pmatrix} = \begin{pmatrix}
\frac{\sigma_0}{\sigma_{\kappa}} & 0 \\ 0 & \frac{\sigma_{\kappa}}{\sigma_0}
\end{pmatrix}, & \zeta \in (- \infty, \zeta_0), \\
\begin{pmatrix}
1 & - \e^x\sigma_0 \\ 0 & 1
\end{pmatrix}\begin{pmatrix}
1 & \e^x\sigma_{\kappa} \\ 0 & 1 \end{pmatrix} = \begin{pmatrix}
1 & \e^x\sigma_{\kappa}-\e^x\sigma_0 \\ 0 & 1
\end{pmatrix}, & \zeta \in (\zeta_0, \infty), \\ 
\begin{pmatrix}
1 & 0 \\ - \e^{-x}\sigma_0^{-1} & 1
\end{pmatrix}\begin{pmatrix}
1 & 0 \\ \e^{-x}\sigma_{\kappa}^{-1} & 1 \end{pmatrix} = \begin{pmatrix}
1 &0 \\ \e^{-x}\sigma_{\kappa}^{-1}-\e^{-x}\sigma_0^{-1} & 1
\end{pmatrix}, & \zeta \in \zeta_0+i\mathbb{R}.
\end{cases}
\end{align}
For the asymptotic condition, notice that
$$-\log(1+\e^{x+t\zeta_0})+x = -t\zeta_0+O(\e^{-x-t\zeta_0}).
$$
Since $x=x_0\kappa^{3\alpha/2}$, for $\kappa$ large enough the asymptotic conditions of $\Psi_{\kappa}$ and $\Psi_{0}$ imply that
\begin{align*}
\Delta (\zeta) &= I+\frac{\e^{-\frac{t\zeta_0}{2}\sigma_3}\E[\Psi_{\kappa}^{(1)}-\Psi_{0}^{(1)}]\E^{-1}\e^{\frac{t\zeta_0}{2}\sigma_3}}{\zeta}+O \left(\frac{1}{\zeta^2}\right),
\end{align*}
as claimed. \hfill $\square$

\begin{remark}\label{rmk:psicc}
A straightforward consequence of the previous calculation is that, from the definition 
$$\Delta^{(1)}: = \e^{-\frac{t\zeta_0}{2}\sigma_3}\E[\Psi_{\kappa}^{(1)}-\Psi_{0}^{(1)}]\E^{-1}\e^{\frac{t\zeta_0}{2}\sigma_3}$$
it follows that $[\Psi_{0}^{(1)}]_{21} = [\Psi_{cc}^{(1)}]_{21} + \Delta^{(1)}_{21}\e^{-t\zeta_0}$. This fact will be helpful in the proof of Theorem \ref{theo3}.
\end{remark}
\subsection{Small norm for the error problem}

Let $J_{\Delta}$ denote the jumps in the Riemann-Hilbert problem for $\Delta$. Then $J_{\Delta}-I$ takes the following form
\begin{align*}
J_{\Delta}-I &= \e^x(\sigma_{\kappa}-\sigma_0)\Psi_{0,+}(\zeta)E_{12}\Psi_{0,+}(\zeta)^{-1} & \zeta \in (\zeta_0, \infty) \\
J_{\Delta}-I &= \e^{-x}(\sigma_{\kappa}^{-1}-\sigma_0^{-1})\Psi_{0,+}(\zeta)E_{21}\Psi_{0,+}(\zeta)^{-1} & \zeta \in \zeta_0+i\mathbb{R} \\
J_{\Delta}-I &= \frac{\sigma_0}{\sigma_{\kappa}} \Psi_{0,+}(\zeta)E_{11}\Psi_{0,+}(\zeta)^{-1}+\frac{\sigma_{\kappa}}{\sigma_{0}} \Psi_{0,+}(\zeta)E_{22}\Psi_{0,+}(\zeta)^{-1}-I & \zeta \in (-\infty,\zeta_0).
\end{align*}

In order to show that the jumps $J_{\Delta}$ are close to identity, we need to show one more auxiliary result. For the next steps, in order to simplify notation, we define 
\begin{equation}
\tilde{\Psi}(\zeta):= \e^{-\frac{x}{2}\sigma_3}\Psi_0(\zeta)\e^{\frac{x}{2}\sigma_3} = \e^{s^{3/2}\left(\frac{V(z_0)}{2}\right)\sigma_3}\E\Psi_{cc}(\zeta)= \e^{-\frac{\log(1+\e^{x+t\zeta_0})}{2}\sigma_3}\E\Psi_{cc}(\zeta),
\label{eq:auxSN}
\end{equation}
where $\Psi_{cc}(\zeta)$ is the solution to the Riemann-Hilbert problem \ref{RHPCC}. With this notation in hands, we prove the following lemma:

\begin{lemma}
Let $t_0>0$ fixed and $\alpha$ satisfying Assumption \ref{asump1}. Then, 
\begin{align*}
\Psi_{0,+}(\zeta)E_{21}\Psi_{0,+}(\zeta)^{-1} &= O(\e^{-\frac{2\sqrt{2}}{3}|\zeta-\zeta_0|^{3/2}} ), & \zeta \in \zeta_0+i\mathbb{R} \\
\Psi_{0,+}(\zeta)E_{12}\Psi_{0,+}(\zeta)^{-1} &=  O(\e^{-\frac{4}{3}|\zeta-\zeta_0|^{3/2}} ) & \zeta > \zeta_0\\
\Psi_{0,+}(\zeta)E_{11}\Psi_{0,+}(\zeta)^{-1} &=  O(\max\{ |\zeta-\zeta_0|^{1/2} \e^{-t\zeta_0}, |\zeta-\zeta_0|^{-1/2} \e^{t\zeta_0}\}), & \zeta<\zeta_0.
\end{align*}
uniformly in $x = x_0\kappa^{3\alpha/2}$ in $t \in [t_0, 1/t_0]$. 
\end{lemma}

\textbf{Proof:} Notice that the desired quantities can be expressed by means of $\tilde{\Psi}$ as follows
\begin{align}
\Psi_{0,+}(\zeta)E_{11}\Psi_{0,+}(\zeta)^{-1} &= \e^{\frac{x}{2}\sigma_3}\tilde{\Psi}_+(\zeta)E_{11}\tilde{\Psi}_+(\zeta)^{-1} \e^{-\frac{x}{2}\sigma_3}= \begin{pmatrix}
\tilde{\Psi}_{11}\tilde{\Psi}_{22} & -\tilde{\Psi}_{11}\tilde{\Psi}_{12}\e^x\\
\tilde{\Psi}_{21}\tilde{\Psi}_{22}\e^{-x} & -\tilde{\Psi}_{21}\tilde{\Psi}_{12}
\end{pmatrix} \label{eq:lemma46.1}\\
\Psi_{0,+}(\zeta)E_{21}\Psi_{0,+}(\zeta)^{-1} &= \e^{x}\e^{\frac{x}{2}\sigma_3}\tilde{\Psi}_+(\zeta)E_{21}\tilde{\Psi}_+(\zeta)^{-1}\e^{-\frac{x}{2}\sigma_3} = \begin{pmatrix}
\tilde{\Psi}_{12}\tilde{\Psi}_{22} \e^x&-\tilde{\Psi}_{12}^2\e^{2x}\\
\tilde{\Psi}_{22}^2 & -\tilde{\Psi}_{12}\tilde{\Psi}_{22}\e^x
\end{pmatrix} \label{eq:lemma46.2}\\
\Psi_{0,+}(\zeta)E_{12}\Psi_{0,+}(\zeta)^{-1} &= \e^{-x}\e^{\frac{x}{2}\sigma_3}\tilde{\Psi}_+(\zeta)E_{12}\tilde{\Psi}_+(\zeta)^{-1}\e^{-\frac{x}{2}\sigma_3} = \begin{pmatrix}-
\e^{-x}\tilde{\Psi}_{11}\tilde{\Psi}_{21} &\tilde{\Psi}_{11}^2\\
-\e^{-2x}\tilde{\Psi}_{21}^2 & \e^{-x}\tilde{\Psi}_{11}\tilde{\Psi}_{21}
\end{pmatrix}.\label{eq:lemma46.3}
\end{align}
In the notation of Section \ref{sec:model}, $\tilde{\Psi}$ can be expressed as follows
$$\begin{pmatrix}
\tilde{\Psi}_{11}(\zeta) & \tilde{\Psi}_{12}(\zeta)\\
\tilde{\Psi}_{21}(\zeta) & \tilde{\Psi}_{22}(\zeta)
\end{pmatrix} = \begin{pmatrix}
S_{11} s^{1/4} \e^{-s^{3/2}g(z)} & S_{12} s^{1/4} \e^{s^{3/2}(g(z)+V(z_0))}\\
S_{21} s^{-1/4} \e^{-s^{3/2}(g(z)+V(z_0))} &S_{22} s^{-1/4} \e^{s^{3/2}g(z)}
\end{pmatrix}.
$$
Now we can apply the small norm results from \cite{CC2019} listed in Section \ref{sec:model} entry-wise to the matrices \eqref{eq:lemma46.1}-\eqref{eq:lemma46.3} defined above. First, take $\zeta \in \zeta_0 + i \mathbb{R}$. Then,
\begin{align*}
|\tilde{\Psi}_{12}^2(\zeta)\e^{2x}| &\leq |S_{12}^2 s^{1/2}| |\e^{s^{3/2}(2g(z)+V(z_0)-V(z))}| |\e^{s^{3/2}(V(z_0)+V(z))}|\e^{2x}\\
&\leq |\zeta-\zeta_0|^{1/2} 2\e^{-\frac{2\sqrt{2}}{3}|\zeta-\zeta_0|^{3/2}}\left|\frac{\e^{2x}}{(1+\e^{x+t\zeta})(1+\e^{x+t\zeta_0})}\right| \leq 2\e^{-2t\zeta_0}|\im\zeta|^{1/2} \e^{-\frac{2\sqrt{2}}{3}|\im\zeta|^{3/2}},\\
|\tilde{\Psi}_{22}^2(\zeta)| &\leq |S_{22}^2 s^{-1/2}| |\e^{s^{3/2}(2g(z)+V(z_0)-V(z))}| |\e^{s^{3/2}(V(z)-V(z_0))}|\\
&\leq |\zeta-\zeta_0|^{-1/2} 2\e^{-\frac{2\sqrt{2}}{3}|\zeta-\zeta_0|^{3/2}}\left|\frac{(1+\e^{x+t\zeta_0})}{(1+\e^{x+t\zeta})}\right| \leq k |\im\zeta|^{-1/2} \e^{-\frac{2\sqrt{2}}{3}|\im\zeta|^{3/2}},\\
|\tilde{\Psi}_{22}(\zeta)\tilde{\Psi}_{12}(\zeta)\e^x| &\leq |S_{22}S_{12} | |\e^{s^{3/2}(2g(z)+V(z_0)-V(z))}| |\e^{s^{3/2}V(z)}|\e^x\\
&\leq 2\e^{-\frac{2\sqrt{2}}{3}s^{3/2}|z-\zeta_0|^{3/2}}\left|\frac{\e^x}{1+\e^{x+t\zeta}}\right|\leq  k\e^{-\frac{2\sqrt{2}}{3}|\im\zeta|^{3/2}}.
\end{align*}
Now take $\zeta >\zeta_0$. Analogous calculations show that
\begin{align*}
|\tilde{\Psi}_{11}^2(\zeta)| &\leq |S_{11}^2 s^{1/2}| |\e^{-s^{3/2}(2g(z)+V(z_0)-V(z))}| |\e^{s^{3/2}(V(z_0)-V(z))}|\\
&\leq |\zeta-\zeta_0|^{1/2} \e^{-\frac{4}{3}(\zeta-\zeta_0)^{3/2}}\left|\frac{1+\e^{x+t\zeta}}{1+\e^{x+t\zeta_0}}\right| \leq |\zeta-\zeta_0|^{1/2} \e^{-\frac{4}{3}(\zeta-\zeta_0)^{3/2}}\e^{t(\zeta-\zeta_0)},\\
|\tilde{\Psi}_{21}^2(\zeta)\e^{-2x}| &\leq |S_{21}^2 s^{-1/2}| |\e^{-s^{3/2}(2g(z)+V(z_0)-V(z))}| |\e^{-s^{3/2}(V(z_0)+V(z))}|\e^{-2x}\\
&\leq |\zeta-\zeta_0|^{-1/2} \e^{-\frac{4}{3}(\zeta-\zeta_0)^{3/2}}(1+\e^{x+t\zeta_0})(1+\e^{x+t\zeta})\e^{-2x}\leq 2 |\zeta-\zeta_0|^{-1/2} \e^{-\frac{4}{3}(\zeta-\zeta_0)^{3/2}}\e^{t(\zeta+\zeta_0)},\\
|\tilde{\Psi}_{21}(\zeta)\tilde{\Psi}_{11}(\zeta)\e^{-x}| &\leq |S_{21}S_{11}| |\e^{-s^{3/2}(2g(z)+V(z_0)-V(z))}| |\e^{-s^{3/2}V(z)}|\e^{-x}\\
&\leq |\zeta-\zeta_0|^{-1/2} \e^{-\frac{4}{3}(\zeta-\zeta_0)^{3/2}}(\e^{-x}+\e^{t\zeta})\leq k|\zeta-\zeta_0|^{-1/2} \e^{-\frac{4}{3}(\zeta-\zeta_0)^{3/2}}\e^{t\zeta}.
\end{align*}
At last, consider $\zeta < \zeta_0$. Then
\begin{align*}
|\tilde{\Psi}_{11}(\zeta)\tilde{\Psi}_{22}(\zeta)| &\leq |S_{11}S_{22}| \leq k,\\
|\tilde{\Psi}_{11}(\zeta)\tilde{\Psi}_{12}(\zeta)\e^{x}| &\leq |S_{11}S_{12}s^{1/2}| \e^{s^{3/2}V(z_0)} \e^x = |(\zeta-\zeta_0)^{1/2}| \frac{\e^x}{1+\e^{x+t\zeta_0}} \leq k |\zeta-\zeta_0|^{1/2},\\
|\tilde{\Psi}_{21}(\zeta)\tilde{\Psi}_{22}(\zeta)\e^{-x}| &\leq |S_{21}S_{22}s^{-1/2}| \e^{-s^{3/2}V(z_0)} \e^{-x}= |(\zeta-\zeta_0)^{-1/2}| (\e^{-x}+\e^{t\zeta_0})\leq k |(\zeta-\zeta_0)^{-1/2}|,\\
|\tilde{\Psi}_{21}(\zeta)\tilde{\Psi}_{12}(\zeta)| &\leq |S_{21}S_{12}| \leq k,
\end{align*}
where the constant $k>0$ is independent of $x$ and uniform in $t \in [t_0, 1/t_0]$. $\square$

\begin{remark} \label{rmk:exp}
The following fact will be very useful for the next results. For any fixed $\nu \in (0,\frac{1}{2})$ let $\kappa$ be large enough so that $\kappa^{\nu} \geq 1$. Then, $|\zeta-\zeta_0| \leq \kappa^{\nu}$ implies that $|\zeta| \leq |\zeta-\zeta_0|+|\zeta_0| \leq 2 \kappa^{\nu}$. Therefore, in the regime $|\zeta-\zeta_0|\leq \kappa^{\nu}$ the expansion $h_{\kappa}(\zeta) = t\zeta+O(|\zeta|/\kappa)$ still holds. Moreover, $|\zeta| = O(|\zeta-\zeta_0|)$, and the Taylor series for the exponential gives that
$$1-\e^{t\zeta-h_{\kappa}(\zeta)} = O\left(\frac{|\zeta|^2}{\kappa}\right)= O\left(\frac{|\zeta-\zeta_0|^2}{\kappa}\right),
$$
which has order $O(\kappa^{2\nu-1})$.
\end{remark}

\begin{lemma}\label{lemma:smal1}
Let $\zeta \in \tilde{\Sigma}_0=(\zeta_0, \infty)$ and $\nu \in(0,\frac{1}{2})$. Fix $t_0, \tilde{x}_0 \in(0,1)$. There exist a real constant $m$ and an appropriate choice of constant $M:=M(t_0)$ such that
$$\norm{J_{\Delta}-I}_{L^1\cap L^2 \cap L^{\infty}(\tilde{\Sigma}_0)} = M \max\{\kappa^{2\nu-1}, \e^{-m \kappa^{3\nu/2}}\},$$
uniformly in $x \geq \tilde{x}_0$ and $t \in [t_0, 1/t_0]$. 
\end{lemma}
\textbf{Proof:} Notice that
\begin{align*}
\left|\e^x(\sigma_{\kappa}-\sigma_0)\right| &= \e^x \frac{\e^{x+t\zeta}-\e^{x+h_{\kappa}}}{(1+\e^{x+t\zeta})(1+\e^{x+h_{\kappa}})}\leq \e^{-h_{\kappa}} -\e^{-t\zeta}.
\end{align*}
Set $I_{\kappa} = \{\zeta \in \tilde{\Sigma}_0 : \,|\zeta-\zeta_0|\leq \kappa^{\nu}\}$ and take $\zeta \in \tilde{\Sigma}_0\backslash I_{\kappa}$. We evaluate the norm entry-wise. For instance, for the entry $12$, Assumption \ref{asump3} implies that
\begin{align*}
||J_{\Delta}-I||_{L^1} \leq & M \int_{\kappa^{\nu}}^{\infty}|\e^{c (z+\zeta_0)^{3/2-\epsilon}}-\e^{-t(z+\zeta_0)}||z^{1/2}\e^{-\frac{4}{3}z^{3/2}}\e^{tz}| \dd z \leq \tilde{M}(t_0)\e^{-m\kappa^{3\nu/2}}\\
||J_{\Delta}-I||_{L^2} \leq & M \left( \int_{\kappa^{\nu}}^{\infty}|\e^{c (z+\zeta_0)^{3/2-\epsilon}}-\e^{-t(z+\zeta_0)}||z\e^{-\frac{8}{3}z^{3/2}}\e^{2tz}| \dd z\right)^{1/2} \leq \tilde{M}\e^{-m\kappa^{3\nu/2}}\\
||J_{\Delta}-I||_{L^{\infty}} \leq & M \supp_{[\kappa^{\nu}, \infty]}|(\e^{c (z+\zeta_0)^{3/2-\epsilon}}-\e^{-t(z+\zeta_0)})z^{1/2}\e^{-\frac{4}{3}z^{3/2}}\e^{tz}| \leq \tilde{M}\e^{-m\kappa^{3\nu/2}}
\end{align*}
It is analogous for the other entries. Thus, there exist a real constant $m$ and an appropriate choice of constant $M:=M(t_0)$ depending on $t_0$ and uniform in $x \leq \tilde{x}_0$, such that
$$||J_{\Delta}-I||_{L^1\cap L^2 \cap L^{\infty}(\tilde{\Sigma}_0\backslash I_{\kappa})} \leq M\e^{-m\kappa^{3\nu/2}}.
$$
For $\zeta \in I_{\kappa}$, we no longer have the exponential decay and rely, instead, in the polynomial decay of $h_{\kappa}-t\zeta$. From Remark \ref{rmk:exp} we have
\begin{align*}
||J_{\Delta}-I||_{L^1} \leq & M \int_{I_{\kappa}}|1-\e^{t\zeta-h_{\kappa}}||\zeta-\zeta_0|^{1/2}|\e^{-t\zeta}\e^{-\frac{4}{3}(\zeta-\zeta_0)^{3/2}}\e^{t(\zeta-\zeta_0)}| \dd z\\
\leq & M\frac{\kappa^{2\nu}}{\kappa} \e^{-t\zeta_0}\int_0^{\kappa^{\nu}}|z^{1/2}\e^{-\frac{4}{3}z^{3/2}}| \dd z = \tilde{M}\kappa^{2\nu-1}\\
||J_{\Delta}-I||_{L^2} \leq & M \kappa^{2\nu-1}\left( \int_0^{\kappa^{\nu}}|z\e^{-\frac{8}{3}z^{3/2}}| \dd z\right)^{1/2} \leq \tilde{M}\kappa^{2\nu-1}\\
||J_{\Delta}-I||_{L^{\infty}} \leq & M \kappa^{2\nu-1} \sup_{[0,\kappa^{\nu}]}|z^{1/2}\e^{-\frac{4}{3}z^{3/2}}| \leq \tilde{M}\kappa^{2\nu-1}.
\end{align*}
Altogether, with an appropriate choice of constant $M>0$, one obtains
$$||J_{\Delta}-I||_{L^1\cap L^2 \cap L^{\infty}( I_{\kappa})} \leq M \kappa^{2\nu-1},
$$
and the result follows. $\square$

\begin{lemma}\label{lemma:smal2}
Let $\zeta \in \tilde{\Sigma}_1\cup \tilde{\Sigma}_3=\zeta_0+i\mathbb{R}$ and $\nu \in(0,\frac{1}{2})$. Fix $t_0, \tilde{x}_0 \in(0,1)$. There exist a real constant $m$ and an appropriate choice of constant $M:=M(t_0)$ such that
$$\norm{J_{\Delta}-I}_{L^1\cap L^2 \cap L^{\infty}(\tilde{\Sigma}_1\cup \tilde{\Sigma}_3)} = M \max\{\kappa^{2\nu-1}, \e^{-m \kappa^{3\nu/2}}\}$$
uniformly in $x \geq \tilde{x}_0$ and $t \in [t_0, 1/t_0]$.
\end{lemma}

\textbf{Proof:} Take $I_{\kappa} = \{\zeta \in \tilde{\Sigma}_1\cup \tilde{\Sigma}_3: \,|\zeta-\zeta_0|\leq \kappa^{\nu}\}$. By Remark \ref{rmk:exp},
\begin{align*}
\left|\e^{-x}(\sigma_{\kappa}^{-1}-\sigma_0^{-1})\right| &= |\e^{t\zeta}|O\left(\frac{|\im\zeta|^2}{\kappa} \right).
\end{align*}
For the entry $12$, it follows that
\begin{align*}
||J_{\Delta}-I||_{L^1} \leq & 2\e^{-t\zeta_0}O\left(\frac{|\im\zeta|^2}{\kappa} \right) \int_0^{\kappa^{\nu}}z^{1/2}\e^{-2\sqrt{2}z^{3/2}/3} \dd z = \tilde{M}\kappa^{2\nu-1}\\
||J_{\Delta}-I||_{L^2} \leq & 2\e^{-t\zeta_0}\kappa^{2\nu-1}\left( \int_0^{\kappa^{\nu}}z\e^{-4\sqrt{2}z^{3/2}/3} \dd z\right)^{1/2} = \tilde{M}\kappa^{2\nu-1}\\
||J_{\Delta}-I||_{L^{\infty}} \leq & 2\e^{-t\zeta_0}\kappa^{2\nu-1} \sup_{[0,\kappa^{\nu}]}|z^{1/2}\e^{-2\sqrt{2}z^{3/2}/3}| = \tilde{M}\kappa^{2\nu-1}.
\end{align*}
Proceeding analogously for the other entries we obtain that for an appropriate choice of constant $M:=M(t_0)$ depending on $t_0$ and uniform on $x \geq \tilde{x}_0$,
$$||J_{\Delta}-I||_{L^1\cap L^2 \cap L^{\infty}( I_{\kappa})} \leq M\kappa^{2\nu-1}.
$$
Now consider $\zeta \in [\tilde{\Sigma}_1\cup \tilde{\Sigma}_3] \backslash I_{\kappa}$. For the entry $12$, we have
\begin{align*}
||J_{\Delta}-I||_{L^1} \leq & 2\e^{-2t\zeta_0}\int|\e^{t\zeta}-\e^{h_{\kappa}(\zeta)}||\im \zeta|^{1/2}\e^{-\frac{2\sqrt{2}}{3}|\im \zeta|^{3/2}}| \dd \zeta \\
\leq & M\int_{\kappa^{\nu}}^{\infty}(\e^{t\zeta_0}+\e^{\re h_{\kappa}(\zeta)})|y|^{1/2}\e^{-\frac{2\sqrt{2}}{3}y^{3/2}} \dd y \leq \tilde{M}\e^{-m\kappa^{3\nu/2}}\\
||J_{\Delta}-I||_{L^2} \leq & M \left( \int_{\kappa^{\nu}}^{\infty}(\e^{t\zeta_0}+\e^{\re h_{\kappa}(\zeta)})^2|y|\e^{-\frac{4\sqrt{2}}{3}y^{3/2}} \dd y\right)^{1/2} \leq \tilde{M}\e^{-m\kappa^{3\nu/2}}\\
||J_{\Delta}-I||_{L^{\infty}} \leq & M \supp_{[\kappa^{\nu}, \infty]}|(\e^{t\zeta_0}+\e^{\re h_{\kappa}(\zeta)})|y|^{1/2}\e^{-\frac{2\sqrt{2}}{3}y^{3/2}}| \leq \tilde{M}\e^{-m\kappa^{3\nu/2}},
\end{align*}
and the same estimate follows analogously for the remaining entries. Therefore, there exist a real constant $m$ and an appropriate choice of constant $M:=M(t_0)$ such that
$$||J_{\Delta}-I||_{L^1\cap L^2 \cap L^{\infty}([\tilde{\Sigma}_1\cup \tilde{\Sigma}_3] \backslash I_{\kappa})} \leq M\e^{-m\kappa^{3\nu/2}},
$$
as claimed. $\square$

\begin{lemma} \label{lemma:smal3}
Let $\zeta \in \tilde{\Sigma}_2 =(-\infty,\zeta_0)$ and $\nu \in(0,\frac{2}{7})$. Fix $t_0, \tilde{x}_0 \in(0,1)$ and let $\eta>0$ be the constant in Assumption \ref{asump3}. There exist a real constant $m$ and an appropriate choice of constant $M:=M(t_0, \eta)$, such that
$$\norm{J_{\Delta}-I}_{L^1\cap L^2 \cap L^{\infty}(\tilde{\Sigma}_2)} = M \max\{\kappa^{\frac{7}{2}\nu-1}, \e^{-m \kappa^{3\nu/2}}\},$$
uniformly for $t \in [t_0, 1/t_0]$ and $ \kappa^{\tilde{\alpha}}\geq x \geq \tilde{x}_0$ for any positive $\tilde{\alpha}$ such that $\tilde{\alpha}=o(\nu)$.
\end{lemma}

\textbf{Proof:} The starting point is the following new expression for the jump matrix
\begin{align*}
J_{\Delta}-I &= \left(\frac{\sigma_0}{\sigma_{\kappa}}-\frac{\sigma_{\kappa}}{\sigma_0}\right) \Psi_{0,+}(\zeta)E_{11}\Psi_{0,+}(\zeta)^{-1}+\left(\frac{\sigma_{\kappa}}{\sigma_{0}} -1\right)I.
\end{align*}
Notice that 
\begin{align*}
\left(\frac{\sigma_{\kappa}}{\sigma_{0}} -1\right) &= \left(\e^{t\zeta-h_{\kappa}(\zeta)}-1\right)\left(1+\e^{-x-h_{\kappa}(\zeta)}\right)^{-1},\\
 \left(\frac{\sigma_0}{\sigma_{\kappa}} -1\right) &= \left(\e^{h_{\kappa}(\zeta)-t\zeta}-1\right)\left(1+\e^{-x-t\zeta}\right)^{-1}.
\end{align*}
Because $h_{\kappa}(\zeta)$ is real-valued in the real line, we have that $|\left(1+\e^{-x-h_{\kappa}(\zeta)}\right)^{-1}| \leq 1$, $|\left(1+\e^{-x-t\zeta}\right)^{-1}| \leq 1$. Take $I_{\kappa} = \{\zeta \in \tilde{\Sigma}_2: \,|\zeta-\zeta_0|\leq \kappa^{\nu}\}$. By Remark \ref{rmk:exp}
$$ \left(\frac{\sigma_{\kappa}}{\sigma_{0}} -1\right) = O\left(\frac{|\zeta-\zeta_0|^2}{\kappa} \right), \qquad \left(\frac{\sigma_0}{\sigma_{\kappa}} -1\right) = O\left(\frac{|\zeta-\zeta_0|^2}{\kappa} \right).
$$
Moreover, by triangular inequality we obtain
$$\left| \frac{\sigma_0}{\sigma_{\kappa}}-\frac{\sigma_{\kappa}}{\sigma_0} \right| \leq \left| \frac{\sigma_0}{\sigma_{\kappa}}-1\right|+ \left|\frac{\sigma_{\kappa}}{\sigma_0} -1 \right|. 
$$
The norm can be computed entry-wise. For the entry $12$, where $|\Psi_{0,+}(\zeta)E_{11}\Psi_{0,+}(\zeta)^{-1}| \leq k(t) |\zeta-\zeta_0|^{1/2}$, the change in variables $z=\zeta_0-\zeta$ leads to
\begin{align*}
||J_{\Delta}-I||_{L^1} \leq & \frac{k_1}{\kappa} \int_0^{\kappa^{\nu}}z^2 \dd z+ \frac{k_2}{\kappa} \int_0^{\kappa^{\nu}}z^{5/2} \dd z = \tilde{k}_1 \kappa^{3\nu-1}+ \tilde{k}_2 \kappa^{\frac{7}{2}\nu-1},\\
||J_{\Delta}-I||_{L^2} \leq & \frac{k_1}{\kappa} \left(\int_0^{\kappa^{\nu}}z^4 \dd z\right)^{1/2}+ \frac{k_2}{\kappa} \left(\int_0^{\kappa^{\nu}}z^{5} \dd z\right)^{1/2} = \tilde{k}_1 \kappa^{\frac{5}{2}\nu-1}+ \tilde{k}_2 \kappa^{3\nu-1},\\
||J_{\Delta}-I||_{L^{\infty}} \leq & \frac{k_1}{\kappa} \sup_{[0,\kappa^{\nu}]} |z^2|+ \frac{k_2}{\kappa} \sup_{[0,\kappa^{\nu}]}|z^{5/2}| = \tilde{k}_1 \kappa^{\frac{5}{2}\nu-1}+ \tilde{k}_2 \kappa^{3\nu-1}.
\end{align*}
Proceeding analogously for the other entries we obtain that for an appropriate choice of constant $M:=M(t_0)$,
$$||J_{\Delta}-I||_{L^1\cap L^2 \cap L^{\infty}( I_{\kappa})} \leq M\kappa^{\frac{7}{2}\nu-1}.
$$
Now consider $\zeta \in \tilde{\Sigma}_2 \backslash I_{\kappa}$. If $|\zeta-\zeta_0| \geq \kappa^{\nu}$, then $|\zeta| \geq |\zeta-\zeta_0|-|\zeta_0| \geq \kappa^{\nu}-1$, and
\begin{align*}
\left|\frac{\sigma_{\kappa}}{\sigma_{0}} -1\right| +\left|\frac{\sigma_0}{\sigma_{\kappa}} -1\right| & \leq 2\left|\e^{x+t\zeta}-\e^{x+h_{\kappa}(\zeta)}\right| \leq 2 \e^{x-(t+\eta)|\zeta|}.
\end{align*}
Therefore, entry $12$ gives us
\begin{align*}
||J_{\Delta}-I||_{L^1} \leq & k_1\e^{x}\int_{\kappa^{\nu}-1}^{\infty}(z+\zeta_0)^{1/2}\e^{-(t+\eta)z}\dd z + k_2\e^{x}\int_{\kappa^{\nu}-1}^{\infty}\e^{-(t+\eta)z}\dd z \\
\leq & \tilde{k}_1\e^{x} \e^{-(t+\eta)(\kappa^{\nu}-1)} + \frac{k_2}{t+\eta}\e^{x}\e^{-(t+\eta)\kappa^{\nu}}\e^{t+\eta} \leq \tilde{M}\e^{-m\kappa^{\nu}},\\
||J_{\Delta}-I||_{L^2} \leq & k_1\e^{x}\left(\int_{\kappa^{\nu}-1}^{\infty}(z+\zeta_0)\e^{-2(t+\eta)z}\dd z\right)^2 + k_2\e^{x}\left(\int_{\kappa^{\nu}-1}^{\infty}\e^{-2(t+\eta)z}\dd z\right)^2 \\
\leq & \tilde{k}_1\e^{x} \e^{-(t+\eta)(\kappa^{\nu}-1)} + \frac{k_2}{\sqrt{2(t+\eta)}}\e^{x}\e^{-(t+\eta)\kappa^{\nu}}\e^{t+\eta} \leq \tilde{M}\e^{-m\kappa^{\nu}},\\
||J_{\Delta}-I||_{L^{\infty}} \leq & k_1\e^{x}\sup_{[\kappa^{\nu}-1, \infty]}(z+\zeta_0)^{1/2}\e^{-(t+\eta)z}+ k_2\e^{x}\sup_{[\kappa^{\nu}-1, \infty]}\e^{-(t+\eta)z} \leq \tilde{M}\e^{-m\kappa^{\nu}}.
\end{align*}
The same estimate follows analogously for the other entries. Consequently, there exist a real constant $m$ depending on $t$ and $\eta$ and an appropriate choice of constant $M:=M(t_0, \eta)$ such that
$$||J_{\Delta}-I||_{L^1\cap L^2 \cap L^{\infty}(\tilde{\Sigma}_2 \backslash I_{\kappa})} \leq M\e^{-m\kappa^{\nu}},
$$
as claimed. $\square$

\begin{remark}
In our particular case, $\kappa = n^{2/3}$ and $x=x_0 n^{\alpha}$. A close inspection of the proof shows us that actually the previous Lemma reads as
$$\norm{J_{\Delta}-I}_{L^1\cap L^2 \cap L^{\infty}(\tilde{\Sigma}_2)} = M \max\{\kappa^{\frac{7}{2}\nu-1}, \e^{-m \kappa^{3\nu/2}}\},$$
for $\nu \in(\frac{3}{2}\alpha,\frac{2}{7})$ uniformly in $x=x_0 n^{\alpha}$, for $\alpha$ under Case 1 in Assumption \ref{asump1}, and
$$\norm{J_{\Delta}-I}_{L^1\cap L^2 \cap L^{\infty}(\tilde{\Sigma}_2)} = M \max\{\kappa^{\frac{5}{2}\nu-1}, \e^{-m \kappa^{3\nu/2}}\},$$
for $\nu \in(\frac{3}{2}\alpha,\frac{2}{5})$ uniformly in $x=x_0 n^{\alpha}$, for $\alpha$ under Case 2 in Assumption \ref{asump1}.
\end{remark}

\begin{theorem}\label{theo:smallnorm}
The solution $\Delta(\zeta)$ to the Riemann-Hilbert problem from Lemma \ref{RHPDelta} exists uniquely. Let $ \kappa^{\tilde{\alpha}}\geq x \geq \tilde{x}_0$. Setting $\tau:=1-\frac{7}{2}\nu \in (0,1)$ for $\tilde{\alpha}\in (0, \frac{4}{21})$ ($\tau:=1-\frac{5}{2}\nu \in (0,1)$ for $\tilde{\alpha}\in (0, \frac{2}{9})$ and $h(\zeta)=t\zeta+O(|\zeta|^3/\kappa^2)$) the following estimate holds
$$\norm{\Delta-I}_{L^2 \cap L^{\infty}(\tilde{\Sigma})} = O(\kappa^{-\tau}),
$$
and $\Delta(\zeta)$ admits an integral representation
\begin{equation}
\Delta(\zeta) = I +\frac{1}{2\pi \i} \int_{\tilde{\Sigma}} \frac{\Delta(z)(J_{\Delta}(z)-I)}{z-\zeta}\dd z.
\label{eq:intdelta}
\end{equation}
Moreover, the following asymptotic expansion holds
\begin{equation}
\Delta(z) = I +O\left(\frac{1}{\kappa^{\tau}(1+|z|)} \right),
\label{eq:deltaasym}
\end{equation}
for $\kappa$ large enough and uniformly in $|z| \leq \kappa^{\nu}$.
\end{theorem}
\textbf{Proof:} Given the $L^2$ and $L^{\infty}$ estimates in Lemmas \ref{lemma:smal1}-\ref{lemma:smal3}, the existence and the norm estimate follows by standard arguments in small norm theory. Now we prove the asymptotic formula \eqref{eq:deltaasym}. Notice that $J_{\Delta}$ is $C^{\infty}$ on $\tilde{\Sigma}$, therefore Holder continuous. Consequently, $\Delta$ extends continuously to boundary values $\Delta_{\pm}$ and
$$\norm{\Delta}_{L^{\infty}(\mathbb{C}\backslash \tilde{\Sigma})} \leq M_{\kappa} := \max \{\norm{\Delta_+}_{L^{\infty}(\tilde{\Sigma})}, \norm{\Delta_-}_{L^{\infty}( \tilde{\Sigma})} \}.
$$
For any $s \in \tilde{\Sigma}$ and $\epsilon>0$ fixed, take the arcs $\partial B_{\epsilon}^{\pm}(s)$ and set $\tilde{\Sigma}^{\pm} = [\tilde{\Sigma}\backslash B_{\epsilon}(s)] \cup \partial B_{\epsilon}^{\pm}(s)$. The integral representation still holds under this deformation, and sending $z \to s$,
\begin{align*}
\Delta_{\pm}(\zeta) &= I +\frac{1}{2\pi i} \int_{\tilde{\Sigma}^{\pm}} \frac{\Delta(z)(J_{\Delta}(z)-I)}{z-\zeta}\dd z\\
|\Delta_{\pm}(\zeta)| &\leq 1 +\frac{1}{\pi \epsilon} \norm{\Delta_-}_{L^{\infty}(\tilde{\Sigma}^{\pm})}\norm{J_{\Delta}-I}_{L^1(\tilde{\Sigma}^{\pm})} \leq 1+\frac{1}{\pi \epsilon} M_{\kappa} \norm{J_{\Delta}-I}_{L^1(\tilde{\Sigma}^{\pm})},
\end{align*}
which implies $M_{\kappa} \leq (1-\frac{1}{\pi \epsilon} \norm{J_{\Delta}-I}_{L^1(\tilde{\Sigma}^{\pm})})^{-1}$. Given the $L^1$ estimates from Lemmas \ref{lemma:smal1}-\ref{lemma:smal3} we have that for $\kappa$ large enough, $M_{\kappa} \leq 2$. Therefore, by the integral representation \eqref{eq:intdelta}, we obtain
$$|\Delta_{\pm}-I| = O(\kappa^{-\tau}).
$$
For the decay in $s$, take $I_s=\{\zeta \in \tilde{\Sigma}^- : |\zeta-s|\geq |s|/2\}$. By triangular inequality, 
\begin{align*}
\left|\int_{\tilde{\Sigma}^{-}} \frac{\Delta(z)(J_{\Delta}(z)-I)}{z-\zeta}\dd z\right| \leq &\left|\int_{I_s} \frac{\Delta(z)(J_{\Delta}(z)-I)}{z-\zeta}\dd z\right| + \left|\int_{\tilde{\Sigma}^{-}\backslash I_s} \frac{\Delta(z)(J_{\Delta}(z)-I)}{z-\zeta}\dd z\right|.
\end{align*}
By previous calculations,
\begin{align*}
\left|\int_{I_s} \frac{\Delta(z)(J_{\Delta}(z)-I)}{z-\zeta}\dd z\right| & \leq 2 M_{\kappa} \sup_{\zeta \in I_s}\frac{1}{|\zeta-s|} \norm{J_{\Delta}-I}_{L^1(I_s)} = O\left(\frac{1}{\kappa^{\tau}|s|}\right),\\
\left|\int_{\tilde{\Sigma}^{-}\backslash I_s} \frac{\Delta(z)(J_{\Delta}(z)-I)}{z-\zeta}\dd z\right| & \leq \frac{4}{\epsilon} \norm{J_{\Delta}-I}_{L^1(\tilde{\Sigma}^{-}\backslash I_s)},
\end{align*}
Therefore, $\norm{J_{\Delta}-I}_{L^1(\tilde{\Sigma}^{-}\backslash I_s)} = o(1)$ uniformly for $|s| \leq \kappa^{\nu}$, and the result follows. $\square$

\section{The Riemann-Hilbert approach for Orthogonal Polynomials}
\label{sec:RHapproach}

The starting problem is the following. Let $\Y(z)= \Y(z; n, x, t)$ depending on parameters $x, n \in \mathbb{R}$ and  $t>0$ be the unique $2 \times 2$ matrix-valued function such that
\begin{problem}\label{RHP1}\hfill
\begin{enumerate}
\item $ \Y(z):= \Y(z; n, x, t)$ is analytic on $\mathbb{C}\setminus \mathbb R$, with continuous boundary values $\Y_\pm$ satisfying the jump condition
\begin{equation}
\Y_+(z)=
\Y_-(z)
\begin{pmatrix}
1 & \omega_n(z) \\ 0 & 1
\end{pmatrix},  z\in \mathbb R,
\end{equation}
where $\omega_n(z):=\omega_n(z;x) = \e^{-nV(z)}\sigma_n(z)$ for $\sigma_n(z)$ given by Equation \eqref{def:sigma}.
\item As $z\to \infty$,
\begin{equation}
\Y(z)=\left(I+\dfrac{\Y^{(1)}}{z}+O\left(\frac{1}{z^2}\right)\right)
z^{n \sigma_3},
\end{equation}
where $\Y^{(1)}:= \Y^{(1)}(n,x,t)$.
\end{enumerate}
\end{problem}

The formulation of this problem goes back to the works of Fokas in '92 \cite{fokas1992}, when it was proved the existence and uniqueness of the solution. Moreover, it connects to orthogonal polynomials in the following way. Let $\uppi_k(z):=\uppi_k^{(n,x)}(z)$ be the monic orthogonal polynomial of degree $k$ with respect to the weight $\omega_n(z)$ defined in Section \ref{sec:intro}. Then,
\begin{equation}
Y(z) = \begin{pmatrix}
\uppi_n^{(n,x)}(z) & \int_{\mathbb{R}}\frac{\uppi_n^{(n,x)}(s) \omega_n(s)}{s-z} \frac{\dd s}{2 \pi \i} \\
-2 \pi \i \upgamma^{(n)}_{n-1}(x)^2\uppi_{n-1}^{(n,x)}(z) & 
-\int_{\mathbb{R}}\frac{\upgamma^{(n)}_{n-1}(x)^2\uppi_{n-1}^{(n,x)}(s)\omega_n(s)}{s-z} \dd s
\end{pmatrix},
\label{eq:RHPYsol}
\end{equation}
is the unique solution to the Riemann-Hilbert problem \ref{RHP1}. From Equation \eqref{eq:RHPYsol} it is straightforward that
\begin{equation}
\upgamma^{(n)}_{n-1}(x)^2 = -\frac{1}{2\pi \i}[Y^{(1)}(n,x)]_{21}.
\label{defgamma}
\end{equation}

Moreover, the Christoffel–Darboux kernel defined in Equation \eqref{eq:defCD} can also be reformulated by means of the Riemann-Hilbert problem. In fact, Equation \eqref{eq:RHPYsol} together with definition \eqref{eq:defCD} gives that, for all $\lambda, \mu \in \mathbb{R}$,
\begin{equation}
K_n^Q(\lambda, \mu; x) = \frac{1}{2\pi \i (\lambda-\mu)} \begin{pmatrix}
0 & 1
\end{pmatrix} Y_+(\mu)^{-1}Y_+(\lambda) \begin{pmatrix}
1 \\ 0
\end{pmatrix},
\end{equation}
and in the confluent limit $\mu \to \lambda$ one obtains that for all $\lambda \in \mathbb{R}$,
\begin{equation}
K_n^Q(\lambda, \lambda; x) = \frac{1}{2\pi \i} \begin{pmatrix}
0 & 1
\end{pmatrix} Y_+(\lambda)^{-1}Y_+'(\lambda) \begin{pmatrix}
1 \\ 0
\end{pmatrix}.
\label{CDconfluent}
\end{equation}

\begin{remark}
    The asymptotic behavior of the solution to this problem had been extensively studied for other choices of weight. For instance, in \cite{deift1999} and \cite{deift1999b} they found asymptotics for general perturbations of the Gaussian weight through this formulation.
\end{remark}

For reasons that will become clearer later, we work with
$$\tilde{\Y}(z) = \Y(z) \e^{-\frac{x}{2}\sigma_3}.
$$

The new Riemann-Hilbert problem reads
\begin{problem}\hfill
\begin{enumerate}
\item $ \tilde{\Y}(z)$ is analytic on $\mathbb{C}\setminus \mathbb R$, with continuous boundary values $\tilde{\Y}_\pm$ satisfying the jump condition
\begin{equation}
\tilde{\Y}_+(z)=
\tilde{\Y}_-(z)
\begin{pmatrix}
1 & \omega_n(z)\e^x \\ 0 & 1
\end{pmatrix},  z\in \mathbb R
\end{equation}
\item As $z\to \infty$,
\begin{equation}
\tilde{\Y}(z)=\left(I+O\left(\frac{1}{z}\right)\right)
z^{n \sigma_3}\e^{-\frac{x}{2}\sigma_3}.
\end{equation}
\end{enumerate}
\end{problem}

The next steps in the asymptotic study of $\tilde{\Y}(z)$ are quite standard, and follow the same reasoning as Section 4 of \cite{deift1999}. For this reason, we save details in the discussion of the first transformations. The main differences from previous works will appear in the construction of the approximate solutions, known as parametrices. The first transformation aims at a better control of the behavior at infinity. Set
\begin{equation}
    \T(z):= \T(z;n) = \e^{-n\ell_V \sigma_3}\tilde{\Y}(z) \e^{n(\phi(z)-\frac{1}{2}V(z))\sigma_3}.
\end{equation}

Then, $\T(z)$ solves the following Riemann-Hilbert

\begin{problem}\label{RHPT}\hfill
\begin{enumerate}
\item $ \T(z)$ is analytic on $\mathbb{C}\setminus \mathbb R$, with continuous boundary values $\T_\pm$ satisfying the jump condition
\begin{equation}
\T_+(z)=
\T_-(z)
\begin{pmatrix}
\e^{n(\phi_+(z)-\phi_-(z))} & \e^x\sigma_n(z)\e^{-n(\phi_+(z)+\phi_-(z))} \\ 0 & \e^{-n(\phi_+(z)-\phi_-(z))}
\end{pmatrix},  \quad z\in \mathbb R
\end{equation}
\item As $z\to \infty$,
\begin{equation}
\T(z)=\left[I+\dfrac{T^{(1)}}{z}+O\left(\frac{1}{z^2}\right)\right]\e^{-\frac{x}{2}\sigma_3}.
\end{equation}
\end{enumerate}
\end{problem}

Because of the properties of $\phi$, we see that the jump behaves as follows
\begin{align*}
    J_{\T} = \left\lbrace \begin{array}{ll}
         \begin{pmatrix}
\e^{n(\phi_+(z)-\phi_-(z))} & \e^x\sigma_n(z) \\ 0 & \e^{-n(\phi_+(z)-\phi_-(z))}
\end{pmatrix}, \hspace{1cm}& z \in (-a,0) \\
   \begin{pmatrix}
1 & \e^x\sigma_n(z)\e^{-n(\phi_+(z)+\phi_-(z))} \\ 0 & 1
\end{pmatrix} ,     & z \in \mathbb{R}/[-a,0].
    \end{array}\right.
\end{align*}

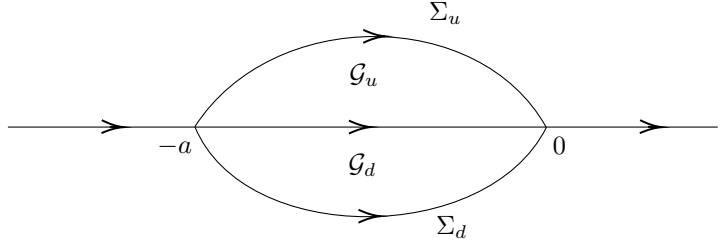
\begin{figure}[H]
\centering
\begin{tikzpicture}[x=0.75pt,y=0.75pt,yscale=-1,xscale=1]

\draw    (167.6,162.45) -- (525.6,162.45) ;
\draw    (261.9,162.24) .. controls (301.09,101.31) and (405.38,101.72) .. (439.14,162.66) ;
\draw    (261.9,162.24) .. controls (286.62,222.77) and (408.99,222.77) .. (439.14,162.66) ;
\draw    (353.83,116.64) -- (355.75,116.64) ;
\draw [shift={(357.75,116.64)}, rotate = 180] [color={rgb, 255:red, 0; green, 0; blue, 0 }  ][line width=0.75]    (10.93,-3.29) .. controls (6.95,-1.4) and (3.31,-0.3) .. (0,0) .. controls (3.31,0.3) and (6.95,1.4) .. (10.93,3.29)   ;
\draw    (346.6,162.45) -- (348.52,162.45) ;
\draw [shift={(350.52,162.45)}, rotate = 180] [color={rgb, 255:red, 0; green, 0; blue, 0 }  ][line width=0.75]    (10.93,-3.29) .. controls (6.95,-1.4) and (3.31,-0.3) .. (0,0) .. controls (3.31,0.3) and (6.95,1.4) .. (10.93,3.29)   ;
\draw    (351.42,207.43) -- (353.34,207.43) ;
\draw [shift={(355.34,207.43)}, rotate = 180] [color={rgb, 255:red, 0; green, 0; blue, 0 }  ][line width=0.75]    (10.93,-3.29) .. controls (6.95,-1.4) and (3.31,-0.3) .. (0,0) .. controls (3.31,0.3) and (6.95,1.4) .. (10.93,3.29)   ;
\draw [line width=0.75]    (218.96,162.33) -- (223.8,162.45) ;
\draw [shift={(225.8,162.5)}, rotate = 181.4] [color={rgb, 255:red, 0; green, 0; blue, 0 }  ][line width=0.75]    (10.93,-3.29) .. controls (6.95,-1.4) and (3.31,-0.3) .. (0,0) .. controls (3.31,0.3) and (6.95,1.4) .. (10.93,3.29)   ;
\draw [line width=0.75]    (490.46,162.33) -- (495.3,162.45) ;
\draw [shift={(497.3,162.5)}, rotate = 181.4] [color={rgb, 255:red, 0; green, 0; blue, 0 }  ][line width=0.75]    (10.93,-3.29) .. controls (6.95,-1.4) and (3.31,-0.3) .. (0,0) .. controls (3.31,0.3) and (6.95,1.4) .. (10.93,3.29)   ;

\draw (378.76,98.03) node [anchor=north west][inner sep=0.75pt]    {$\Sigma _{u}$};
\draw (382.25,206.18) node [anchor=north west][inner sep=0.75pt]    {$\Sigma _{d}$};
\draw (338.64,129.17) node [anchor=north west][inner sep=0.75pt]    {$\mathcal{G}_{u}$};
\draw (441.14,166.06) node [anchor=north west][inner sep=0.75pt]    {$0$};
\draw (338.14,174.67) node [anchor=north west][inner sep=0.75pt]    {$\mathcal{G}_{d}$};
\draw (241.8,166.4) node [anchor=north west][inner sep=0.75pt]    {$-a$};

\end{tikzpicture}
\caption{Opening of lenses.}
\label{fig:open}
\end{figure}

Moreover, since $\e^x\sigma_n(z)=O(\e^{x})$ and $\e^{-n\phi(z)} = O(\e^{-n})$ for $z \in \mathbb{R}/[-a,0]$, it follows that $J_{\T} = I+o(1)$ as $n \to \infty$ in this interval. On the other hand, for $z \in (-a,0)$, the jump oscillates. Due to this behavior, we perform an opening of lenses. Set
\begin{equation}
    \S(z) = \left\lbrace \begin{array}{ll}
         \T(z)\begin{pmatrix}
1 & 0\\ -\e^{-x}\sigma_n(z)^{-1} \e^{2n\phi(z)}  & 1
\end{pmatrix}, \hspace{1cm}& z \in \mathcal{G}_u \\
\T(z)\begin{pmatrix}
1 & 0\\ \e^{-x}\sigma_n(z)^{-1} \e^{2n\phi(z)}  & 1
\end{pmatrix}, \hspace{1cm}& z \in \mathcal{G}_d \\
  \T(z),    & \text{otherwise},
    \end{array}\right.
\end{equation}
where $\mathcal{G}_u$ and $\mathcal{G}_d$ are the regions depicted in Figure \ref{fig:open}.

Then, $\S(z)$ solves the following problem
\begin{problem}\label{RHPT}\hfill
\begin{enumerate}
\item $\S(z)$ is analytic on $\mathbb{C}\setminus \mathbb R$, with continuous boundary values $\S_\pm$ satisfying the jump condition
\begin{equation}
\S_+(z)=
\S_-(z)
\left\lbrace \begin{array}{ll}
\begin{pmatrix}
0 & \e^x\sigma_n(z)\\ -\e^{-x}\sigma_n(z)^{-1} & 0
\end{pmatrix}, \hspace{1cm}& z \in (-a,0), \\
\begin{pmatrix}
1 & 0\\ \e^{-x}\sigma_n(z)^{-1} \e^{2n\phi(z)}  & 1
\end{pmatrix},    & z \in \Sigma_u \cup \Sigma_d,\\
\begin{pmatrix}
1 & \e^x\sigma_n(z)\e^{-2n\phi_+(z)} \\ 0 & 1
\end{pmatrix},    & z \in \mathbb{R}/[-a,0].
    \end{array}\right.
\end{equation}
\item As $z\to \infty$,
\begin{equation}
\S(z)=\left[I+\dfrac{\S^{(1)}}{z}+O\left(\frac{1}{z^2}\right)\right]\e^{-\frac{x}{2}\sigma_3}.
\end{equation}
\end{enumerate}
\end{problem}

Now, as $n \to \infty$, all jumps converge to identity, except for $z \in (-a,0)$. Therefore, the problem can be approximated. to a global parametrix away of the points $\{-a,0\}$, where we should build local solutions. The problem for the global parametrix is given as follows
\begin{problem}\label{RHPG}\hfill
\begin{enumerate}
\item $ \G(z)$ is analytic on $\mathbb{C}\setminus [-a,0]$, with continuous boundary values $\G_\pm$ satisfying the jump condition
\begin{equation}
\G_+(z)=
\G_-(z)
\begin{pmatrix}
0 & \e^x\sigma_n(z)\\ -\e^{-x}\sigma_n(z)^{-1} & 0
\end{pmatrix}, \qquad z \in (-a,0).
\end{equation}
\item As $z\to \infty$,
\begin{equation}
\G(z)=\left[I+\dfrac{\G^{(1)}}{z}+O\left(\frac{1}{z^2}\right)\right]\e^{-\frac{x}{2}\sigma_3}.
\label{eq:asymptG}
\end{equation}
\end{enumerate}
\end{problem}

For $z \in \mathbb{C}/[-a,0]$ we define the following auxiliary function
\begin{equation}
    \g(z) = \frac{((z+a)z)^{1/2}}{2 \pi} \int_{-a}^0 \frac{\log \sigma_n(s)}{\sqrt{|s|(s+a)}} \frac{\d s}{s-z}.
    \label{eq:defg}
\end{equation}
This function has the following immediate properties:
\begin{itemize}
    \item As $z \to \infty$, $\g(z) = \g_0+O(z^{-1})$, where
    $$ \g_0 = -\frac{1}{2 \pi} \int_{-a}^0 \frac{\log \sigma_n(s)}{\sqrt{|s|(s+a)}} \d.$$
    \item for $-a<z<0$,
    $$ \g_+(z)+\g_-(z) = -\log \sigma_n(z).$$
\end{itemize}

Set $\M(z) = \e^{\g_0\sigma_3} \G(z) \e^{-\g(z)\sigma_3}\e^{\frac{x}{2}\sigma_3}$. Then, $\M(z)$ solves the following Riemann-Hilbert problem
\begin{problem}\label{RHPM}\hfill
\begin{enumerate}
\item $ \M(z)$ is analytic on $\mathbb{C}\setminus [-a,0]$, with continuous boundary values $\G_\pm$ satisfying the jump condition
\begin{equation}
\M_+(z)=
\M_-(z)
\begin{pmatrix}
0 & 1\\ -1 & 0
\end{pmatrix}.
\end{equation}
\item As $z\to \infty$,
\begin{equation}
\M(z)=I+\dfrac{\M^{(1)}}{z}+O\left(\frac{1}{z^2}\right).
\end{equation}
\end{enumerate}
\end{problem}
The solution to this last problem is standard in Riemann-Hilbert Theory (see, for instance, \cite{deift1999}), and is given by
$$\M(z) = U_0 \left(\frac{z}{z+a}\right)^{\sigma_3/4} U_0^{-1}.$$

Such solution is not well-behaved in the neighborhood of the endpoints of the support of the equilibrium measure. Therefore we must build local solutions around $\{-a,0\}$. But first we need some asymptotic estimates on the auxiliary $\g$-function. Take $\alpha$ under one of the cases in Assumption \ref{asump1} and consider the following lemma,

\begin{lemma}
\label{lemma:GG21}
Take the Laplace-type integral
\begin{equation}
F(n):= \int_0^a g(s) \ln(1+\e^{x-n^{2/3}f(s)}) \dd s, \quad a \in (0, \infty].
\end{equation}
Suppose $f$ is $\mathcal{C}^{\infty}$ in a neighborhood of the origin, with unique global minimum on $[0,a]$ at $s=0$ with $f(0)=0$, $f'(0)>0$, $g \in L^1(0,a)$, and for some $\delta>0$ it is of the form $g(s) = \dfrac{\tilde{g}(s)}{s^{1/2}}$ for $0<s<\delta$, where $\tilde{g} \in \mathcal{C}^{\infty}$ in a neighborhood of the origin. Let $x=x_0n^{\alpha}$ for $\alpha$ under Assumption \ref{asump1}. Then $F(n)$ assumes an expansion of the form
\begin{equation}
F(n) = n^{-\frac{1}{3}+\frac{3\alpha}{2}} \hat{g}(0)\frac{4x_0^{3/2}}{3}+n^{-\frac{1}{3}-\frac{\alpha}{2}} x_0^{-1/2}\hat{g}(0)F_1^{(0)} +
n^{-1+\frac{5\alpha}{2}}\hat{g}'(0)\frac{4x_0^{5/2}}{15}+O(n^{-1/3}), \quad n \to \infty,
\label{eq:GG21}
\end{equation}
where $F_1^{(0)} = \int_0^{x} \left((1+z/x)^{-1/2}  +(1-z/x)^{-1/2}\right) \ln(1+\e^{-z}) \dd z$ and the function $\hat{g}$ satisfies $ \hat{g}(f(s)) = f(s)^{1/2}g(s),$ for $|s|$ sufficiently small.
\end{lemma}

\textbf{Proof:} The arguments are similar to Appendix A of \cite{GG21}. The main difference in the analysis is that, as $x \to \infty$, the argument of the logarithm has to be analyzed more carefully. We first show that the result holds for $f(s)=s$. Notice that the greater contribution comes from the origin: since $\alpha< 2/3$, for $n$ large enough there is $m>0$ such that
$$F(n):= \int_0^{\delta} g(s) \ln(1+\e^{x-n^{2/3}s}) \dd s + O(\e^{-mn^{2/3}}).
$$
For the first integral, we can use the Taylor expansion of the function $\tilde{g}$ and rewrite the integral as
\begin{align*}
\int_0^{\delta} g(s) \ln(1+\e^{x-n^{2/3}s}) \dd s =& \tilde{g}(0)\int_0^{\delta} s^{-1/2} \ln(1+\e^{x-n^{2/3}s}) \dd s + \tilde{g}'(0)\int_0^{\delta} s^{1/2} \ln(1+\e^{x-n^{2/3}s}) \dd s + E_3(n),
\end{align*}
where
$$|E_3(n)| \leq \frac{1}{2}\sup_{s \in (0,\delta)}|\tilde{g}''(s)|\left|\int_0^{\delta} s^{3/2} \ln(1+\e^{x-n^{2/3}s}) \dd s\right| = O(n^{\frac{9\alpha}{2}-\frac{7}{3}}).
$$
For the first of the remaining integrals, notice that
\begin{align*}
\int_0^{\delta} s^{-1/2} \ln(1+\e^{x-n^{2/3}s}) \dd s 
=&\frac{1}{n^{1/3}}\left[-\int_{-x}^0 \frac{z}{\sqrt{z+x}}\dd z +\int_0^{x} \left(\frac{1}{\sqrt{z+x}}+\frac{1}{\sqrt{x-z}}  \right) \ln(1+\e^{-z}) \dd z \right.\\
&\left.+\int_{x}^{\delta n^{2/3}-x}\frac{1}{\sqrt{z+x}}\ln(1+\e^{-z}) \dd z\right] = \frac{4x^{3/2}}{3n^{1/3}}+O(n^{-1/3}x^{-1/2}).
\end{align*}
Analogously,
\begin{align*}
\int_0^{\delta} s^{1/2} \ln(1+\e^{x-n^{2/3}s}) \dd s =&\frac{1}{n}\left[-\int_{-x}^0 z\sqrt{z+x}\dd z +\int_0^{x} \left(\sqrt{z+x}+\sqrt{x-z}  \right) \ln(1+\e^{-z}) \dd z \right.\\
&\left.+\int_{x}^{\delta n^{2/3}-x}\sqrt{z+x}\ln(1+\e^{-z}) \dd z\right] = \frac{4x^{5/2}}{15n}+O(n^{-1}x^{1/2}).
\end{align*}
Altogether,
\begin{align*}
F(n) =& n^{-\frac{1}{3}+\frac{3\alpha}{2}} \tilde{g}(0)\frac{4x_0^{3/2}}{3}+n^{-\frac{1}{3}-\frac{\alpha}{2}} x_0^{-1/2}\tilde{g}(0)F_1^{(0)} + n^{-\frac{1}{3}} \tilde{g}(0)F_1^{(1)} +
n^{-1+\frac{5\alpha}{2}}\tilde{g}'(0)\frac{4x_0^{5/2}}{15}+n^{-1+\frac{\alpha}{2}} x_0^{1/2}\tilde{g}'(0)F_2^{(0)} \\
&+ n^{-1} \tilde{g}'(0)F_2^{(1)}+ O(n^{\frac{9\alpha}{2}-\frac{7}{3}}),
\end{align*}
where
\begin{align*}
F_1^{(0)} =& \int_0^{x} \left((1+z/x)^{-1/2}  +(1-z/x)^{-1/2}\right) \ln(1+\e^{-z}) \dd z & F_1^{(1)} =&\int_{x}^{\delta n^{2/3}-x}(z+x)^{-1/2}\ln(1+\e^{-z}) \dd z\\
F_2^{(0)} =& \int_0^{x} \left((1+z/x)^{1/2}  +(1-z/x)^{1/2}\right) \ln(1+\e^{-z}) \dd z & F_2^{(1)} =&\int_{x}^{\delta n^{2/3}-x}(z+x)^{1/2}\ln(1+\e^{-z}) \dd z.
\end{align*}
The general result then follows by the Inverse Function Theorem. $\square$\\

Applied to our $\g$-function \eqref{eq:defg}, this lemma leads to the following estimate:
\begin{lemma}
\label{lemma:estimateg}
For any fixed $x_0>0$, the estimate
\begin{align*}
\g_0(z) &= \dfrac{2 \sqrt{t}x_0^{3/2}}{3\pi \sqrt{a}}n^{-\frac{1}{3}+\frac{3\alpha}{2}}+ O(n^{-\frac{1}{3}-\frac{\alpha}{2}}).
\end{align*}
is valid uniformly for $x = x_0 n^{\alpha}$, $\alpha\in [\epsilon, \frac{2}{9}-\epsilon]$, as $n \to \infty$. Moreover, the estimate
\begin{align*}
\g(z) &= \left(\frac{z}{z+a}\right)^{-1/2}\dfrac{2 \sqrt{t}x_0^{3/2}}{3\pi\sqrt{a}}n^{-\frac{1}{3}+\frac{3\alpha}{2}}+ O(n^{-\frac{1}{3}-\frac{\alpha}{2}}).
\end{align*}
is valid for $z$ on compacts of $\mathbb{C}\backslash [-a,0]$ and for $x = x_0 n^{\alpha}$ as $n \to \infty$.
\end{lemma}
\textbf{Proof:} With the change of variables $s \mapsto -s$, one has
\begin{align*}
I_{\g_0} &= -\int_0^a \dfrac{\log (1+\e^{x-n^{2/3}Q(-s)})}{\sqrt{s}\sqrt{a-s}} \dd s, &
I_{\q} &= \int_0^a \dfrac{\log (1+\e^{x-n^{2/3}Q(-s)})}{\sqrt{s}\sqrt{a-s}} \dfrac{\dd s}{s+z}.
\end{align*}
Applying Lemma \ref{lemma:GG21}, one has
\begin{align*}
\g(z) &= \dfrac{((z+a)z)^{\frac{1}{2}}}{2\pi} I_{\g} = \dfrac{2\sqrt{t}}{3\pi \sqrt{a}} \dfrac{\sqrt{z+a}}{\sqrt{z}}n^{-\frac{1}{3}+\frac{3\alpha}{2}}x_0^{\frac{3}{2}}+ O(n^{-\frac{1}{3}-\frac{\alpha}{2}}), \\
\g_0(z) &= -\dfrac{1}{2\pi} I_{\g_0} = \dfrac{2 \sqrt{t}}{3\pi \sqrt{a}}n^{-\frac{1}{3}+\frac{3\alpha}{2}}x_0^{\frac{3}{2}}+ O(n^{-\frac{1}{3}-\frac{\alpha}{2}}),
\end{align*}
as claimed. $\square$\\

\begin{corollary} \label{corol:expMG}
Under the same assumptions as Lemma \ref{lemma:estimateg}, it follows that for $\gamma = \frac{1}{3}-\frac{3\alpha}{2}$,
\begin{align*}
\M(z) \e^{\frac{-x}{2}\sigma_3}\G^{-1}(z) &= I+O(n^{-\gamma}).
\end{align*}
\end{corollary}
\textbf{Proof:} A straightforward calculation shows that
\begin{align*}
\M(z) \e^{-\frac{x}{2}\sigma_3}\G^{-1}(z) &= \M(z)\e^{-\g(z)\sigma_3}M^{-1}(z)\e^{\g_0\sigma_3} \\
&= \M(z)\left[I-\g(z)\sigma_3+ \dfrac{\g^2(z)\sigma_3^2}{2}+\cdots\right]\M^{-1}(z)\left[ I+\g_0\sigma_3+\dfrac{\g_0^2\sigma_3^2}{2}+\cdots \right] \\
&= I-n^{-\gamma}x_0^{3/2} \dfrac{2\sqrt{t}}{3 \pi \sqrt{a}}\left[\sigma_3-\dfrac{1}{m^{1/2}(z)}\M(z)\sigma_3\M^{-1}(z)\right] + O(n^{-\beta}),
\end{align*}
where $\gamma = \frac{1}{3}-\frac{3\alpha}{2}$ and $\beta = \min \left\{\frac{2}{3}-3\alpha, \frac{1}{3}+\frac{\alpha}{2}\right\}$. $\square$

\subsection{Local parametrix around $-a$}
\label{sec:paramA}

Fix $\delta>0$ and take $B_{\delta}(-a)$ a ball around $-a$. The parametrix solution $P^{(a)}(z)$ must solve the following problem
\begin{problem}\label{RHPPa}\hfill
\begin{enumerate}
\item $ P^{(a)}(z)$ is analytic on $B_{\delta}(-a)\setminus (\mathbb R\cup \Sigma_u \cup \Sigma_d)$, where $\Sigma_u, \Sigma_d$ are the same as in Figure \ref{fig:open}, with boundary values satisfying the jump condition
\begin{equation}
P^{(a)}_+(z)=
P^{(a)}_-(z)
\left\lbrace \begin{array}{ll}
\begin{pmatrix}
0 & \e^{x}\sigma_n(z)\\ -\e^{-x}\sigma_n(z)^{-1} & 0
\end{pmatrix}, \hspace{1cm}& z \in (-a,0)\cap B_{\delta}(-a), \\
\begin{pmatrix}
1 & 0\\ \e^{-x}\sigma_n(z)^{-1} \e^{2n\phi(z)}  & 1
\end{pmatrix},    & z \in (\Sigma_u \cup \Sigma_d) \cap B_{\delta}(-a),\\
\begin{pmatrix}
1 & \e^x\sigma_n(z)\e^{-2n\phi_+(z)} \\ 0 & 1
\end{pmatrix},    & z \in (-\infty, -a) \cap B_{\delta}(-a).
 \end{array}\right.
\end{equation}
\item As $n\to \infty$,
\begin{equation}
P^{(a)}(z) \G^{-1}(z)=I+o(1).
\end{equation}
\item As $z\to -a$, the solution remains bounded.
\end{enumerate}
\end{problem}

For $x=x_0 n^{\alpha}$, $\alpha$ under Assumption \ref{asump1}, one has the existence of a constant $m>0$ such that $|\e^{x-n^{2/3}Q(z)}|\leq \e^{-n^{2/3} m}$ for all $z \in B_{\delta}(-a)$. Moreover, $\sigma_n(z)$ is analytic in $B_{\delta}(-a)$ and one can conjugate $\sigma_n(z)$ out of the jumps, as follows. Set
$$L(z) = P^{(a)}(z)\e^{\frac{1}{2}\log \sigma_n(z)\sigma_3}\e^{\frac{x}{2}\sigma_3}\e^{-n\phi(z)\sigma_3}.
$$

Then, $L(z)$ solves the following Riemann-Hilbert problem

\begin{enumerate}
\item $ L(z)$ is analytic on $B_{\delta}(-a)\setminus (\mathbb R\cup \Sigma_u \cup \Sigma_d)$, with continuous boundary values $L_\pm$ satisfying the jump condition
\begin{equation}
L_+(z)=
L_-(z)
\left\lbrace \begin{array}{ll}
\begin{pmatrix}
0 & 1\\ -1 & 0
\end{pmatrix}, \hspace{1cm}& z \in (-a,0)\cap B_{\delta}(-a), \\
\begin{pmatrix}
1 & 0\\ 1  & 1
\end{pmatrix},    & z \in (\Sigma_u \cup \Sigma_d) \cap B_{\delta}(-a),\\
\begin{pmatrix}
1 & 1 \\ 0 & 1
\end{pmatrix},    & z \in (-\infty, -a) \cap B_{\delta}(-a).
 \end{array}\right.
\end{equation}
\item As $n\to \infty$,
\begin{equation}
L(z) =[I+o(1)]\G(z)\e^{\frac{x}{2}\sigma_3}\e^{\frac{1}{2}\log \sigma_n(z)\sigma_3}\e^{-n\phi(z)\sigma_3},
\end{equation}
where $\e^{\frac{1}{2}\log \sigma_n(z)\sigma_3} = I+O(\e^{-n^{2/3}m})$.
\item As $z\to -a$, the solution remains bounded.
\end{enumerate}

In the local variable $\zeta = n^{2/3}\varphi(z)$, we look for $2\times 2$ matrix-valued function $\Psi_{\Ai}(\zeta)$ such that

\begin{enumerate}
\item $\Psi_{\Ai}(\zeta)$ is analytic on $\mathbb{C}\setminus (\mathbb R\cup \e^{2\pi/3}\mathbb{R}_- \cup \e^{4\pi/3}\mathbb{R}_-)$, with continuous boundary values $\Psi_{\Ai,\pm}$ satisfying the jump condition
\begin{equation}
\Psi_{\Ai,+}(\zeta)=
\Psi_{\Ai,-}(\zeta)
\left\lbrace \begin{array}{ll}
\begin{pmatrix}
0 & 1\\ -1 & 0
\end{pmatrix}, \hspace{1cm}& \zeta \in (0, \infty), \\
\begin{pmatrix}
1 & 0\\ 1  & 1
\end{pmatrix},    & \zeta \in \e^{2\pi/3}\mathbb{R}_- \cup \e^{4\pi/3}\mathbb{R}_-,\\
\begin{pmatrix}
1 & 1 \\ 0 & 1
\end{pmatrix},    & z \in (-\infty, 0).
 \end{array}\right.
\end{equation}
\item As $\zeta\to \infty$,
\begin{equation}
\Psi_{\Ai}(\zeta) =\zeta^{\sigma_3/4}U_0[I+O(\zeta^{-3/2})]\e^{-\frac{2}{3}\zeta^{3/2}\sigma_3}.
\end{equation}
\item As $\zeta\to 0$, the solution remains bounded.
\end{enumerate}

The solution is given by means of the Airy function and its derivatives. Here the precise formula is omitted, but we refer to \cite{GG21} for further discussions. The solution to the Riemann-Hilbert problem \ref{RHPPa} is given by
\begin{align*}
P^{(a)}(z)= E(z)\Psi_{\Ai}(n^{2/3}\varphi(z))\e^{-\frac{1}{2}\log \sigma_n(z)\sigma_3}\e^{n\phi(z)\sigma_3}\e^{-\frac{x}{2}\sigma_3},\\
E(z) = \M(z)U_0^{-1}(n^{2/3}\varphi(z))^{-\sigma_3/4}.
\end{align*}

Moreover, the asymptotic condition for $\Psi_{\Ai}$ as $\zeta\to \infty$ implies that
\begin{align*}
P^{(a)}(z) =& (I+O(n^{-1}))\M(z)\e^{-\frac{x}{2}\sigma_3}+O(\e^{-n^{2/3}\eta})= (I+O(n^{-\gamma}))\G(z).
\end{align*}

\subsection{Local parametrix around $0$}
\label{sec:param0}

Take $\mathcal{U}^0$ a neighborhood of the origin. The parametrix $P^{(0)}(z)$ must solve the following problem

\begin{problem}\label{RHPP0}\hfill
\begin{enumerate}
\item $P^{(0)}(z)$ is analytic on $\mathcal{U}^0\setminus (\mathbb R\cup \Sigma_u \cup \Sigma_d)$, where $\Sigma_u, \Sigma_d$ are the same as in Figure \ref{fig:open}, with boundary values satisfying the jump condition
$$
P_+^{(0)}(z)=
P_-^{(0)}(z)
\times 
\begin{cases}
\begin{pmatrix}
1 & \e^x\sigma_n(z) \e^{-2n\phi_+(z)}\\ 0 & 1
\end{pmatrix}, & z\in (0, \infty)\cap \mathcal{U}^0, \\
\begin{pmatrix}
1 & 0 \\ \e^{-x}\sigma_n(z)^{-1} \e^{2n\phi(z)} & 1
\end{pmatrix}, & z\in (\Sigma_d \cup \Sigma_u)\cap \mathcal{U}^0, \\
\begin{pmatrix}
0 & \e^x\sigma_n \\ -\e^{-x}\sigma_n(z)^{-1} & 0
\end{pmatrix}, & z\in (-\infty,0)\cap \mathcal{U}^0, 
\end{cases}
$$
\item As $n\to \infty$, for $z \in \partial \mathcal{U}^0$
$$
P^{(0)}(z)\G(z)^{-1}=\left(I+o(1)\right).
$$
\item The solution remains bounded as $z \to 0$.
\end{enumerate}
\end{problem}

Set $ L(z) = P^{(0)}(z)\e^{-n\phi(z)\sigma_3}$. Then, $L(z)$ solves the following Riemann-Hilbert problem
\begin{enumerate}
\item $L(z)$ has boundary values related by the jump conditions
$$
L_+(z)=
L_-(z)
\times 
\begin{cases}
\begin{pmatrix}
1 & \e^x\sigma_n(z) \\ 0 & 1
\end{pmatrix}, & z\in (0, \infty)\cap \mathcal{U}^0, \\
\begin{pmatrix}
1 & 0 \\ \e^{-x}\sigma_n(z)^{-1} & 1
\end{pmatrix}, & z\in (\Sigma_d \cup \Sigma_u)\cap \mathcal{U}^0, \\
\begin{pmatrix}
0 & \e^x\sigma_n \\ -\e^{-x}\sigma_n(z)^{-1} & 0
\end{pmatrix}, & z\in (-\infty,0)\cap \mathcal{U}^0. 
\end{cases}
$$
\item As $n\to \infty$, for $z \in \delta U$
$$
L(z)=\left(I+o(1)\right)\G(z)\e^{-n\phi(z)\sigma_3}.
$$
\item The solution remains bounded as $z \to 0$.
\end{enumerate}

The results of Section \ref{sec:model} imply that the solution is given by $L(z) = E_n(z) \tilde{\Psi}_n(n^{2/3}\varphi(z)),$ where
\begin{align*}
E_n(z)=& \M(z)U_0[n^{2/3}\varphi(z)]^{-\sigma_3/4}\E^{-1}\e^{-\frac{x}{2}\sigma_3}\e^{\frac{\log(1+\e^{x+t\zeta_0})}{2}\sigma_3},\\
\tilde{\Psi}_n(\zeta=n^{2/3}\varphi(z))=& \Phi_{\kappa = n^{2/3}}(\zeta; h_{\kappa}(\zeta) = -\kappa Q(\varphi^{-1}(\zeta/\kappa))).
\end{align*}

The asymptotic behavior for $P^{(0)}(z)$ then becomes
\begin{align*}
P^{(0)}(z) =&  \left(I+O(n^{-1/3})\right)\M(z)\e^{-\frac{x}{2}\sigma_3} = \left(I+O(n^{-\gamma})\right)\G(z).
\end{align*}

\subsection{Small norm for the orthogonal polynomials RHP}

Now set
\begin{align*}
\R(z)= \S(z) \begin{cases}
[P^{(0)}(z)]^{-1}, \hspace{0.5cm} & z \in \mathcal{U}^0,\\ 
[P^{(a)}(z)]^{-1},  & z \in B_{\delta}(-a), \\ 
\G(z)^{-1}, & \text{ elsewhere}. \end{cases}
\end{align*}

\begin{lemma}
The matrix-valued function $\R(z)$ solves the following Riemann-Hilbert problem
\begin{enumerate}
\item $ \R(z)$ is analytic on $\mathbb{C}\setminus \Sigma_{\R}$, where $\Sigma_R = \partial \mathcal{U}^0\cup\partial B_{\delta}(-a) \cup \Sigma_{\S}\backslash (\mathcal{U}^0 \cup B_{\delta}(-a)\cup[-a,0])$ with continuous boundary values $\R_\pm$ satisfying the jump condition
\begin{equation}
\R_+(z)=
\R_-(z) \begin{cases}
\G_+(z)J_{\S} \G_+(z)^{-1}, & z \in \Sigma_{\R} \backslash (\mathcal{U}^0 \cup B_{\delta}(-a)),\\
P^{(0)}(z)\G(z)^{-1}, & z \in \mathcal{U}^0,\\
P^{(a)}(z)\G(z)^{-1}, & z \in B_{\delta}(-a).
\end{cases}.
\end{equation}
\item As $z\to \infty$,
\begin{equation}
\R(z)=I+\frac{\R^{(1)}}{z}+O(z^{-2}).
\label{eq:asymptR}
\end{equation}
\end{enumerate}
\end{lemma}
\textbf{Proof:} By construction, $\R(z)$ is analytic for all $z \in\Sigma_{\S}\cup \partial \mathcal{U}^0\cup\partial B_{\delta}(-a)$ Moreover, since $S(z)$ has the same jumps as $P^{(0)}(z)$ in the interior of $\mathcal{U}^0$ and the same jumps as $P^{(a)}(z)$ in the interior of $B_{\delta}(-a)$, it follows that $\R(z)$ is analytic in $\mathcal{U}^0 \cup B_{\delta}(-a)$. In $\partial \mathcal{U}^0$ we have
\begin{align*}
\R_+(z) =& \S(z)\G(z)^{-1} = \S(z)[P^{(0)}(z)]^{-1}P^{(0)}(z)\G(z)^{-1} = \R_-(z) P^{(0)}(z)\G(z)^{-1},
\end{align*}
and $J_{\R}(z) = P^{(0)}(z)\G(z)^{-1}$. Analogously, for $z \in B_{\delta}(-a)$ we obtain  $J_{\R}(z) = P^{(a)}(z)\G(z)^{-1}$. Finally, for $z \in \Sigma_{\S}\backslash (\overline{\mathcal{U}^0 \cup B_{\delta}(-a)})$, 
\begin{align*}
\R_+(z) =& \S_+(z)\G_+(z)^{-1} = \S_-(z)\G_-(z)^{-1} \G_-(z)J_{\S} \G_+(z)^{-1} = \R_-(z) \G_-(z)J_{\S} \G_+(z)^{-1}.
\end{align*}
Now the claimed jump follows from the fact that $\G_+(z)=\G_-(z)J_{\S}$ for $z \in (-a,0)$ and $\G_+(z)=\G_-(z)$ for $z \in \Sigma_{\S}\backslash (\overline{\mathcal{U}^0 \cup B_{\delta}(-a)}\cup[-a,0])$. For the asymptotic condition, as $z \to \infty$
\begin{align*}
\R(z)= \S(z)\G(z)^{-1} = I + \frac{\R^{(1)}}{z}+O(z^{-2}),
\end{align*}
where $\R^{(1)} = \S^{(1)}-\G^{(1)}$. $\square$\\

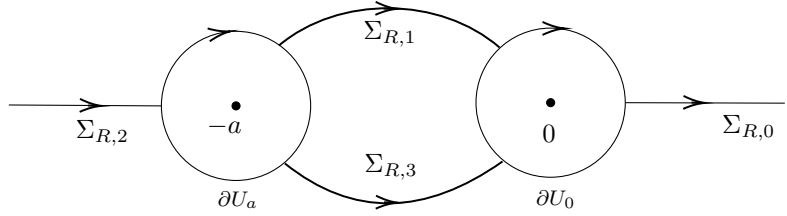
\begin{figure}[H]
\centering
\begin{tikzpicture}[x=0.75pt,y=0.75pt,yscale=-1,xscale=1]

\draw   (90.4,151.14) .. controls (90.4,130.35) and (107.26,113.5) .. (128.04,113.5) .. controls (148.83,113.5) and (165.68,130.35) .. (165.68,151.14) .. controls (165.68,171.93) and (148.83,188.78) .. (128.04,188.78) .. controls (107.26,188.78) and (90.4,171.93) .. (90.4,151.14) -- cycle ;
\draw    (13.45,150.68) -- (90.4,151.14) ;
\draw   (249.07,149.52) .. controls (249.07,128.73) and (265.93,111.88) .. (286.71,111.88) .. controls (307.5,111.88) and (324.35,128.73) .. (324.35,149.52) .. controls (324.35,170.3) and (307.5,187.16) .. (286.71,187.16) .. controls (265.93,187.16) and (249.07,170.3) .. (249.07,149.52) -- cycle ;
\draw    (324.35,149.52) -- (404.82,149.76) ;
\draw  [fill={rgb, 255:red, 0; green, 0; blue, 0 }  ,fill opacity=1 ] (284.69,149.52) .. controls (284.69,148.4) and (285.59,147.49) .. (286.71,147.49) .. controls (287.83,147.49) and (288.74,148.4) .. (288.74,149.52) .. controls (288.74,150.63) and (287.83,151.54) .. (286.71,151.54) .. controls (285.59,151.54) and (284.69,150.63) .. (284.69,149.52) -- cycle ;
\draw  [fill={rgb, 255:red, 0; green, 0; blue, 0 }  ,fill opacity=1 ] (126.02,151.14) .. controls (126.02,150.02) and (126.92,149.11) .. (128.04,149.11) .. controls (129.16,149.11) and (130.07,150.02) .. (130.07,151.14) .. controls (130.07,152.26) and (129.16,153.17) .. (128.04,153.17) .. controls (126.92,153.17) and (126.02,152.26) .. (126.02,151.14) -- cycle ;
\draw [line width=0.75]    (152.67,180.17) .. controls (186.95,207.68) and (223.45,206.68) .. (262.45,179.18) ;
\draw [line width=0.75]    (149.67,120.67) .. controls (179.95,95.18) and (230.95,95.68) .. (260.95,121.68) ;
\draw [line width=0.75]    (202.67,199.83) -- (207.5,199.95) ;
\draw [shift={(209.5,200)}, rotate = 181.4] [color={rgb, 255:red, 0; green, 0; blue, 0 }  ][line width=0.75]    (10.93,-3.29) .. controls (6.95,-1.4) and (3.31,-0.3) .. (0,0) .. controls (3.31,0.3) and (6.95,1.4) .. (10.93,3.29)   ;
\draw [line width=0.75]    (357.75,149.47) -- (362.59,149.59) ;
\draw [shift={(364.58,149.64)}, rotate = 181.4] [color={rgb, 255:red, 0; green, 0; blue, 0 }  ][line width=0.75]    (10.93,-3.29) .. controls (6.95,-1.4) and (3.31,-0.3) .. (0,0) .. controls (3.31,0.3) and (6.95,1.4) .. (10.93,3.29)   ;
\draw [line width=0.75]    (202.5,102) -- (207.33,102.12) ;
\draw [shift={(209.33,102.17)}, rotate = 181.4] [color={rgb, 255:red, 0; green, 0; blue, 0 }  ][line width=0.75]    (10.93,-3.29) .. controls (6.95,-1.4) and (3.31,-0.3) .. (0,0) .. controls (3.31,0.3) and (6.95,1.4) .. (10.93,3.29)   ;
\draw [line width=0.75]    (120.95,114.18) -- (123.07,113.83) ;
\draw [shift={(125.04,113.5)}, rotate = 170.57] [color={rgb, 255:red, 0; green, 0; blue, 0 }  ][line width=0.75]    (10.93,-3.29) .. controls (6.95,-1.4) and (3.31,-0.3) .. (0,0) .. controls (3.31,0.3) and (6.95,1.4) .. (10.93,3.29)   ;
\draw [line width=0.75]    (293.55,112.04) ;
\draw [shift={(293.55,112.04)}, rotate = 180] [color={rgb, 255:red, 0; green, 0; blue, 0 }  ][line width=0.75]    (10.93,-3.29) .. controls (6.95,-1.4) and (3.31,-0.3) .. (0,0) .. controls (3.31,0.3) and (6.95,1.4) .. (10.93,3.29)   ;
\draw [line width=0.75]    (53.37,150.97) -- (58.2,151.09) ;
\draw [shift={(60.2,151.14)}, rotate = 181.4] [color={rgb, 255:red, 0; green, 0; blue, 0 }  ][line width=0.75]    (10.93,-3.29) .. controls (6.95,-1.4) and (3.31,-0.3) .. (0,0) .. controls (3.31,0.3) and (6.95,1.4) .. (10.93,3.29)   ;

\draw (112.28,155.73) node [anchor=north west][inner sep=0.75pt]    {$-a$};
\draw (278.17,192.4) node [anchor=north west][inner sep=0.75pt]  [font=\footnotesize]  {$\partial U_{0}$};
\draw (281.28,159.73) node [anchor=north west][inner sep=0.75pt]    {$0$};
\draw (372,154.4) node [anchor=north west][inner sep=0.75pt]    {$\Sigma _{R,0}$};
\draw (191,106.9) node [anchor=north west][inner sep=0.75pt]    {$\Sigma _{R,1}$};
\draw (46,156.9) node [anchor=north west][inner sep=0.75pt]    {$\Sigma _{R,2}$};
\draw (191.5,173.9) node [anchor=north west][inner sep=0.75pt]    {$\Sigma _{R,3}$};
\draw (117.67,192.9) node [anchor=north west][inner sep=0.75pt]  [font=\footnotesize]  {$\partial U_{a}$};

\end{tikzpicture}
\caption{Contour for the Riemann-Hilbert problem $\R(z)$.}
\end{figure}

\begin{lemma}
\label{lemma:smallJR}
Let $t_0 \in (0,1)$ and $x_0>0$ a real positive fixed constant. Take $\alpha$ under Assumption \ref{asump1} and set $\gamma=\frac{1}{3}-\frac{3 \alpha}{2}$. Then, there exist $m>0$ such that 
\begin{align*}
\norm{J_{\R}-I}_{L^1\cap L^2\cap L^{\infty}(\partial \mathcal{U}^0)\cup\partial B_{\delta}(-a))} =& O(n^{-\gamma})\\
\norm{J_{\R}-I}_{L^1\cap L^2\cap L^{\infty}(\Sigma_{\R} \backslash(\partial \mathcal{U}^0 \cup \partial B_{\delta}(-a))} =& O(\e^{-mn}).
\end{align*}
uniformly in $t \in [t_0, 1/t_0]$ and $x=x_0 n^{\alpha}$.
\end{lemma}
\textbf{Proof:} Take $z \in \partial \mathcal{U}^0$. From Section \ref{sec:param0} it follows that
$$|J_{\R}(z)-I| = |P^{(0)}(z)\G_+(z)^{-1}-I| = O(n^{-\gamma}).
$$
Now, because $\mathcal{U}^0$ is a bounded set, this estimate implies that
$$\norm{J_{\R}-I}_{L^1\cap L^2\cap L^{\infty}(\partial \mathcal{U}^0)} = O(n^{-\gamma}).
$$
Analogously, for $z \in B_{\delta}(-a)$ from Section \ref{sec:paramA} it follows that
$$|J_{\R}(z)-I| = |P^{(0)}(z)\G_+(z)^{-1}-I| = O(n^{-\gamma}).
$$
Once again, because $B_{\delta}(-a)$ is bounded, the claimed decay order follows. At last, we look at $z \in \Sigma_{\R}\backslash(\partial \mathcal{U}^0 \cup \partial B_{\delta}(-a))$. Pick $\tilde{\delta}>0$ such that $B_{\tilde{\delta}}(0) \subset \mathcal{U}^0$. By Lemma \ref{lemma:GG21}, there exist $\tilde{m}, m'>0$ such that
\begin{align}
\re \phi_+(z) &\geq \tilde{m}, \qquad z \in \mathbb{R}\backslash (-a-\delta, \tilde{\delta}), \label{eq:proof1}\\
\re \phi_+(z) &\leq -m', \qquad z \in \Sigma_{\R,1}\cup \Sigma_{\R,3}.
\end{align}
Moreover, recall that $\sigma_n$ is bounded in the real line and $\sigma_n(z)^{-1}=O(\e^{cn^{2/3}})$ for $z \in \Sigma_{\R,1}\cup \Sigma_{\R,3}$.
Take, for instance, $z \in \Sigma_{\R,0}$. Then,
\begin{align*}
|J_{\R}(z)-I| =& |\G_+(z)J_{\S} \G_+(z)^{-1}-I| = |\sigma_n(z)\e^{-2n\phi_+(z)}\G_+(z)\e^{\frac{x}{2}\sigma_3}E_{12} \e^{-\frac{x}{2}\sigma_3}\G_+(z)^{-1}| \\
\leq& |\sigma_n(z)||\e^{-2n\phi_+(z)}||\G_+(z)\e^{\frac{x}{2}\sigma_3}E_{12} \e^{-\frac{x}{2}\sigma_3}\G_+(z)^{-1}|.
\end{align*} 
Recall that $\G(z) = \e^{-\g_0\sigma_3}U_0 (\frac{z}{z+a})^{\sigma_3/4}U_0^{-1}\e^{\g(z)\sigma_3}\e^{-\frac{x}{2}\sigma_3}$, and for all $z \in (\delta, \infty) \supset \Sigma_{\R,0}$ it follows that $|z/(z+a)| \leq 1$ and $|(z+a)/z| \leq (\delta+a)/\delta$. Consequently, $|(\G(z)\e^{\frac{x}{2}\sigma_3})^{\pm 1}|$ is bounded for all $z \in \Sigma_{\R,0}$, and Equation \eqref{eq:proof1} together with item 3 from Lemma \ref{lemma:GG21} implies the existence of constants $M,\hat{m}>0$ such that
\begin{align*}
\norm{J_{\R}(z)-I}_{L^1(\Sigma_{\R,0})} \leq & M \int_{\delta}^{\infty} |\e^{-2n\phi_+(z')}| \dd z' =O(\e^{-\hat{m} n}),\\
\norm{J_{\R}(z)-I}_{L^2(\Sigma_{\R,0})} \leq & M \left(\int_{\delta}^{\infty} |\e^{-4n\phi_+(z')}| \d z'\right)^{1/2} =O(\e^{-\hat{m} n}),\\
\norm{J_{\R}(z)-I}_{L^{\infty}(\Sigma_{\R,0})} \leq & M \sup_{z \in (\delta,\infty)} |\e^{-2n\phi_+(z)}| = O(\e^{-\hat{m} n}).
\end{align*}
The same reasoning applies to $z \in \Sigma_{\R,2}$. Finally, for $z \in \Sigma_{\R,1} \cup \Sigma_{\R,3}$,
\begin{align*}
|J_{\R}(z)-I| =& |\sigma_n(z)^{-1}\e^{2n\phi(z)}\G_+(z)\e^{\frac{x}{2}\sigma_3}E_{21} \e^{-\frac{x}{2}\sigma_3}\G_+(z)^{-1}| \\
\leq& |\sigma_n(z)^{-1}||\e^{2n\phi(z)}||\G_+(z)\e^{\frac{x}{2}\sigma_3}E_{21} \e^{-\frac{x}{2}\sigma_3}\G_+(z)^{-1}|.
\end{align*} 
Together with the fact that $|(\G(z)\e^{\frac{x}{2}\sigma_3})^{\pm 1}|$ is bounded for all $z$ away from $[-a,0]$ and that $\sigma_n(z)^{-1}=O(\e^{cn^{2/3}})$ for some constant $c>0$ and $z \in \Sigma_{\R,1}\cup \Sigma_{\R,3}$, it implies the existence of constants $M, \hat{m}>0$ such that
\begin{align*}
\norm{J_{\R}(z)-I}_{L^1(\Sigma_{\R,1}\cup \Sigma_{\R,3})} \leq & M \e^{cn^{2/3}} \int_{\Sigma_{\R,1}\cup \Sigma_{\R,3}} |\e^{2n\phi(z')}| \d z' = O(\e^{-\hat{m} n}),\\
\norm{J_{\R}(z)-I}_{L^2(\Sigma_{\R,1}\cup \Sigma_{\R,3})} \leq & M\e^{cn^{2/3}} \left(\int_{\Sigma_{\R,1}\cup \Sigma_{\R,3}} |\e^{4n\phi(z')}| \d z'\right)^{1/2} =O(\e^{-\hat{m} n}),\\
\norm{J_{\R}(z)-I}_{L^{\infty}(\Sigma_{\R,1}\cup \Sigma_{\R,3})} \leq & M\e^{cn^{2/3}} \sup_{z \in \Sigma_{\R,1}\cup \Sigma_{\R,3}} |\e^{2n\phi(z)}| = O(\e^{-\hat{m} n}).
\end{align*}
Therefore, there exist a constant $m>0$ such that
$$\norm{J_{\R}-I}_{L^1\cap L^2\cap L^{\infty}(\Sigma_{\R} \backslash(\partial\mathcal{U}^0 \cup \partial B_{\delta}(-a))} = O(\e^{-mn}),
$$
as claimed. $\square$\\

\begin{lemma}
\label{lemma:Rint}
Under the same assumptions as Lemma \ref{lemma:smallJR}, it follows that
$$\R(z) = I+O(n^{-\gamma}), \qquad \text{and} \quad \R'(z) = O(n^{-\gamma}),
$$
uniformly in  $x = x_0n^{\alpha}$ and $t \in [t_0, 1/t_0]$. Moreover, the term $\R^{(1)}$ in the asymptotic expansion \eqref{eq:asymptR} can be expressed as
\begin{equation}
\R_1(t,n) = -\frac{1}{2\pi \i} \int_{\partial \mathcal{U}^0 \cup \partial B_{\delta}(-a)}(J_{\R}(s)-I)\dd s +O(n^{-2\gamma}).
\end{equation}
\end{lemma}

\textbf{Proof:} Because of the previous Lemma, it is straightforward from small norm theory that for all $z \in \mathbb{C}\backslash \Sigma_{\R}$ the matrix-valued function $\R(z)$ admits the following representation
\begin{align*}
\R(z) = I +\frac{1}{2 \pi \i} \int_{\Sigma_{\R}} \frac{\R_-(s)(J_{\R}(s)-I)}{s-z}\dd s.
\end{align*}
%
By a reasoning analogous to the one in Theorem \ref{theo:smallnorm}, it follows that $|\R_-(z)|^{\pm 1}$ is bounded, so there exist $M>0$ such that $|\R_-(z)|^{\pm 1} \leq M$. For the term $\R^{(1)}$, the asymptotic expansion for $z \to \infty$ gives
$$\R(z) = I -\frac{1}{2 \pi \i z} \int_{\Sigma_{\R}}\R_-(s)(J_\R(s)-I) \dd s +O(z^{-2}).
$$
By Cauchy-Schwartz, setting $\tilde{\Sigma}_R:=\Sigma_{\R}\backslash \partial \mathcal{U}^0 \cup \partial B_{\delta}(-a)$
$$\left| \int_{\tilde{\Sigma}_R}\R_-(s)(J_{\R}(s)-I) \dd s \right| \leq \norm{\R_-(s)}_{L^{\infty}(\tilde{\Sigma})}\norm{J_{\R}(s)-I}_{L^1(\tilde{\Sigma})} = O(\e^{-mn}).
$$
Moreover,
$$\int_{\Sigma_{\R} \backslash \tilde{\Sigma}_{\R}}\R_-(s)(J_{\R}(s)-I) \dd s = \int_{\Sigma_{\R} \backslash \tilde{\Sigma}_{\R}}(J_{\R}(s)-I) \dd s +\underbrace{\int_{\Sigma_{\R} \backslash \tilde{\Sigma}_{\R}}(\R_-(s)-I)(J_{\R}(s)-I) \dd s}_{O(n^{-2\gamma})},
$$
and the result follows. \hfill $\square$

\section{Multiplicative statistics}
\label{sec:multiest}

In this section we apply the previous Riemann-Hilbert results in order to prove the main theorems. The starting point is the connection between the multiplicative statistics for the point process and the Christoffel-Darboux kernel, which is given by the following result.
\begin{lemma}
Let $L_n^Q(x)$ be the multiplicative statistics defined in Section \ref{sec:intro}. Then, for any $x \geq 0$, it holds that
$$\log L_n^Q(x) = - \int_{-\infty}^x\int_{\mathbb{R}} K_n^Q(\lambda, \lambda;x') \frac{\omega_n(\lambda)}{1+\e^{-x'+n^{2/3}Q(\lambda)}}\dd \lambda \dd x'.
$$
\end{lemma}

Different approaches for the proof can be found in \cite{CG2023} and \cite{GG21}. Together with Equation \eqref{CDconfluent}, which connects the kernel to the Riemann-Hilbert problem studied in the previous sections,  this lemma leads to the main results of this work.

\subsection{Proof of Proposition \ref{theo1}}

In order to simplify notation, set $\tilde{\Xi}(z) = I+ \e^{-x'}\sigma_n(z)^{-1}\e^{2n\phi(z)}E_{21}\chi_{(-a,0)}(z)$. Unfolding the transformations in the Riemann-Hilbert analysis, one gets
$${\Y}_+(z) = \e^{n\ell_V \sigma_3}\R_+(z)\P_+(z) \tilde{\Xi}_+(z)\e^{-n\left(\phi_+(z)-\frac{V(z)}{2}\right)\sigma_3}\e^{\frac{x'}{2}\sigma_3},
$$
where 
$$\P(z) = \left\lbrace \begin{array}{ll}
P^{(0)}(z), \hspace{0.5cm} & z \in \mathcal{U}^0,\\
P^{(a)}(z), \hspace{0.5cm} & z \in B_{\delta}(-a),\\
\G(z), \hspace{0.5cm} & \text{elsewhere}.
\end{array}\right.
$$

Then, using Equation \eqref{CDconfluent} we obtain
\begin{align*}
K_n^Q(\lambda, \lambda):= K_n^Q(\lambda, \lambda; x') =&\frac{1}{2\pi \i} \e^{x'}\e^{-2n\left(\phi_+(\lambda)-\frac{V(\lambda)}{2}\right)} \left(A(\lambda)+B(\lambda)+C(\lambda)\right),
\end{align*}
where
\begin{align*}
A(\lambda) =& \left[\tilde{\Xi}_+^{-1}(\lambda)[\P_+(\lambda)]^{-1}[\R_+(\lambda)]^{-1}\R_+'(\lambda)\P_+(\lambda) \tilde{\Xi}_+(\lambda)\right]_{21}\\
B(\lambda) =& \left[\tilde{\Xi}_+^{-1}(\lambda)[\P_+(\lambda)]^{-1}\P_+'(\lambda) \tilde{\Xi}_+(\lambda)\right]_{21}\\
C(\lambda) =& \left[\tilde{\Xi}_+^{-1}(\lambda)\tilde{\Xi}_+'(\lambda)\right]_{21} = \e^{-x'}\e^{2n\phi_+(\lambda)}\frac{2n\phi_+'(\lambda)\sigma_n(\lambda)-\sigma_n'(\lambda)}{\sigma_n(\lambda)^2}\chi_{(-a,0)}(\lambda).
\end{align*}

Notice that
$$(1+\e^{-x'+n^{2/3}Q(\lambda)})(1+\e^{x'-n^{2/3}Q(\lambda)})=2+\e^{-x'+n^{2/3}Q(\lambda)}+\e^{x'-n^{2/3}Q(\lambda)} > \e^{-x'+n^{2/3}Q(\lambda)}
$$
and therefore
$$\left|\frac{\omega_n(\lambda)}{1+\e^{-x'+n^{2/3}Q(\lambda)}}\right| < \e^{-nV(\lambda)} \e^{x'-n^{2/3}Q(\lambda)}.
$$

Now take the partition of the real line $I_1 = (-\infty, -a-\delta)$, $I_2 = (-a-\delta, -a +\delta)$, $I_3 = (-a+\delta, -\hat{\delta})$, $I_4 = (-\hat{\delta}, \tilde{\delta})$ and $I_5 = (\tilde{\delta}, \infty)$. 

\begin{lemma}
Let $t_0>0$ and $\alpha$ under Assumption \ref{asump1}. Then, there exist $m>0$ such that
$$ \int_{-\infty}^x\int_{I_1 \cup I_3 \cup I_5} K_n^Q(\lambda, \lambda) \frac{\omega_n(\lambda)}{1+\e^{-x'+n^{2/3}Q(\lambda)}}\dd \lambda \dd x' = O(\e^{-m n^{2/3}}),
$$
uniformly for $x = x_0 n^{\alpha}$ and $t \in [t_0, 1/t_0]$.
\end{lemma}

\textbf{Proof:} By the asymptotic condition \eqref{eq:asymptG}, we have that $(\G(\lambda)\e^{\frac{x'}{2}\sigma_3})^{\pm 1}$ decay as $\lambda \to \pm \infty$. Moreover, the explicit formulation
$$ \G(\lambda) = \e^{-\g_0\sigma_3} U_0 \left(\frac{\lambda}{\lambda+a}\right)^{\sigma_3/4}U_0^{-1}\e^{\g(\lambda)\sigma_3}\e^{-\frac{x'}{2}\sigma_3},
$$
together with decay of $\e^{\pm \g_0}$ and $\e^{\pm \g(\lambda)}$ in $n$ (see Lemma \ref{lemma:estimateg}), shows that away from a neighborhood of $\{-a, 0\}$, $|\G(\lambda)\e^{\frac{x'}{2}\sigma_3}|^{\pm 1}$ is bounded in $\lambda$ and in $n$, and it is uniform in $x' \leq x$ for $x = x_0 n^{\alpha}$, $\alpha$ under Assumption \ref{asump1}, and $t \in [t_0, 1/t_0]$. The previous formulation also gives
\begin{align*}
\G'(\lambda) =& \frac{1}{4}\e^{-\g_0\sigma_3} U_0 \begin{pmatrix}\lambda^{-\frac{3}{4}}(\lambda+a)^{-\frac{1}{4}}-\lambda^{\frac{1}{4}}(\lambda+a)^{-\frac{5}{4}} & 0\\
0 & \lambda^{-\frac{1}{4}}(\lambda+a)^{-\frac{3}{4}}-\lambda^{-\frac{5}{4}}(\lambda+a)^{\frac{1}{4}}\end{pmatrix}U_0^{-1}\e^{\g(\lambda)\sigma_3}\e^{-\frac{x'}{2}\sigma_3}+\\
& \e^{-\g_0\sigma_3} U_0 \left(\frac{\lambda}{\lambda+a}\right)^{\sigma_3/4}U_0^{-1}\g'(\lambda)\sigma_3\e^{\g(\lambda)\sigma_3}\e^{-\frac{x'}{2}\sigma_3},
\end{align*}
and the same reasoning as before implies the boundedness of $|\G'(\lambda)\e^{\frac{x'}{2}\sigma_3}|$ in $\lambda$ and in $n$. By Lemma \ref{lemma:Rint} and asymptotic condition \eqref{eq:asymptR}, one has that $\R(\lambda)$ and $\R'(\lambda)$ are bounded as $\lambda\to \pm \infty$ and are also bounded in $n$ as $n \to \infty$. Therefore, there exist $M>0$ such that
$$|\R(\lambda)^{\pm 1}|, |\R'(\lambda)|, |\G(\lambda)\e^{\frac{x'}{2}\sigma_3}|^{\pm 1}, |\G'(\lambda)\e^{\frac{x'}{2}\sigma_3}| \leq M,
$$
where the bound is uniform in $x' \leq x$ for $x = x_0 n^{\alpha}$, $\alpha$ under Assumption \ref{asump1} and $t \in [t_0, 1/t_0]$. Take $\lambda \in I_1 \cup I_5$. Then, $\tilde{\Xi}(\lambda) = \I$, and
\begin{align*}
\int_{I_1 \cup I_5} K_n^Q(\lambda, \lambda) \frac{\omega_n(\lambda)}{1+\e^{-x'+n^{2/3}Q(\lambda)}}\dd \lambda =& \frac{1}{2\pi\i}\int_{I_1 \cup I_5} \e^{x'}\{A(\lambda)+B(\lambda)\} \frac{\sigma_n(\lambda)\e^{-2n\phi_+(\lambda)}}{1+\e^{-x'+n^{2/3}Q(\lambda)}}\dd \lambda.
\end{align*}
From Lemma \ref{lemma:GG21} one has $\re \phi_+(\lambda)>0$ for all $\lambda \in \mathbb{R}\backslash [-a,0]$, and $\phi(\lambda)$ has the same decay as $V(\lambda)$ as $\lambda \to \infty$, which assures the convergence of the integral. Moreover,
\begin{align*}
A(\lambda) =& \e^{-x'}\left[\e^{-\frac{x'}{2}\sigma_3}[\G_+(\lambda)]^{-1}[\R_+(\lambda)]^{-1}\R_+'(\lambda)\G_+(\lambda) \e^{\frac{x'}{2}\sigma_3}\right]_{21},\\
B(\lambda) =& \e^{-x'}\left[\e^{-\frac{x'}{2}\sigma_3}[\G_+(\lambda)]^{-1}\G_+'(\lambda) \e^{\frac{x'}{2}\sigma_3}\right]_{21},
\end{align*}
and there exist constants $M, m>0$ such that
\begin{align*}
\left|\int_{I_1 \cup I_5} K_n^Q(\lambda, \lambda) \frac{\omega_n(\lambda)}{1+\e^{-x'+n^{2/3}Q(\lambda)}}\dd \lambda\right| \leq & \tilde{M}\int_{I_1 \cup I_5} \e^{-2n\re\phi_+(\lambda)}\e^{x'-n^{2/3}Q(\lambda)}\dd \lambda \leq M\e^{-n m + x'}.
\end{align*}
Integrating in $x'$, one obtains
\begin{align*}
\left|\int_{-\infty}^{x}\int_{I_1 \cup I_5} K_n^Q(\lambda, \lambda) \frac{\omega_n(\lambda)}{1+\e^{-x'+n^{2/3}Q(\lambda)}}\dd \lambda\right| \leq & M\e^{-n m + x} = O(\e^{-nm}),
\end{align*}
where the last equality comes from the fact that $x = O(n^{\alpha})$ where $\alpha <2/9$. Now take $\lambda \in I_3$. The claimed bounds for $|\G(\lambda)\e^{\frac{x'}{2}\sigma_3}|^{\pm 1}, |\G'(\lambda)\e^{\frac{x'}{2}\sigma_3}|, |\R(\lambda)|^{\pm 1}$ and $|\R'(\lambda)|$ still hold. Set $\tilde{A}(\lambda) = \e^{-\frac{x'}{2}\sigma_3}[\G_+(\lambda)]^{-1}[\R_+(\lambda)]^{-1}\R_+'(\lambda)\G_+(\lambda)\e^{\frac{x'}{2}\sigma_3}$ and $\tilde{B}(\lambda)=\e^{-\frac{x'}{2}\sigma_3}[\G_+(\lambda)]^{-1}\G_+'(\lambda)\e^{\frac{x'}{2}\sigma_3}$. In this interval,
\begin{align*}
A(\lambda) =&\e^{-x'}\left( \sigma_n(\lambda)^{-1}\e^{2n\phi_+(\lambda)}\{ [\tilde{A}(\lambda)]_{22}-[\tilde{A}(\lambda)]_{11}\}-\sigma_n(\lambda)^{-2}\e^{4n\phi_+(\lambda)}[\tilde{A}(\lambda)]_{12} + [\tilde{A}(\lambda)]_{21}\right)
\\
B(\lambda) =& \e^{-x'}\left(\sigma_n(\lambda)^{-1}\e^{2n\phi_+(\lambda)}\{ [\tilde{B}(\lambda)]_{22}-[\tilde{B}(\lambda)]_{11}\}-\sigma_n(\lambda)^{-2}\e^{4n\phi_+(\lambda)}[\tilde{B}(\lambda)]_{12} + [\tilde{B}(\lambda)]_{21}\right).
\end{align*}
Because $Q(\lambda)$ is strictly positive, we can suppose without loss of generality that $Q(\lambda) \geq Q(\tilde{\delta})$ for all $\lambda \in I_3$. Moreover, $\phi_+(\lambda)$ is purely imaginary. Consequently,

\begin{align*}
\int_{I_3} K_n^Q(\lambda, \lambda) \frac{\omega_n(\lambda)}{1+\e^{-x'+n^{2/3}Q(\lambda)}}\dd \lambda =& \frac{1}{2\pi \i}\int_{I_3} \frac{\sigma_n^{-1}(\lambda)\{ [\tilde{A}(\lambda)]_{22}-[\tilde{A}(\lambda)]_{11}+ [\tilde{B}(\lambda)]_{22}-[\tilde{B}(\lambda)]_{11}\}}{2+\e^{x'-n^{2/3}Q(\lambda)}+\e^{-x'+n^{2/3}Q(\lambda)}}\dd \lambda \\
&-
\int_{I_3} \{ [\tilde{A}(\lambda)]_{12}+ [\tilde{B}(\lambda)]_{12}\}\frac{\sigma_n(\lambda)^{-2}\e^{2n\phi_+(\lambda)}}{2+\e^{x'-n^{2/3}Q(\lambda)}+\e^{-x'+n^{2/3}Q(\lambda)}}\dd \lambda \\
&+ 
\int_{I_3}
\{ [\tilde{A}(\lambda)]_{21}+ [\tilde{B}(\lambda)]_{21}+C(\lambda)\e^{x'}\} \frac{\e^{-2n\phi_+(\lambda)}}{2+\e^{x'-n^{2/3}Q(\lambda)}+\e^{-x'+n^{2/3}Q(\lambda)}}\dd \lambda.
\end{align*}
Consequently, there exist $M>0$ such that
\begin{align*}
\left|\int_{I_3} K_n^Q(\lambda, \lambda) \frac{\omega_n(\lambda)}{1+\e^{-x'+n^{2/3}Q(\lambda)}}\dd \lambda\right| \leq& M \int_{I_3}\left[ 1+(1+\e^{x'-n^{2/3}Q(\lambda)})+2n|\phi_+'(\lambda)|+n^{2/3}|Q'(\lambda)|\right] \times\\
&(1+\e^{x'- n^{2/3}Q(\lambda)}) \e^{x'- n^{2/3}Q(\lambda)}\dd \lambda.
\end{align*}
Moreover, $Q'$ and $\phi'$ are continuous functions, therefore bounded in compact sets. Consequently, there exist a constant $\tilde{M}>0$ such that
\begin{align*}
\left|\int_{I_3} K_n^Q(\lambda, \lambda) \frac{\omega_n(\lambda)}{1+\e^{-x'+n^{2/3}Q(\lambda)}}\dd \lambda\right| \leq& \tilde{M}|I_3|(1+\e^{x'- n^{2/3}Q(\tilde{\delta})})\e^{x'- n^{2/3}Q(\tilde{\delta})}n = O(\e^{x'-n^{2/3}m}).
\end{align*}
Integrating in $x'$, one obtains
\begin{align*}
\left|\int_{-\infty}^{x}\int_{I_3} K_n^Q(\lambda, \lambda) \frac{\omega_n(\lambda)}{1+\e^{-x'+n^{2/3}Q(\lambda)}}\dd \lambda\right| \leq & M\e^{x-n^{2/3} m} = O(\e^{-n^{2/3}m}),
\end{align*}
where the last equality comes from the fact that $x = O(n^{\alpha})$ where $\alpha <2/9<2/3$. $\square$\\

\begin{lemma}
Let $t_0>0$ and $\alpha$ under Assumption \ref{asump1}. Then, there exist $m>0$ such that
$$ \int_{-\infty}^x\int_{I_2} K_n^Q(\lambda, \lambda) \frac{\omega_n(\lambda)}{1+\e^{-x'+n^{2/3}Q(\lambda)}}\dd \lambda \dd x' = O(\e^{-m n^{2/3}}),
$$
uniformly for $x = x_0 n^{\alpha}$ and $t \in [t_0, 1/t_0]$.
\end{lemma}

\textbf{Proof:} For $\lambda \in B_{\delta}(-a)$, the parametrix solution is given by means of Airy functions in the local variable $\zeta = n^{2/3}\varphi(\lambda)$. Here we split the analysis into two parts. Fix $M>0$. For $|\zeta| \leq M$, we use that $\Psi_{Ai}(\zeta)$ and $\Psi_{Ai}'(\zeta)$ are continuous, therefore bounded in compact sets. From the expression
\begin{align*}
E(\lambda) =&U_0 \lambda^{\sigma_3/4} \begin{pmatrix}
0 & -i\left(\frac{(n^{2/3}\varphi(\lambda))}{\lambda+a}\right)^{1/4} \\
-i\left(\frac{n^{2/3}\varphi(\lambda)}{\lambda+a}\right)^{1/4} & 0
\end{pmatrix},
\end{align*}
it is clear that $E(\lambda)$ grows as most as $n^{1/6}$ as $n \to \infty$ for all $\lambda \in B_{\delta}(-a)$. Moreover,
\begin{align*}
E'(\lambda) =& \M'(\lambda)U_0^{-1}(n^{2/3}\varphi(\lambda))^{-\sigma_3/4}+\M(\lambda)U_0^{-1}(n^{2/3}\varphi(\lambda))^{-\sigma_3/4}\begin{pmatrix}
-\frac{\varphi'(\lambda)}{4\varphi(\lambda)}& 0 \\ 0 & \frac{\varphi'(\lambda)}{4\varphi(\lambda)}
\end{pmatrix} = O(n^{5/6}).
\end{align*}
Consequently,
$$P^{(a)}(\lambda)= \underbrace{E(\lambda)\Psi_{Ai}(n^{2/3}\varphi(\lambda))}_{O(n^{1/6})}\underbrace{\e^{-\frac{1}{2}\log\sigma_n(\lambda)\sigma_3}}_{I+O(\e^{-n^{2/3}m})}\e^{n\phi(\lambda)\sigma_3}\e^{-\frac{x'}{2}\sigma_3}
$$
and it follows that $P^{(a)}_+(\lambda)\e^{\frac{x'}{2}\sigma_3}\e^{-n\phi_+(\lambda)\sigma_3}=O(n^{1/6})$ for $|n^{2/3}\varphi(\lambda)| \leq M$ and $[P^{(a)}(\lambda)]'=O(n^{5/6})\e^{-\frac{x'}{2}\sigma_3}\e^{n\phi+(\lambda)\sigma_3}$. For $|\zeta| \geq M$ we can use
$$P_+(\lambda)=(I+O(n^{-\gamma}))\G(\lambda).
$$
As $\lambda \to 0$, the fact that $\varphi$ conformal together with $\varphi(0)=0$ implies the existence of $c>0$ such that $|\lambda+a| \geq c/n^{2/3}$. Therefore, in this set it holds that $\G(\lambda) \e^{\frac{x'}{2}\sigma_3} = O(n^{1/6})$ and $\G'(\lambda) \e^{\frac{x'}{2}\sigma_3} = O(n^{5/6})$. Denote $a^*=c/n^{2/3}$. Then,
\begin{align*}
\left|\int_{-a-\delta}^{-a-a^*} K_n^Q(\lambda, \lambda) \frac{\omega_n(\lambda)}{1+\e^{-x'+n^{2/3}Q(\lambda)}}\dd \lambda\right| \leq & \int_{-a-\delta}^{-a-a^*} O(n) \e^{-2n\re\phi_+(\lambda)+x'-n^{2/3}Q(\lambda)}\dd \lambda,\\
\left|\int_{-a-a^*}^{-a} K_n^Q(\lambda, \lambda) \frac{\omega_n(\lambda)}{1+\e^{-x'+n^{2/3}Q(\lambda)}}\dd \lambda\right| \leq & \int_{-a-a^*}^{-a} O(n) \e^{x'-n^{2/3}Q(\lambda)}\dd \lambda,\\
\therefore
\left|\int_{-a-\delta}^{-a} K_n^Q(\lambda, \lambda) \frac{\omega_n(\lambda)}{1+\e^{-x'+n^{2/3}Q(\lambda)}}\dd \lambda\right| \leq & M\e^{x-n^{2/3} m}.
\end{align*}
Analogously, for $\lambda\in [-a,-a+a^*]$ one has
\begin{align*}
|A(\lambda)| \leq & n^{1/3}M\e^{-x'+2n\phi_+(\lambda)}\{1+\sigma_n^{-1}(\lambda)+\sigma_n^{-2}(\lambda)\}\\
|B(\lambda)| \leq & nM\e^{-x'+2n\phi_+(\lambda)}\{1+\sigma_n^{-1}(\lambda)+\sigma_n^{-2}(\lambda)\},
\end{align*}
and, for $\lambda\in [-a+a^*,-a+\delta]$ one has
\begin{align*}
|A(\lambda)| \leq & n^{1/3}M\e^{-x'}\{1+\e^{2n\phi_+(\lambda)}\sigma_n^{-1}(\lambda)+\e^{4n\phi_+(\lambda)}\sigma_n^{-2}(\lambda)\}\\
|B(\lambda)| \leq & nM\e^{-x'}\{1+\e^{2n\phi_+(\lambda)}\sigma_n^{-1}(\lambda)+\e^{4n\phi_+(\lambda)}\sigma_n^{-2}(\lambda)\}.
\end{align*}
Consequently, because $\phi_+$ is purely imaginary and $Q(\lambda)$ is strictly positive, there are constants $M, m>0$ such that
\begin{align*}
\left|\int_{-a}^{-a+\delta} K_n^Q(\lambda, \lambda) \frac{\omega_n(\lambda)}{1+\e^{-x'+n^{2/3}Q(\lambda)}}\dd \lambda\right| \leq & n\tilde{M}\int_{-a}^{-a+\delta} \e^{x'-n^{2/3}Q(\lambda)}\{1+\sigma_n^{-1}(\lambda)+\sigma_n^{-2}(\lambda)\}\dd \lambda,\\
&+ \int_{-a}^{-a+\delta} \left|\frac{2n\phi_+'(\lambda)}{\sigma_n(\lambda)}-\frac{\sigma_n'(\lambda)}{\sigma_n^2(\lambda)} \right| \e^{x'-n^{2/3}Q(\lambda)}\dd \lambda\leq M\e^{x'-n^{2/3} m}.
\end{align*}
Integrating in $x'$, the claimed estimate follows. $\square$\\

At last, we show that the only relevant contribution comes from a subset of $I_4$.

\begin{lemma}
Let $t_0>0$ and $\alpha$ under Assumption \ref{asump1}. Then, there exist $\tilde{\epsilon}, \bar{\epsilon}, m>0$ such that
$$ \int_{-\infty}^x\int_{I_4\backslash [-\tilde{\epsilon}n^{\alpha-\frac{2}{3}}, \bar{\epsilon}n^{\frac{2}{3}\alpha-\frac{2}{3}}]} K_n^Q(\lambda, \lambda) \frac{\omega_n(\lambda)}{1+\e^{-x'+n^{2/3}Q(\lambda)}}\dd \lambda \dd x' = O(\e^{-m n^{\alpha}}),
$$
uniformly for $x = x_0 n^{\alpha}$ and $t \in [t_0, 1/t_0]$.
\end{lemma}

\textbf{Proof}: Take $c>0$ such that $|\lambda| \geq c/n^{2/3}$ implies $|\zeta| \geq M$ for some constant $M>0$. Then
\begin{align*}
A(\lambda) =& \tilde{\Xi}_+^{-1}(\lambda)\e^{\frac{x'}{2}\sigma_3}\underbrace{\e^{-\frac{x'}{2}\sigma_3}\G(\lambda)^{-1}}_{O(n^{1/6})}[\R_+(\lambda)]^{-1}\R_+'(\lambda)\underbrace{\G(\lambda)\e^{\frac{x'}{2}\sigma_3}}_{O(n^{1/6})}\e^{-\frac{x'}{2}\sigma_3} \tilde{\Xi}_+(\lambda),\\
B(\lambda) =& \tilde{\Xi}_+^{-1}(\lambda)\e^{\frac{x'}{2}\sigma_3}\underbrace{\e^{-\frac{x'}{2}\sigma_3}\G(\lambda)^{-1}}_{O(n^{1/6})}[I+O(n^{-\gamma})]\underbrace{\G'(\lambda)\e^{\frac{x'}{2}\sigma_3}}_{O(n^{5/6})}\e^{-\frac{x'}{2}\sigma_3} \tilde{\Xi}_+(\lambda).
\end{align*}
Consequently, for $\lambda\in(-\hat{\delta},-c/n^{2/3})$, there are constants $m_1, m_2, m_3>0$ such that
\begin{align*}
A(\lambda)+B(\lambda) =& \left[ \tilde{\Xi}_+^{-1}(\lambda)\e^{\frac{x'}{2}\sigma_3}O(n)\e^{-\frac{x'}{2}\sigma_3} \tilde{\Xi}_+(\lambda) \right]_{21} = \e^{-x'}O(n)\{m_1 +m_2\sigma_n^{-1}\e^{2n\phi_+(\lambda)}+m_3\sigma_n^{-2}\e^{4n\phi_+(\lambda)}\}.
\end{align*}
Because $Q(\lambda) \sim -t \lambda$, there exist $\tilde{m}, \tilde{\epsilon}>0$ such that
$$n^{2/3}Q(-\tilde{\epsilon}n^{\alpha-\frac{2}{3}}):= \tilde{m} n^{\alpha}+O(n^{2\alpha-\frac{2}{3}})> x_0 n^{\alpha}.
$$
Therefore, $x-n^{2/3}Q(-\tilde{\epsilon}n^{\alpha-\frac{2}{3}})\leq x_0 n^{\alpha}-\tilde{m}n^{\alpha}+O(n^{2\alpha-\frac{2}{3}})<0$. Consequently, using that $\re \phi_+ = 0$ in this interval, we have that for some constants $M, \hat{m}>0$ it holds that
\begin{align*}
\left|\int_{-\hat{\delta}}^{-\tilde{\epsilon}n^{\alpha-\frac{2}{3}}} \e^{x'}\e^{-2n\phi_+(\lambda)} [A(\lambda)+B(\lambda)] \frac{\sigma_n(\lambda)}{1+\e^{-x'+n^{2/3}Q(\lambda)}}\dd \lambda\right|\leq & M\int_{-\hat{\delta}}^{-\tilde{\epsilon}n^{\alpha-\frac{2}{3}}} O(n)\{1 +\sigma_n^{-1}+\sigma_n^{-2}\}\e^{x'-n^{2/3}Q(\lambda)}\dd \lambda\\
\leq& M\e^{\hat{m}(x'-\tilde{m}n^{\alpha})}.
\end{align*}
As a result, we have that
\begin{align*}
\left|\int_{-\infty}^x\int_{-\hat{\delta}}^{-\tilde{\epsilon}n^{\alpha-\frac{2}{3}}} \e^{x}\e^{-2n\phi_+(\lambda)} [A(\lambda)+B(\lambda)] \frac{\sigma_n(\lambda)}{1+\e^{-x+n^{2/3}Q(\lambda)}}\dd \lambda \dd x\right| \leq M \e^{\bar{m}(x-\tilde{m}n^{\alpha})} \leq M \e^{-mn^{\alpha}},
\end{align*}
for some constants $m, M>0$. Analogously, for $\lambda\in(c/n^{2/3}, \tilde{\delta})$, we have
\begin{align*}
A(\lambda)+B(\lambda) =& \left[\e^{\frac{x'}{2}\sigma_3}O(n)\e^{-\frac{x'}{2}\sigma_3} \right]_{21} = \e^{-x'}O(n).
\end{align*}
Because $\phi(\lambda) \sim c_{\phi} \lambda^{3/2}$ as $\lambda \to 0$, there exist $\bar{m}, \bar{\epsilon}>0$ such that
$$n\phi(\bar{\epsilon}n^{\frac{2}{3}\alpha-\frac{2}{3}}):= \bar{m} n^{\alpha}+O(n^{2\alpha-\frac{2}{3}})> x_0 n^{\alpha}.
$$
Therefore, we have that for some constants $M, \hat{m}>0$ it holds that
\begin{align*}
\left|\int_{\bar{\epsilon}n^{\frac{2}{3}\alpha-\frac{2}{3}}}^{\tilde{\delta}} \e^{x'}\e^{-2n\phi_+(\lambda)} [A(\lambda)+B(\lambda)] \frac{\sigma_n(\lambda)}{1+\e^{-x'+n^{2/3}Q(\lambda)}}\dd \lambda\right|\leq & M\int_{\bar{\epsilon}n^{\frac{2}{3}\alpha-\frac{2}{3}}}^{\tilde{\delta}} O(n)\e^{-2n\phi(\lambda)+x'-n^{2/3}Q(\lambda)}\dd \lambda\\
\leq& M\e^{\hat{m}(x'-\bar{m}n^{\alpha})}.
\end{align*}
As a result, we have that
\begin{align*}
\left|\int_{-\infty}^x\int_{\bar{\epsilon}n^{\frac{2}{3}\alpha-\frac{2}{3}}}^{\tilde{\delta}} \e^{x}\e^{-2n\phi(\lambda)} [A(\lambda)+ B(\lambda)] \frac{\sigma_n(\lambda)}{1+\e^{-x+n^{2/3}Q(\lambda)}}\dd \lambda \dd x\right| \leq M \e^{\hat{m}(x-\bar{m}n^{\alpha})} \leq M \e^{-mn^{\alpha}}.
\end{align*}
At last, for $C(\lambda)$ we have
\begin{align*}
\left|\int_{-\hat{\delta}}^{-\tilde{\epsilon}n^{\alpha-\frac{2}{3}}} \frac{2n\phi_+'(\lambda)\sigma_n(\lambda)-\sigma_n'(\lambda)}{\sigma_n(\lambda)(1+\e^{-x+n^{2/3}Q(\lambda)})}\dd \lambda\right|\leq & \left|\int_{-\hat{\delta}}^{-\tilde{\epsilon}n^{\alpha-\frac{2}{3}}} \frac{n^{2/3}Q'(\lambda) \e^{x'-n^{2/3}Q(\lambda)}\sigma_n(\lambda)}{1+\e^{-x'+n^{2/3}Q(\lambda)}}\dd \lambda\right|\\
&+\left|\int_{-\hat{\delta}}^{-\tilde{\epsilon}n^{\alpha-\frac{2}{3}}}\frac{2n\phi_+'(\lambda)}{1+\e^{-x'+n^{2/3}Q(\lambda)}}\dd \lambda\right| \\
\leq & n \tilde{M}\int_{-\hat{\delta}}^{-\tilde{\epsilon}n^{\alpha-\frac{2}{3}}}\e^{x'-n^{2/3}Q(\lambda)}\dd \lambda \leq M \e^{\hat{m}(x'-\tilde{m}n^{\alpha})}.
\end{align*}
Consequently,
\begin{align*}
\left|\int_{-\infty}^x\int_{\bar{\epsilon}n^{\frac{2}{3}\alpha-\frac{2}{3}}}^{\tilde{\delta}} \e^{x'}\e^{-2n\phi(\lambda)} C(\lambda)\frac{\sigma_n(\lambda)}{1+\e^{-x'+n^{2/3}Q(\lambda)}}\dd \lambda \dd x'\right| \leq M \e^{\hat{m}(x-\tilde{m}n^{\alpha})} \leq M \e^{-mn^{\alpha}},
\end{align*}
and the result follows. $\square$\\

Therefore, the relevant contribution comes from $(-\tilde{\epsilon}n^{\alpha-\frac{2}{3}}, \bar{\epsilon}n^{\frac{2}{3}\alpha-\frac{2}{3}})$. 

\subsection{Proof of Theorem \ref{theo2}}

For simplicity, let us denote $\mathcal{K}:= [-\tilde{\epsilon}n^{\alpha-\frac{2}{3}}, \bar{\epsilon}n^{\frac{2}{3}\alpha-\frac{2}{3}}]$. In order to simplify notation, define $\tilde{\Delta}:=\E^{-1}\e^{\frac{t\zeta_0}{2}\sigma_3}\Delta(\zeta)\e^{-\frac{t\zeta_0}{2}\sigma_3}\E$ and
$$\bar{\sigma}(\zeta):= \frac{\sigma_{\kappa}(\zeta)}{1+\e^{-x-h_{\kappa}(\zeta)}}.
$$

Notice that in the relevant interval, $|\zeta| \leq M n^{\alpha} \leq M n^{2\nu/3}$, and therefore Remark \ref{rmk:exp} can be applied for $\kappa=n^{2/3}$. This will be used as follows:
\begin{align*}
\frac{\sigma_{\kappa}(\zeta)}{1+\e^{-x-h_{\kappa}(\zeta)}}=&\frac{\sigma_{\kappa}(\zeta)}{\sigma_{0}(\zeta)}\frac{1+\e^{-x-t\zeta}}{1+\e^{-x-h_{\kappa}(\zeta)}}\frac{\sigma_{0}(\zeta)}{1+\e^{-x-t\zeta}}\\
=& \frac{\sigma_{0}(\zeta)}{1+\e^{-x-t\zeta}}\times \left\lbrace\begin{array}{ll}
1+O(n^{2/3(1-2\alpha)}), \qquad & \alpha<4/21\\
1+O(n^{2/3(2-3\alpha)}), \qquad & 4/21 \leq \alpha<2/9
\end{array}\right..
\end{align*}

\begin{proposition}
Set $\kappa=n^{2/3}$, $-{\delta}_2n^{\alpha} = -n^{2/3}\varphi(-\tilde{\epsilon}n^{\alpha-2/3})$ and ${\delta}_3n^{2\alpha/3} = -n^{2/3}\varphi(-\bar{\epsilon}n^{2\alpha/3-2/3})$. Then
\begin{align*}
&\int_{\mathcal{K}}K_n^Q(\lambda, \lambda) \frac{\omega_n(\lambda)}{1+\e^{-x+n^{2/3}Q(\lambda)}}\dd \lambda = \mathcal{I}_1+\mathcal{I}_2+\mathcal{I}_3,\\
\mathcal{I}_1 =& \int_{-{\delta}_2n^{\alpha}}^{\delta_3n^{2\alpha/3}} \bar{\sigma}(\zeta) \left[\Xi(\zeta)^{-1}\Psi_{cc}(\zeta)^{-1}\frac{\dd}{\dd\zeta}\{\Psi_{cc}(\zeta)\Xi(\zeta)\}\right]_{21}\dd \zeta\\
\mathcal{I}_2 =& \int_{-{\delta}_2n^{\alpha}}^{\delta_3n^{2\alpha/3}} \bar{\sigma}(\zeta)\left[\Xi(\zeta)^{-1}\Psi_{cc}(\zeta)^{-1} \tilde{\Delta}(\zeta)^{-1}F_1(\lambda) \tilde{\Delta}(\zeta)\Psi_{cc}(\zeta)\Xi(\zeta)\right]_{21} \frac{\dd \zeta}{n^{2/3}\varphi'(\lambda)},
\\
\mathcal{I}_3 =& \int_{-{\delta}_2n^{\alpha}}^{\delta_3n^{2\alpha/3}} \bar{\sigma}(\zeta)\left[\Xi(\zeta)^{-1}\Psi_{cc}(\zeta)^{-1}\E^{-1}\e^{\frac{t\zeta_0}{2}\sigma_3}\Delta(\zeta)^{-1}\frac{\dd \Delta(\zeta)}{\dd\zeta}\e^{-\frac{t\zeta_0}{2}\sigma_3}\E\Psi_{cc}(\zeta)\Xi(\zeta)\right]_{21} \dd \zeta,
\end{align*}
where
$$F_1(\lambda) =f(\lambda)\sigma_3+ [n^{2/3}\varphi(\lambda)]^{\sigma_3/4}\left(\frac{\lambda}{\lambda+a}\right)^{-\sigma_3/4}U_0^{-1}\R_+^{-1}(\lambda)\R'(\lambda)U_0\left(\frac{\lambda}{\lambda+a}\right)^{\sigma_3/4}[n^{2/3}\varphi(\lambda)]^{-\sigma_3/4}.
$$
\end{proposition}

\textbf{Proof}: For the term $A(\lambda)$ and $\zeta = n^{2/3}\varphi(\lambda)$ we have
\begin{align*}
A(\lambda) &= \e^{2n\phi_+(\lambda)-x}\left[\Xi(\zeta)^{-1}\Psi_{cc}(\zeta)^{-1}\tilde{\Delta}(\zeta)^{-1}[n^{2/3}\varphi(\lambda)]^{\sigma_3/4}\left(\frac{\lambda}{\lambda+a}\right)^{-\sigma_3/4} U_0^{-1} \R^{-1}(\lambda)\right. \\
&\hspace{2cm}\left. \R'(\lambda)U_0\left(\frac{\lambda}{\lambda+a}\right)^{\sigma_3/4}[n^{2/3}\varphi(\lambda)]^{-\sigma_3/4}\tilde{\Delta}(\zeta)\Psi_{cc}(\zeta)\Xi(\zeta)\right]_{21}.
\end{align*}
For the term $B(\lambda)$ and $\kappa=n^{2/3}$ we start by evaluating
\begin{align*}
[P_+(\lambda)]^{-1}P_+'(\lambda)=&\e^{-n\phi_+(\lambda)\sigma_3}\Phi_{\kappa}(\zeta)^{-1} E_n^{-1} E_n'\Phi_{\kappa}(\zeta)\e^{n \phi(\lambda)\sigma_3}+\e^{-n\phi_+(\lambda)\sigma_3}\Phi_{\kappa}(\zeta)^{-1} \Phi_{\kappa}(\zeta)'\e^{n \phi(\lambda)\sigma_3}+n\phi'(\lambda)\sigma_3,
\end{align*}
where $'$ denotes the derivative with respect to $\lambda$. For the first term, notice that
\begin{align*}
E_n'(\lambda) = \underbrace{\frac{1}{4}\left(\frac{a}{\lambda(\lambda+a)}-\frac{n^{2/3}\varphi'(\lambda)}{\zeta}\right)}_{{f}(\lambda)}E_n(\lambda) \e^{-\frac{t\zeta_0}{2}\sigma_3}\E \sigma_3 \E^{-1}\e^{\frac{t\zeta_0}{2}\sigma_3},
\end{align*}
and, consequently, $E_n^{-1} E_n' = {f}(\lambda)\e^{-\frac{t\zeta_0}{2}\sigma_3}\E \sigma_3 \E^{-1}\e^{\frac{t\zeta_0}{2}\sigma_3}$. Therefore,
\begin{align*}
\Phi_{\kappa}(\zeta)^{-1} E_n^{-1} E_n'\Phi_{\kappa}(\zeta) =& {f}(\lambda) \bar{\Xi}_{n^{2/3}}(\zeta)^{-1}\e^{\frac{x}{2}\sigma_3}\Psi_{cc}(\zeta)^{-1}\tilde{\Delta}(\zeta)^{-1}\sigma_3\tilde{\Delta}(\zeta)\Psi_{cc}(\zeta)\e^{-\frac{x}{2}\sigma_3}\bar{\Xi}_{n^{2/3}}(\zeta),
\end{align*}
and
\begin{align*}
\begin{split}\left[\tilde{\Xi}_+^{-1}(\lambda)\e^{-n\phi_+(\lambda)\sigma_3}\Phi_{\kappa}(\zeta)^{-1} E_n^{-1} E_n'\Phi_{\kappa}(\zeta)\e^{n \phi(\lambda)\sigma_3} \tilde{\Xi}_+(\lambda)\right]_{21}=\\
f(\lambda)\e^{2n\phi_+(\lambda)-x}\left[\Xi(\zeta)^{-1}\Psi_{cc}(\zeta)^{-1}\tilde{\Delta}(\zeta)^{-1} \sigma_3\right.  \left.\tilde{
\Delta}(\zeta)\Psi_{cc}(\zeta)\Xi(\zeta)\right]_{21}\end{split}
\end{align*}
where $\Xi(\zeta) = I+\sigma_{n^{2/3}}(\zeta)^{-1}\chi_{(-\infty, \zeta_0)}(\zeta)E_{21}$. For the second term we have
\begin{align*}
\left[\tilde{\Xi}_+^{-1}(\lambda)\e^{-n\phi_+(\lambda)\sigma_3}\Phi_{\kappa}(\zeta)^{-1} \Phi_{\kappa}'(\zeta)\e^{n \phi(\lambda)\sigma_3} \tilde{\Xi}_+(\lambda)\right]_{21}=\\
\e^{2n\phi_+(\lambda)-x}\left[\Xi(\zeta)^{-1}\Psi_{cc}(\zeta)^{-1}\E^{-1}\e^{\frac{t\zeta_0}{2}\sigma_3}\Delta(\zeta)^{-1}\Delta'(\zeta)\e^{-\frac{t\zeta_0}{2}\sigma_3}\E\Psi_{cc}(\zeta)\Xi(\zeta)\right]_{21}\\
+ \e^{2n\phi_+(\lambda)-x}\left[\Xi(\zeta)^{-1}\Psi_{cc}(\zeta)^{-1}\Psi_{cc}'(\zeta)\Xi(\zeta)\right]_{21}+\e^{2n\phi_+(\lambda)-x}[\sigma_n(\lambda)^{-1}]'|_{(0,\varphi^{-1}(\zeta_0/n^{2/3}))},
\end{align*}
and the last term reads
$$\left[\tilde{\Xi}_+^{-1}(\lambda)n\phi'(\lambda)\sigma_3 \tilde{\Xi}_+(\lambda)\right]_{21} = -2n\phi'(\lambda) \sigma_n(\lambda)^{-1}\e^{2n\phi_+(\lambda)-x}\chi_{(-a,0)}(\lambda).
$$
Summing and subtracting $\left[\Xi(\zeta)^{-1}\Xi'(\zeta)\right]_{21}$, we obtain
\begin{align*}
\left[\Xi(\zeta)^{-1}\Psi_{cc}(\zeta)^{-1}\Psi_{cc}'(\zeta)\Xi(\zeta)\right]_{21} = \left[\Xi(\zeta)^{-1}\Psi_{cc}(\zeta)^{-1}\{\Psi_{cc}(\zeta)\Xi(\zeta)\}'\right]_{21}-[\sigma_n^{-1}(\lambda)]'\chi_{(-a,\varphi^{-1}(\zeta_0/n^{2/3}))}(\lambda).
\end{align*}
For the term $C(\lambda)$,
\begin{align*}
\int_{-\tilde{\epsilon}n^{\alpha-\frac{2}{3}}}^0 \e^{x}\e^{-2n\phi_+(\lambda)} C(\lambda) \frac{\sigma_n(\lambda)}{1+\e^{-x+n^{2/3}Q(\lambda)}}\dd \lambda &=
\int_{-\tilde{\epsilon}n^{\alpha-\frac{2}{3}}}^0 \left[2n\phi_+'(\lambda)-\frac{\sigma_n'(\lambda)}{\sigma_n(\lambda)}\right]\frac{1}{1+\e^{-x+n^{2/3}Q(\lambda)}}\dd \lambda.
\end{align*}
Thus, summing $A$, $B$ and $C$ we obtain
\begin{align*}
&\int_{\mathcal{K}} \e^{x}\e^{-2n\phi_+(\lambda)} (A(\lambda)+ B(\lambda)+C(\lambda)) \frac{\sigma_n(\lambda)}{1+\e^{-x+n^{2/3}Q(\lambda)}}\dd \lambda =\\
&\int_{\mathcal{K}} \frac{\sigma_n(\lambda)}{1+\e^{-x+n^{2/3}Q(\lambda)}}\left\{\left[\Xi(\zeta)^{-1}\Psi_{cc}(\zeta)^{-1}\E^{-1}\e^{\frac{t\zeta_0}{2}\sigma_3}\Delta(\zeta)^{-1}\Delta'(\zeta)\e^{-\frac{t\zeta_0}{2}\sigma_3}\E\Psi_{cc}(\zeta)\Xi(\zeta)\right]_{21}\right.\\
&+ \left[\Xi(\zeta)^{-1}\Psi_{cc}(\zeta)^{-1}\tilde{\Delta}^{-1} F_1(\lambda)\tilde{\Delta}\Psi_{cc}(\zeta)\Xi(\zeta)\right]_{21}\left.+ \left[\Xi(\zeta)^{-1}\Psi_{cc}(\zeta)^{-1}\{\Psi_{cc}(\zeta)\Xi(\zeta)\}'\right]_{21}\right\}\dd \lambda,
\end{align*}
where $'$ denotes the derivative with respect to $\lambda$, $\zeta=n^{2/3}\varphi(\lambda)$ and
$$F_1(\lambda) =f(\lambda)\sigma_3+ [n^{2/3}\varphi(\lambda)]^{\sigma_3/4}\left(\frac{\lambda}{\lambda+a}\right)^{-\sigma_3/4}U_0^{-1}\R_+^{-1}(\lambda)\R'(\lambda)U_0\left(\frac{\lambda}{\lambda+a}\right)^{\sigma_3/4}[n^{2/3}\varphi(\lambda)]^{-\sigma_3/4}.
$$
The change in variables  $\lambda \mapsto \zeta=n^{2/3}\varphi(\lambda)$ gives the claimed result. $\square$\\

The only thing left to do is to prove that $\mathcal{I}_2$ and $\mathcal{I}_3$ decay in $x$. It is not trivial, since the local solution depends on $\Psi_{cc}(\zeta)$, which is not necessarily bounded in $x$ as $x \to \infty$ and on $\E$, which has polynomial growth in $x$. Therefore, we need to investigate carefully the integrals, making use of the solution to the Riemann-Hilbert problem studied by Claeys and Cafasso in \cite{CC2019}.

\begin{lemma}
Let $t_0>0$ and $\alpha$ under Assumption \ref{asump1}. Then
$$\mathcal{I}_3 = \int_{-{\delta}_2n^{\alpha}}^{\delta_3n^{2\alpha/3}} \bar{\sigma}(\zeta)\left[\Xi(\zeta)^{-1}\Psi_{cc}(\zeta)^{-1}\E^{-1}\e^{\frac{t\zeta_0}{2}\sigma_3}\Delta(\zeta)^{-1}\frac{\dd \Delta(\zeta)}{\dd\zeta}\e^{-\frac{t\zeta_0}{2}\sigma_3}\E\Psi_{cc}(\zeta)\Xi(\zeta)\right]_{21} \dd \zeta = O(n^{-2\tau/3})
$$
uniformly for $x = x_0 n^{\alpha}$ and $t \in [t_0, 1/t_0]$.
\end{lemma}

\textbf{Proof:} Set $K=[-{\delta}_2n^{\alpha}/s+1, \delta_3n^{2\alpha/3}/s+1]$. We analyze the integral in further details with the help of the solution $\Psi_{cc}(\zeta)$ from \cite{CC2019}. The change in variables $z = \zeta/s+1$, for $s = x/t$ gives
\begin{align*}
\mathcal{I}_3 =& s\int_K \bar{\sigma}(sz-s)\e^{-s^{3/2}(2g(z)+V(z_0))}\left[\Xi_1(z)^{-1}\hat{S}(z)^{-1}s^{-\sigma_3/4}\e^{\frac{t\zeta_0}{2}\sigma_3}O(\kappa^{-\tau})\e^{-\frac{t\zeta_0}{2}\sigma_3}s^{\sigma_3/4}\hat{S}(z)\Xi_1(z)\right]_{21} \dd z,
\end{align*}
where 
$$\Xi_1(z) = \begin{pmatrix}
1 & 0 \\ \e^{s^{3/2}(2g(z)-V(z)+V(z_0))}(1+O(.))\chi_{(-\infty, z_0)}(z) & 1
\end{pmatrix},
$$
where $O(.) = O(n^{2/3(1-2\alpha)})$ for $0<\alpha<4/21$ and $O(n^{2/3(2-3\alpha)})$ for $4/21 \leq \alpha<2/9$. Recalling the parametrices in Section \ref{sec:model}, we have that for $z \in (z_0+\varepsilon, \infty)\cap K$ and some real constant $M>0$,
\begin{align*}
\mathcal{I}_3|_{(z_0+\varepsilon, \infty)} =&s\int_{z_0+\varepsilon}^{\infty}\chi_K(z) \bar{\sigma}(sz-s)\e^{-s^{3/2}(2g(z)+V(z_0))}\left[\hat{S}(z)^{-1}s^{-\sigma_3/4}\e^{\frac{t\zeta_0}{2}\sigma_3}O(n^{-2\tau/3})\e^{-\frac{t\zeta_0}{2}\sigma_3}s^{\sigma_3/4}\hat{S}(z)\right]_{21} \dd z\\
=& s\int_{z_0+\varepsilon}^{\infty}\chi_K(z) \bar{\sigma}(sz-s)\e^{-s^{3/2}(2g(z)+V(z_0))}\left[U_0(z-z_0)^{-\sigma_3/4}R_{cc}^{-1}(z)s^{-\sigma_3/4}\e^{\frac{t\zeta_0}{2}\sigma_3}O(n^{-2\tau/3}) \right. \\
& \hspace{8cm}\left.\e^{-\frac{t\zeta_0}{2}\sigma_3}s^{\sigma_3/4}R_{cc}(z)(z-z_0 )^{\sigma_3/4} U_0^{-1}\right]_{21} \dd z\\
|\mathcal{I}_3|_{(z_0+\varepsilon, \infty)}|\leq& \left|\int_{z_0+\varepsilon}^{\infty}\bar{\sigma}(sz-s)\sigma_0(z)^{-1} O(n^{-2\tau/3}s^{3/2}|z-z_0|^{1/2})\e^{-\frac{4}{3}s^{3/2}(z-z_0)^{3/2}} \dd z\right|,\\
\leq & M n^{-2\tau/3}\int_{z_0+\varepsilon}^{\infty} s^{3/2}|z-z_0|^{1/2}\e^{-\frac{4}{3}s^{3/2}(z-z_0)^{3/2}} \dd z \leq  M n^{-2\tau/3}\e^{-\frac{4}{3}s^{3/2} \varepsilon^{3/2}}= O(n^{-2\tau/3}).
\end{align*}
In the interval $(z_0-\varepsilon, z_0+\varepsilon)$ 
\begin{align*}
\begin{split}\mathcal{I}_3|_{(z_0-\varepsilon, z_0+\varepsilon)} = s\int_{z_0}^{z_0+\varepsilon} \chi_K(z)\bar{\sigma}(sz-s)\sigma_0(z)^{-1}\left[\tilde{\Xi}_1(z)\Phi_{\Ai}^{cc}(s\mu(z))^{-1}\left(\frac{z-z_0}{\mu(z)}\right)^{-\sigma_3/4}\e^{\frac{t\zeta_0}{2}\sigma_3}O(n^{-2\tau/3}s^{-2}) \right. \\
\left.\e^{-\frac{t\zeta_0}{2}\sigma_3}\left(\frac{z-z_0}{\mu(z)}\right)^{\sigma_3/4}\Phi_{\Ai}^{cc}(s\mu(z))\right]_{21} \dd z + O(n^{-2\tau/3}) = O(n^{-2\tau/3}),\end{split}
\end{align*}
where $\tilde{\Xi}_1(z) = I +(1+O(.))\chi_{(-\infty, z_0)}(z)E_{21}$. At last, for $z \in (-\infty, z_0-\varepsilon)\cap K$, $2g(z)+V(z_0)-V(z)$ is purely imaginary, therefore
\begin{align*}
\mathcal{I}_3|_{(-\infty, z_0-\varepsilon)} =& s\int_{-\infty}^{z_0-\varepsilon}\chi_K(z) \bar{\sigma}(sz-s)\sigma_0(z)^{-1}\e^{-s^{3/2}(2g(z)+V(z_0)-V(z))}\left[\Xi_1(z)^{-1}U_0(z-z_0)^{-\sigma_3/4}R_{cc}^{-1}(z)s^{-\sigma_3/4}\right.\\
& \hspace{2cm} \left.\e^{\frac{t\zeta_0}{2}\sigma_3}O(n^{-2\tau/3}s^{-2})\e^{-\frac{t\zeta_0}{2}\sigma_3}s^{\sigma_3/4}R_{cc}(z)(z-z_0 )^{\sigma_3/4} U_0^{-1}\Xi_1(z)\right]_{21} \dd z\\
|\mathcal{I}_3|\leq & M s^{-1/2}n^{-2\tau/3} \left[\left|\int_{-\infty}^{0}\chi_K(z) \frac{1}{1+\e^{-zst}}(z-z_0)^{1/2} \dd z \right|+ \left|\int_{0}^{z_0-\varepsilon}\chi_K(z) \frac{1}{1+\e^{-zst}}(z-z_0)^{1/2} \dd z \right|\right]\\
&\leq M s^{-1/2}n^{-2\tau/3}\left[\int_{-\infty}^{0}\chi_K(z) \e^{zst}|z-z_0|^{1/2} \dd z + \int_{0}^{z_0-\varepsilon}\chi_K(z) |z-z_0|^{1/2} \dd z\right] = O(s^{-1/2}n^{-2\tau/3}),
\end{align*}
and the result follows. $\square$\\

\begin{lemma}
Let $t_0>0$ and $\alpha$ under Assumption \ref{asump1}. Then
\begin{align*}
\mathcal{I}_{21} = f(\zeta)\int_{\mathcal{K}} \bar{\sigma}_{\kappa}(\zeta)\left[\Xi(\zeta)^{-1}\Psi_{cc}(\zeta)^{-1}\tilde{\Delta}(\zeta)^{-1}
\sigma_3 \tilde{\Delta}(\zeta)\Psi_{cc}(\zeta)\Xi(\zeta)\right]_{21} \frac{\dd \zeta}{n^{2/3}\varphi'(\lambda)} = O(x^3n^{-2/3})
\end{align*}
uniformly for $x = x_0 n^{\alpha}$ and $t \in [t_0, 1/t_0]$.
\end{lemma}
\textbf{Proof:} Take the change in variables $z= \zeta/s+1$, for $s = x/t$. Notice that 
\begin{align*}
f(z) = -\frac{1}{a}+O\left(\frac{s}{n^{2/3}}(z-1)\right).
\end{align*}
In the new variable,
\begin{align*}
\mathcal{I}_{21} =& s\int_K f(z)\bar{\sigma}(sz-s)\e^{-s^{3/2}(2g(z)+V(z_0))}\left[\Xi_1(z)^{-1}\hat{S}(z)^{-1}s^{-\sigma_3/4}\e^{\frac{t\zeta_0}{2}\sigma_3}\Delta(\zeta)^{-1}\e^{-\frac{t\zeta_0}{2}\sigma_3}\E \right. \\
& \left. \sigma_3 \E^{-1}\e^{\frac{t\zeta_0}{2}\sigma_3}\Delta(\zeta)\e^{-\frac{t\zeta_0}{2}\sigma_3}s^{\sigma_3/4}\hat{S}(z)\Xi_1(z)\right]_{21} \frac{\dd z}{n^{2/3}\varphi'(\lambda)},
\end{align*}
where $\Xi_1(z) = \e^{-s^{3/2}(g(z)-V(z)/2+V(z_0)/2) \sigma_3} \tilde{\Xi}_1(z)\e^{s^{3/2}(g(z)-V(z)/2+V(z_0)/2) \sigma_3}$. For $z \in (z_0+\varepsilon, \infty)\cap K$, there exist $M>0$ such that
\begin{align*}
\mathcal{I}_{21}|_{(z_0+\varepsilon, \infty)} =& s\int_{z_0+\varepsilon}^{\infty}f(z)\chi_K(z) \bar{\sigma}(sz-s)\e^{-s^{3/2}(2g(z)+V(z_0))}\left[U_0(z-z_0)^{-\sigma_3/4}R_{cc}^{-1}(z)s^{-\sigma_3/4}\e^{\frac{t\zeta_0}{2}\sigma_3}\Delta(\zeta)^{-1}\e^{-\frac{t\zeta_0}{2}\sigma_3} \right. \\
& \left.\hspace{2cm}\E \sigma_3 \E^{-1}\e^{\frac{t\zeta_0}{2}\sigma_3}\Delta(\zeta)\e^{-\frac{t\zeta_0}{2}\sigma_3}s^{\sigma_3/4}R_{cc}(z)(z-z_0 )^{\sigma_3/4} U_0^{-1}\right]_{21} \dd z\\
|\mathcal{I}_{21}|_{(z_0+\varepsilon, \infty)}|\leq& M s n^{-2/3}\left|\int_{z_0+\varepsilon}^{\infty} s^{5/2}(z-z_0)^{1/2}\e^{-\frac{4}{3}s^{3/2}(z-z_0)^{3/2}} \dd z\right| \leq  M s^{3/2}n^{-2/3}\e^{-\frac{4}{3}s^{3/2} \varepsilon^{3/2}}= O(n^{-2/3}).
\end{align*}
In the interval $(z_0-\varepsilon, z_0+\varepsilon)$ 
\begin{align*}
\mathcal{I}_{21}|_{(z_0-\varepsilon, z_0+\varepsilon)} =& s\int_{z_0-\varepsilon}^{z_0+\varepsilon} f(z) \chi_K(z)\bar{\sigma}(sz-s)\sigma_0(z)^{-1}\left[\tilde{\Xi}_1(z)^{-1}\Phi_{\Ai}^{cc}(s\mu(z))^{-1}\left(\frac{z-z_0}{s\mu(z)}\right)^{-\sigma_3/4}R_{cc}(z)^{-1}\right.\\
&\left.s^{-\sigma_3/4}\e^{\frac{t\zeta_0}{2}\sigma_3}\Delta(\zeta)^{-1}\e^{-\frac{t\zeta_0}{2}\sigma_3}\E
\sigma_3 \E^{-1}\e^{\frac{t\zeta_0}{2}\sigma_3}\Delta(\zeta)\e^{-\frac{t\zeta_0}{2}\sigma_3}s^{\sigma_3/4}R_{cc}(z)\left(\frac{z-z_0}{s\mu(z)}\right)^{\sigma_3/4}\right.\\
&\left.\Phi_{\Ai}^{cc}(s\mu(z))\tilde{\Xi}_1(z)\right]_{21} \frac{\dd z}{n^{2/3} \varphi'(\lambda)}\\
=& O(s^{3}n^{-2/3}),
\end{align*}
where $\tilde{\Xi}_1(z) = I +(1+O(.))\chi_{(-\infty, z_0)}(z)E_{21}$. At last, for $z \in (-\infty, z_0-\varepsilon)\cap K$,
\begin{align*}
\mathcal{I}_{21}|_{(-\infty, z_0-\varepsilon)} =& s\int_{-\infty}^{z_0-\varepsilon}\chi_K(z) \bar{\sigma}(sz-s)\sigma_0(z)^{-1}\e^{-s^{3/2}(2g(z)+V(z_0)-V(z))}\left[\Xi_1(z)^{-1}U_0(z-z_0)^{-\sigma_3/4}R_{cc}^{-1}(z)s^{-\sigma_3/4}\right.\\
&\left.\e^{\frac{t\zeta_0}{2}\sigma_3}\Delta(\zeta)^{-1}\e^{-\frac{t\zeta_0}{2}\sigma_3}\E \sigma_3 \E^{-1}\e^{\frac{t\zeta_0}{2}\sigma_3}\Delta(\zeta)\e^{-\frac{t\zeta_0}{2}\sigma_3}s^{\sigma_3/4}R_{cc}(z)(z-z_0 )^{\sigma_3/4} U_0^{-1}\Xi_1(z)\right]_{21} \frac{\dd z}{n^{2/3}\varphi'(\lambda)},
\end{align*}
The term inside parenthesis has order $s^{3/2}(z-z_0)^{1/2}$. Moreover, $2g(z)+V(z_0)-V(z)$ is purely imaginary in this interval. Consequently, we have that for some constant $M>0$,
\begin{align*}
|\mathcal{I}_{21}|_{(-\infty, z_0-\varepsilon)}| \leq M s^{5/2}n^{-2/3}\left[\left|\int_{-\infty}^{0} \frac{\chi_K(z)}{1+\e^{-zst}}(z-z_0)^{1/2} \dd z \right|+ \left|\int_{0}^{z_0-\varepsilon}\frac{\chi_K(z)}{1+\e^{-zst}}(z-z_0)^{1/2} \dd z \right|\right]\\
\leq M s^{5/2}n^{-2/3}\left[\int_{-\infty}^{0}\chi_K(z) \e^{zst}|z-z_0|^{1/2} \dd z + \int_{0}^{z_0-\varepsilon}\chi_K(z) |z-z_0|^{1/2} \dd z\right] = O(s^{5/2}n^{-2/3}),
\end{align*}
and the result follows. $\square$\\

\begin{lemma}
Let $t_0>0$ and $\alpha$ under Assumption \ref{asump1}. Then
\begin{align*}
\mathcal{I}_{22} =& \int_{\mathcal{K}} \bar{\sigma}_{\kappa}(\zeta) \left[\Xi(\zeta)^{-1}\Psi_{cc}(\zeta)^{-1}\tilde{\Delta}(\zeta)^{-1}
\zeta^{\sigma_3/4}\left(\frac{\lambda}{\lambda+a}\right)^{-\sigma_3/4}U_0^{-1}\R_+^{-1}(\lambda)\R'(\lambda) \right. \\
& \left.U_0\left(\frac{\lambda}{\lambda+a}\right)^{\sigma_3/4}\zeta^{-\sigma_3/4} \tilde{\Delta}(\zeta)\Psi_{cc}(\zeta)\Xi(\zeta)\right]_{21} \frac{\dd \zeta}{n^{2/3}\varphi'(\lambda)} = O(n^{-1-\gamma}x^{9/2})
\end{align*}
uniformly for $x = x_0 n^{\alpha}$ and $t \in [t_0, 1/t_0]$.
\end{lemma}

\textbf{Proof:} First, notice that after the changes in variables $\lambda \mapsto \zeta = n^{2/3} \varphi(\lambda) \mapsto z = \zeta/s+1$ for $s = x/t$,
\begin{align*}
\E \zeta ^{\sigma_3/4}\left(\frac{\lambda}{\lambda+a}\right)^{-\sigma_3/4}U_0^{-1}\R_+^{-1}(\lambda)\R'(\lambda)U_0\left(\frac{\lambda}{\lambda+a}\right)^{\sigma_3/4}\zeta^{-\sigma_3/4} \E^{-1} = \begin{pmatrix}
O(n^{-\frac{1}{3}-\gamma}s^2) & O(n^{-\frac{1}{3}-\gamma}s^4)\\
O(n^{-\frac{1}{3}-\gamma}) & O(n^{-\frac{1}{3}-\gamma}s^2)
\end{pmatrix}.
\end{align*}
Thus, for $K=[-{\delta}_2n^{\alpha}/s+1, \delta_3n^{2\alpha/3}/s+1]$,
\begin{align*}
\mathcal{I}_{22} =& s\int_K \bar{\sigma}(sz-s)\e^{-s^{3/2}(2g(z)+V(z_0))}\left[\Xi_1(z)^{-1}\hat{S}(z)^{-1}s^{-\sigma_3/4}\e^{\frac{t\zeta_0}{2}\sigma_3}\Delta(\zeta)^{-1}\e^{-\frac{t\zeta_0}{2}\sigma_3}\right.\\
&\left.\begin{pmatrix}
O(n^{-\frac{1}{3}-\gamma}s^2) & O(n^{-\frac{1}{3}-\gamma}s^4)\\
O(n^{-\frac{1}{3}-\gamma}) & O(n^{-\frac{1}{3}-\gamma}s^2)
\end{pmatrix}\e^{\frac{t\zeta_0}{2}\sigma_3}\Delta(\zeta)\e^{-\frac{t\zeta_0}{2}\sigma_3}s^{\sigma_3/4}\hat{S}(z)\Xi_1(z)\right]_{21} \frac{\dd z}{n^{2/3}\varphi'(\lambda)}.
\end{align*}
For $z \in (z_0+\varepsilon, \infty)\cap K$, a rough estimate gives that for some constant $M>0$,
\begin{align*}
\mathcal{I}_{22}|_{(z_0+\varepsilon, \infty)} =& s\int_{z_0+\varepsilon}^{\infty}\chi_K(z) \bar{\sigma}(sz-s)\e^{-s^{3/2}(2g(z)+V(z_0))}\left[U_0(z-z_0)^{-\sigma_3/4}R_{cc}^{-1}(z)s^{-\sigma_3/4}\e^{\frac{t\zeta_0}{2}\sigma_3}\Delta(\zeta)^{-1}\e^{-\frac{t\zeta_0}{2}\sigma_3}\right.\\
&\left.\begin{pmatrix}
O(n^{-\frac{1}{3}-\gamma}s^2) & O(n^{-\frac{1}{3}-\gamma}s^4)\\
O(n^{-\frac{1}{3}-\gamma}) & O(n^{-\frac{1}{3}-\gamma}s^2)
\end{pmatrix}\e^{\frac{t\zeta_0}{2}\sigma_3}\Delta(\zeta)\e^{-\frac{t\zeta_0}{2}\sigma_3}s^{\sigma_3/4}R_{cc}(z)(z-z_0 )^{\sigma_3/4} U_0^{-1}\right]_{21} \frac{\dd z}{n^{2/3}\varphi'(\lambda)}\\
|\mathcal{I}_{22}|\leq& M n^{-1-\gamma} s\left|\int_{z_0+\varepsilon}^{\infty} s^{9/2}(z-z_0)^{1/2}\e^{-\frac{4}{3}s^{3/2}(z-z_0)^{3/2}} \dd z\right| \leq M n^{-1-\gamma} s^4 \e^{-\frac{4}{3}s^{3/2} \varepsilon^{3/2}}=O(n^{-1-\gamma}).
\end{align*}
In the interval $(z_0-\varepsilon, z_0+\varepsilon)$ 
\begin{align*}
\mathcal{I}_{22}|_{(z_0-\varepsilon, z_0+\varepsilon)} =& s\int_{z_0-\varepsilon}^{z_0+\varepsilon} \chi_K(z)\bar{\sigma}(sz-s)\sigma_0(z)^{-1}\left[\tilde{\Xi}_1(z)^{-1}\Phi_{\Ai}^{cc}(s\mu(z))^{-1}\left(\frac{z-z_0}{s\mu(z)}\right)^{-\sigma_3/4}R_{cc}^{-1}(z)s^{-\sigma_3/4}\right.\\
&\left.\e^{\frac{t\zeta_0}{2}\sigma_3}\Delta(\zeta)^{-1}\e^{-\frac{t\zeta_0}{2}\sigma_3}\begin{pmatrix}
O(n^{-\frac{1}{3}-\gamma}s^2) & O(n^{-\frac{1}{3}-\gamma}s^4)\\
O(n^{-\frac{1}{3}-\gamma}) & O(n^{-\frac{1}{3}-\gamma}s^2)
\end{pmatrix}\e^{\frac{t\zeta_0}{2}\sigma_3}\Delta(\zeta)\e^{-\frac{t\zeta_0}{2}\sigma_3}s^{\sigma_3/4}R_{cc}(z)\right.\\
&\left.\left(\frac{z-z_0}{s\mu(z)}\right)^{\sigma_3/4}\Phi_{\Ai}^{cc}(s\mu(z))\tilde{\Xi}_1(z)\right]_{21} \frac{\dd z}{n^{2/3} \varphi'(\lambda)}\\
=& s\int_{z_0-\varepsilon}^{z_0+\varepsilon} \chi_K(z)\bar{\sigma}(sz-s)\sigma_0(z)^{-1}\left[\tilde{\Xi}_1(z)^{-1}\Phi_{\Ai}^{cc}(s\mu(z))^{-1}\left(\frac{z-z_0}{\mu(z)}\right)^{-\sigma_3/4}\right.\\
&\left.\begin{pmatrix}
O(n^{-\frac{1}{3}-\gamma}s^2) & O(n^{-\frac{1}{3}-\gamma}s^4)\\
O(n^{-\frac{1}{3}-\gamma}) & O(n^{-\frac{1}{3}-\gamma}s^2)
\end{pmatrix}\left(\frac{z-z_0}{\mu(z)}\right)^{\sigma_3/4}\Phi_{\Ai}^{cc}(s\mu(z))\tilde{\Xi}_1(z)\right]_{21} \frac{\dd z}{n^{2/3} \varphi'(\lambda)}+O(n^{-\gamma}).
\end{align*}
The explicit evaluation of the term within parentheses gives 
\begin{align*}
\mathcal{I}_{22}|_{(z_0-\varepsilon, z_0+\varepsilon)} = & O(n^{-\frac{1}{3}-\gamma}s^5)\int_{z_0-\varepsilon}^{z_0+\varepsilon} \chi_K(z)\bar{\sigma}(sz-s)\sigma_0(z)^{-1}\left(\frac{z-z_0}{\mu(z)}\right)^{-1/2}\\
& \hspace{2cm} 2\pi\left[\Ai(s \mu(z))^2 +w^2\Ai(s \mu(z))\Ai(w^2 s \mu(z))O(.)\right] \frac{\dd z}{n^{2/3} \varphi'(\lambda)}+O(n^{-\gamma}), 
\end{align*}
where $O(.) = O(n^{2/3(1-2\alpha)})$ for $0<\alpha<4/21$ and $O(n^{2/3(2-3\alpha)})$ for $4/21 \leq \alpha<2/9$. Set $\delta>0$ such that $|z-z_0| \leq \delta/s$ implies $s \mu(z)$ bounded. Then, 
\begin{align*}
\mathcal{I}_{22}|_{(z_0-\delta/s, z_0+\delta/s)} \leq & n^{-1-\gamma}s^5 M\int_{z_0-\delta/s}^{z_0+\delta/s} \dd z = O(n^{-1-\gamma}s^4).
\end{align*}
On the complementary set, as $s$ grows, $\Ai(s \mu(z))\mu(z)^{1/4} \sim s^{-1/4}\e^{-3(s \mu(z))^{3/2}/2}$, and
\begin{align*}
\mathcal{I}_{22}|_{(z_0-\varepsilon, z_0-\delta/s) \cup (z_0+\delta/s, z_0+\varepsilon)} \leq & n^{-1-\gamma}s^{9/2} M\int \left(z-z_0\right)^{-1/2} \e^{-3(s \mu(z))^{3/2}} \dd z = O(n^{-1-\gamma}s^{9/2}).
\end{align*}
At last, for $z \in (-\infty, z_0-\varepsilon)\cap K$
\begin{align*}
\mathcal{I}_{22}|_{(-\infty, z_0-\varepsilon)} =& s\int_{-\infty}^{z_0-\varepsilon}\chi_K(z) \bar{\sigma}(sz-s)\sigma_0(z)^{-1}\e^{-s^{3/2}(2g(z)+V(z_0)-V(z))}\left[\Xi_1(z)^{-1}U_0(z-z_0)^{-\sigma_3/4}R_{cc}^{-1}(z)\right.\\
&\left.s^{-\sigma_3/4}\e^{\frac{t\zeta_0}{2}\sigma_3}\Delta(\zeta)^{-1}\e^{-\frac{t\zeta_0}{2}\sigma_3}\begin{pmatrix}
O(n^{-\frac{1}{3}-\gamma}s^2) & O(n^{-\frac{1}{3}-\gamma}s^4)\\
O(n^{-\frac{1}{3}-\gamma}) & O(n^{-\frac{1}{3}-\gamma}s^2)
\end{pmatrix}\e^{\frac{t\zeta_0}{2}\sigma_3}\Delta(\zeta)\e^{-\frac{t\zeta_0}{2}\sigma_3}s^{\sigma_3/4}\right.\\
&\left.R_{cc}(z)(z-z_0 )^{\sigma_3/4} U_0^{-1}\Xi_1(z)\right]_{21} \frac{\dd z}{n^{2/3}\varphi'(\lambda)},
\end{align*}
The term inside parenthesis has order $s^{7/2}n^{-\frac{1}{3}-\gamma}(z-z_0)^{1/2}$. Moreover, $2g(z)+V(z_0)-V(z)$ is purely imaginary in this interval, and, consequently, we have that for some constant $M>0$,
\begin{align*}
|\mathcal{I}_{22}|_{(-\infty, z_0-\varepsilon)}| \leq M s^{9/2}n^{-1-\gamma
}\left[\left|\int_{-\infty}^{0}\chi_K(z) \frac{1}{1+\e^{-zst}}(z-z_0)^{1/2} \dd z \right|+ \left|\int_{0}^{z_0-\varepsilon}\chi_K(z) \frac{1}{1+\e^{-zst}}(z-z_0)^{1/2} \dd z \right|\right]\\
\leq M s^{9/2}n^{-1-\gamma
}\left[\int_{-\infty}^{0}\chi_K(z) \e^{zst}|z-z_0|^{1/2} \dd z + \int_{0}^{z_0-\varepsilon}\chi_K(z) |z-z_0|^{1/2} \dd z\right] = O(s^{9/2}n^{-1-\gamma}) = O(n^{-\gamma}),
\end{align*}
and the result follows. $\square$\\

\subsection{Proof of Theorem \ref{theo3}}

The chain of transformations
$$\Y^{(1)} = \tilde{\Y}^{(1)} = \T^{(1)}\e^{-2n\ell_V} = \S^{(1)}\e^{-2n\ell_V} = (\R^{(1)}+\G^{(1)})\e^{-2n\ell_V},
$$
together with Equation \eqref{defgamma} give
$$\upgamma^{(n)}_{n-1}(x)^2 = -\frac{\e^{-2n\ell_V}}{2\pi \i}[\R^{(1)}+\G^{(1)}]_{21}.
$$
By the construction of the global parametrix we have $[\G^{(1)}]_{21} = -\frac{\i a}{4}\e^{2\q_0}$. The term of order $n^{-\gamma}$ for $\gamma<\frac{1}{3}$ comes from the difference between $\G$ and $\M$. More precisely,
\begin{align*}
J_{\R}(s)-I =& -n^{-\gamma} \frac{2\sqrt{t}x_0^{3/2}}{3\pi\sqrt{a}}\left[\sigma_3-\left(\frac{s}{s-a}\right)^{-1/2}\M(z) \sigma_3 \M^{-1}(z)\right]+ O(n^{-\gamma-\beta}).
\end{align*}
Combined with Theorem \ref{lemma:Rint} this expression gives 
\begin{align*}
\R^{(1)} =& -\frac{1}{2\pi \i} \int_{\partial \mathcal{U}^0 \cup \partial B_{\delta}(-a)} (J_{\R}(s)-I) \dd s + O(n^{-2\gamma})\\
=& \frac{1}{2\pi \i}\frac{2 \sqrt{t}x_0^{3/2}}{3 \pi \sqrt{a}} \int_{\partial \mathcal{U}^0} \sigma_3 - \left(\frac{s}{s-a}\right)^{-1/2}\M(s) \sigma_3 \M^{-1}(s) \dd s + O(n^{-\gamma-\beta}),
\end{align*}
where for the last equality we used the analyticity of the integrand in $B_{\delta}(-a)$. Consequently,
$$[\R^{(1)}]_{21} = n^{-\gamma}\frac{\i\sqrt{at}x_0^{3/2}}{3\pi}+ O(n^{-\gamma-\beta}).
$$
Expanding $[\G^{(1)}]_{21}$ and summing the results, we see that the term of order $n^{-\gamma}$ cancels out. We then look at the second leading term, which comes either from $\g$ is $\frac{1}{9} < \alpha < \frac{2}{9}$ or from the asymptotic expansion for the local parametrix if $0 < \alpha \leq \frac{1}{9}$. In the first case, from Corollary \ref{corol:expMG} we can obtain the sub-leading term of order $n^{-2\gamma}$ gives contribution
$-\frac{2\i tx_0^{3}}{9\pi^2}$, which is cancelled by the expansion of $[\G^{(1)}]_{21}$ once again. Consequently, in both cases we are left with the contribution from the local parametrix, of order $n^{-1/3}$. From Section \ref{sec:param0}, such contribution is given by
\begin{align*}
-\frac{1}{2\pi\i} \int_{\partial \mathcal{U}^0} \left(\frac{s}{s-a}\right)^{-1/2}\frac{\tilde{\Psi}_{n,21}^{(1)}}{2\varphi(s)^{1/2}}\dd s = \frac{\tilde{\Psi}_{n,21}^{(1)}a}{2c_V^{1/2}},
\end{align*}
and we finally obtain
$$\upgamma^{(n)}_{n-1}(x)^2 = \e^{-2n\ell_V} \left(\frac{a}{8\pi}-\frac{\tilde{\Psi}_{n,21}^{(1)}a}{4\pi\i c_V^{1/2}}+O(n^{-\frac{1}{3}-\frac{\alpha}{2}})\right),
$$
and the result follows from Remark \ref{rmk:psicc}.

\bibliographystyle{siam} 
\bibliography{bibliography}

\begin{thebibliography}{10}

\bibitem{guionnet2009}
{\sc G.~W. Anderson, A.~Guionnet, and O.~Zeitouni}, {\em An introduction to
  random matrices}, no.~118 in Cambridge Studies in Advanced Mathematics,
  Cambridge University Press, 2010.

\bibitem{BB2019}
{\sc D.~Betea and J.~Bouttier}, {\em The periodic {S}chur process and free
  fermions at finite temperature}, Math. Phys. Anal. Geom., 22 (2019),
  pp.~Paper No. 3, 47.

\bibitem{BG2016}
{\sc A.~Borodin and V.~Gorin}, {\em Moments match between the {KPZ} equation
  and the {A}iry point process}, SIGMA Symmetry Integrability Geom. Methods
  Appl., 12 (2016), pp.~Paper No. 102, 7.

\bibitem{BCT2022}
{\sc T.~Bothner, M.~Cafasso, and S.~Tarricone}, {\em Momenta spacing
  distributions in anharmonic oscillators and the higher order finite
  temperature {A}iry kernel}, Ann. Inst. Henri Poincar\'{e} Probab. Stat., 58
  (2022), pp.~1505--1546.

\bibitem{Brezin1990}
{\sc E.~Br\'ezin and V.~A. Kazakov}, {\em Exactly solvable field theories of
  closed strings}, Phys. Lett. B, 236 (1990), pp.~144--150.

\bibitem{CC2019}
{\sc M.~Cafasso and T.~Claeys}, {\em A {R}iemann-{H}ilbert approach to the
  lower tail of the {K}ardar-{P}arisi-{Z}hang equation}, Comm. Pure Appl.
  Math., 75 (2022), pp.~493--540.

\bibitem{CCR2021}
{\sc M.~Cafasso, T.~Claeys, and G.~Ruzza}, {\em Airy kernel determinant
  solutions to the kdv equation and integro-differential painlevé equations},
  Comm. Math. Phys., 386 (2021), pp.~1107--1153.

\bibitem{CS2025}
{\sc M.~Cafasso and C.~M. da~Silva~Pinheiro}, {\em Unitary ensembles with a
  critical edge point, their multiplicative statistics and the
  korteweg-de-vries hierarchy}, 2025.

\bibitem{CG2023}
{\sc T.~Claeys and G.~Glesner}, {\em Determinantal point processes conditioned
  on randomly incomplete configurations}, Ann. Inst. Henri Poincar\'e{} Probab.
  Stat., 59 (2023), pp.~2189--2219.

\bibitem{CV2007}
{\sc T.~Claeys and M.~Vanlessen}, {\em Universality of a double scaling limit
  near singular edge points in random matrix models}, Comm. Math. Phys., 273
  (2007), pp.~499--532.

\bibitem{C2012}
{\sc I.~Corwin}, {\em The {K}ardar-{P}arisi-{Z}hang equation and universality
  class}, Random Matrices Theory Appl., 1 (2012), pp.~1130001, 76.

\bibitem{deift1999}
{\sc P.~Deift, T.~Kriecherbauer, K.~T.-R. McLaughlin, S.~Venakides, and
  X.~Zhou}, {\em Strong asymptotics of orthogonal polynomials with respect to
  exponential weights}, Comm. Pure Appl. Math., 52 (1999), pp.~1491--1552.

\bibitem{deift1999b}
{\sc P.~Deift, T.~Kriecherbauer, K.~T.-R. McLaughlin, S.~Venakides, and
  X.~Zhou}, {\em Uniform asymptotics for polynomials orthogonal with respect to
  varying exponential weights and applications to universality questions in
  random matrix theory}, Comm. Pure Appl. Math., 52 (1999), pp.~1335--1425.

\bibitem{fokas1992}
{\sc A.~S. Fokas, A.~R. Its, and A.~V. Kitaev}, {\em The isomonodromy approach
  to matrix models in {$2$}{D} quantum gravity}, Comm. Math. Phys., 147 (1992),
  pp.~395--430.

\bibitem{forrester92}
{\sc P.~J. Forrester}, {\em The spectrum edge of random matrix ensembles},
  Nuclear Phys. B, 402 (1993), pp.~709--728.

\bibitem{GG21}
{\sc P.~Ghosal and G.~L.~F. Silva}, {\em Universality for multiplicative
  statistics of {H}ermitian random matrices and the integro-differential
  {P}ainlev\'e{II} equation}, Comm. Math. Phys., 397 (2023), pp.~1237--1307.

\bibitem{takeuchi2011}
{\sc T.~KA, S.~M, S.~T, and S.~H.}, {\em Growing interfaces uncover universal
  fluctuations behind scale invariance}, Sci Rep.,  (2011).

\bibitem{KPZ1986}
{\sc M.~Kardar, G.~Parisi, and Y.-C. Zhang}, {\em Dynamic scaling of growing
  interfaces}, Phys. Rev. Lett., 56 (1986), pp.~889--892.

\bibitem{DMS2018}
{\sc P.~Le~Doussal, S.~N. Majumdar, and G.~Schehr}, {\em Multicritical edge
  statistics for the momenta of fermions in nonharmonic traps}, Phys. Rev.
  Lett., 121 (2018), p.~030603.

\bibitem{lubinsky2009}
{\sc D.~S. Lubinsky}, {\em A new approach to universality limits involving
  orthogonal polynomials}, Ann. of Math. (2), 170 (2009), pp.~915--939.

\bibitem{mehta2004}
{\sc M.~L. Mehta}, {\em Random matrices}, Elsevier, San Diego, 2004 - Third
  Edition.

\bibitem{Saff_book}
{\sc E.~B. Saff and V.~Totik}, {\em Logarithmic potentials with external
  fields}, vol.~316 of Fundamental Principles of Mathematical Sciences,
  Springer-Verlag, Berlin, 1997.
\newblock Appendix B by Thomas Bloom.

\bibitem{TW1994}
{\sc C.~A. Tracy and H.~Widom}, {\em Level spacing distributions and the
  {B}essel kernel}, Comm. Math. Phys., 161 (1994), pp.~289--309.

\end{thebibliography}


	
	

\end{document}